\documentclass[aps, twocolumn, pre, floatfix, superscriptaddress,longbibliography]{revtex4-1}

\usepackage{amsmath,amssymb,amsthm}
\usepackage{graphicx}

\newcommand{\bra}[1]{\ensuremath{\left\langle#1\right|}}
\newcommand{\ket}[1]{\ensuremath{\left|#1\right\rangle}}
\newcommand{\bracket}[1]{\ensuremath{\left\langle #1 \right\rangle}}

\newcommand{\var}{\operatorname{var}}
\newcommand{\cov}{\operatorname{cov}}

\newcommand{\tr}{\operatorname{tr}}

\graphicspath{{.}{figures/}}

\usepackage{color}

\definecolor{darkblue}{rgb}{0,0,.65}
\definecolor{darkgreen}{rgb}{0.3,0.6,0.3}
\definecolor{cyan1}{rgb}{0.0, 0.6, 0.6}
\usepackage[%
pdfstartview=FitH,%
breaklinks=true,%
bookmarks=true,%
colorlinks=true,%
anchorcolor=black,%
citecolor=red,
filecolor=black,%
menucolor=black,%
urlcolor=darkblue,%
linkcolor=blue,%
]{hyperref}

\usepackage{cleveref}

\begin{document}

\title{Eigenstate thermalization scaling in approaching  the classical limit}

\begin{abstract}

  According to the eigenstate thermalization hypothesis (ETH), the eigenstate-to-eigenstate
  fluctuations of expectation values of local observables should decrease with increasing system
  size.  In approaching the thermodynamic limit --- the number of sites and the particle number
  increasing at the same rate --- the fluctuations should scale as $\sim D^{-1/2}$ with the Hilbert
  space dimension $D$.  Here, we study a different limit --- the classical or semiclassical limit
  --- by increasing the particle number in fixed lattice topologies.  We focus on the paradigmatic
  Bose-Hubbard system, which is quantum-chaotic for large lattices and shows mixed behavior for
  small lattices.  We derive expressions for the expected scaling, assuming ideal eigenstates having
  Gaussian-distributed random components.  We show numerically that, for larger lattices, ETH
  scaling of physical mid-spectrum eigenstates follows the ideal (Gaussian) expectation, but 
  for smaller lattices, the scaling occurs via a different exponent.  We examine several plausible
  mechanisms for this anomalous scaling.
  
\end{abstract}

\newcommand{\maynoothTP}{Department of Theoretical Physics, Maynooth University, Co.\ Kildare, Ireland}

\author{Goran Nakerst}
\affiliation{\maynoothTP}

\author{Masudul Haque}
\affiliation{\maynoothTP}
\affiliation{Max-Planck-Institut f\"{u}r Physik komplexer Systeme, D-01187 Dresden, Germany}

\maketitle

\section{Introduction}
% historical context

During the last decade and half, a considerable amount of research has focused on understanding how
isolated quantum systems can relax and thermalize. A cornerstone of this understanding is the
Eigenstate Thermalization Hypothesis (ETH) \cite{Deutsch1991,Srednicki1994, srednicki1996thermal,
  srednicki1999approach, Rigol_Nature2008, Polkovnikov_RMP2011, Rigol_Srednicki_PRL2012_alternatives,
  Eisert2015, Dalessio2016}.  According to the ETH, the diagonal matrix elements of observables in
the eigenstate basis of the Hamiltonian, the eigenstate expectation values (EEV), coincide locally
with the micro-canonical ensemble.  This means that the EEVs vary smoothly as a function of energy
eigenvalues. The smoothness is quantified as the fluctuations of the EEVs being exponentially small
as a function of the system size.

A standard quantitative statement of ETH states that the matrix elements of an operator $A$ representing a typical physical observable should have the form \cite{srednicki1996thermal, srednicki1999approach}
\begin{equation}
\langle E_{\alpha} \vert A \vert E_{\beta}  \rangle = \delta_{\alpha\beta} f_{A}^{(1)}(\bar{E}) + e^{-S(\bar{E})/2}f_{A}^{(2)}(\bar{E},\omega) R_{\alpha\beta}
\end{equation}
where $S$ is the entropy, $\vert E_{\alpha} \rangle$ is an energy eigenstate with eigenenergy
$E_{\alpha}$, $\bar{E}=(E_{\alpha} + E_{\beta})/2$, and $\omega=E_{\beta} - E_{\alpha}$.  The
$f_{A}^{(1/2)}$ are smooth functions, and $R_{\alpha\beta}$ is a (pseudo) random variable with zero
mean and unit variance.
For systems with finite Hilbert space dimension $D$, the entropy scales as $S\sim \log D$.  Thus a
crucial aspect of ETH is the scaling of the width of the distribution of either diagonal or
off-diagonal matrix elements: when approaching the thermodynamic limit, this width falls off as
$e^{-S/2}\sim D^{-1/2}$, i.e., exponentially with system size.  This scaling can be understood using
the similarity between typical many-body eigenstates and random states \cite{Marquardt_PRE2012,
  Beugeling_scaling_PRE14, Beugeling_offdiag_PRE2015}.  This behavior contrasts sharply with
integrable systems, which do not obey ETH scaling --- the width of diagonal matrix element
distributions generally have power law decay with system size
\cite{Ziraldo_Santoro_relaxation_PRB2013, Ikeda_Ueda_PRE2013_LiebLiniger, Beugeling_scaling_PRE14,
  Alba_PRB2015, ArnabSenArnabDas_PRB16, Magan_randomfreefermions_PRL2016,
  HaqueMcClarty_SYKETH_PRB2019, Mierzejewski_Vidmar_PRL2020}, and the off-diagonal matrix element
generally has a non-gaussian distribution \cite{Beugeling_offdiag_PRE2015,
  HaqueMcClarty_SYKETH_PRB2019, LeBlond_Mallaya_Vidmar_Rigol_PRE2019}.

Evidence from a large number of numerical studies strongly suggests that ETH is satisfied for
typical states of generic nonintegrable systems and for physical observables
\cite{Rigol_Nature2008, Rigol_PRL2009, BiroliKollathLauchli_PRL10, RigolSantos_PRA10,
  SantosRigol_PRE10, Roux_PRA2010, Motohashi2011, Marquardt_PRE2012, BrandinoKonikMussardo_PRB12,
Steinigeweg_Prelovsek_PRE13,
Kim_Ikeda_Huse_PRE2014,
Beugeling_scaling_PRE14,
Sorg_Vidmar_Pollet_HeidrichMeisner_PRA2014, Steinigeweg_Gogolin_Gemmer_PRL2014,
Beugeling_offdiag_PRE2015, Fratus_Srednicki_PRE2015, Nandkishore_Huse_AnnuRev2015,
Khodja_Steinigeweg_Gemmer_PRE2015, 
Dalessio2016, Mondaini_Srednicki_Rigol_PRE2016, Chandran_Burnell_PRB2016, Luitz_Barlev_PRL2016,
Mondaini_Rigol_PRE2017_tfIsing, Sonner_JHEP2017_eth_syk, Powell_PRB2017_dimermodelETH, 
Sagawa_PRL2018_LargeDeviationETH, Dymarsky_Lashkari_Liu_PRE2018, HunterJones_Zhou_JHEP2018_eth_syk,  
HaqueMcClarty_SYKETH_PRB2019, Vidmar_HeidrichM_PRB2019_HolsteinpolaronETH, Khaymovich_Haque_McClarty_PRL2019,  Mierzejewski_Vidmar_PRL2020}.  
In particular, several studies have examined the decrease of the width of matrix element
distributions as the thermodynamic limit is approached, both for chaotic and integrable systems
\cite{Steinigeweg_Prelovsek_PRE13, Ziraldo_Santoro_relaxation_PRB2013,
  Ikeda_Ueda_PRE2013_LiebLiniger, Beugeling_scaling_PRE14, Steinigeweg_Gogolin_Gemmer_PRL2014,
  Beugeling_offdiag_PRE2015, Alba_PRB2015, ArnabSenArnabDas_PRB16, Magan_randomfreefermions_PRL2016,
  Chandran_Burnell_PRB2016, Powell_PRB2017_dimermodelETH, HunterJones_Zhou_JHEP2018_eth_syk,
  Ramazaki_Ueda_PRL2018_mostfewbody, Khemani_Laumann_Chandran_PRB2019_rydberg,
  HaqueMcClarty_SYKETH_PRB2019, Vidmar_HeidrichM_PRB2019_HolsteinpolaronETH, Khaymovich_Haque_McClarty_PRL2019,
  Mierzejewski_Vidmar_PRL2020, Goold_Rigol_PRL2020}.
For lattice systems, the thermodynamic limit involves increasing both the lattice size and the
particle number simultaneously, keeping the average density fixed.  In the present work, we will
instead explore approaching the \emph{classical limit} --- we consider  increasing particle numbers
in fixed lattice topologies.  We investigate the scaling behavior of EEV fluctuations as a function
of Hilbert space dimension, as the classical limit is approached.

We focus on bosonic systems and study the Bose-Hubbard Hamiltonian on fixed numbers of sites.  The
dynamics of these systems in the classical limit have been extensively studied, where they are
described by a discrete nonlinear Schroedinger equation.
%
%% \cite{Hallwood_Burnett_Dunningham_JModOpt2007, Cassidy_thesis2010, Kolovsky_IntJModPhys2016,
%%   Rubeni_Links_Foerster_PRA2016, Rautenberg_Gaertner_PRA2020}.
%
% (CITE? MAYBE NOT NECESSARY).
%
Because of the existence of a classical limit, Bose-Hubbard systems have also been extensively used
as a testbed for semi-classical methods \cite{Polkovnikov_semiclassics_TWA_PRA2003,
  Mahmud_Reinhardt_BHdimer_semiclassical_PRA2005, Hiller_Kottos_Geisel_PRA2006,
  Mossmann_Jung_PRA2006_semiclassical_BHtrimer,  
  Graefe_Korsch_semiclassical_PRA2007, Trimborn_Witthaut_Korsch_PRA2009,
  Polkovnikov_AnnPhys2010_PhaseSpace,
  Trippenbach_BHdimer_PRA2011, Simon_Strunz_PRA2012,
  Simon_Strunz_PRA2014, Veksler_Fishman_semiclassical_NJP2015, Engl_Urbina_Richter_PRE2015,
  Engl_Urbina_KRichter_PhilTrans2016, Grossmann_Strunz_semiclassical_JPA2016,
  Dubertrand_Mueller_NJP2016_semiclassics_spectralstat,
Tomsovic_Schlagheck_Richter_PRA2018_postEhrenfest,
Tomsovic_PRE2018_saddle_BH,  Rammensee_PRL2018, Corney_PRA2019_BHdimer, Schlagheck_PRL2019}.  Quantum dynamics of these
systems have also been compared with the dynamics of the corresponding classical limit
\cite{Milburn_Munro_PRA2000_BHtrimer, Weiss_Teichmann_PRL2008_BHdimer_quantumclassical,
  Viscondi_Furuya__JPhysA2011_BHtrimer, 
  Gertjerenken_Weiss_BHdimer_quantumclassical_PRA2013, Kolovsky_IntJModPhys2016,
  Heinisch_Holthaus_ZNforschungA2016, Rautenberg_Gaertner_PRA2020}.

We consider $N$ particles in $k$ sites, arranged linearly.  We approach the large limit of large
Hilbert space dimension ($D\to\infty$) by keeping $k$ fixed and increasing $N$, as opposed to the
usual thermodynamic limit for which the ratio $N/k$ would be kept fixed. The 2-site system
(Bose-Hubbard dimer, $k=2$) is integrable, and hence we omit this case in the present work on ETH
scaling.  We focus on lattices with sizes from $k=3$ up to $k=10$.  The case of $k=3$ (and to a
lesser extent $k=4$) is particularly interesting: although not integrable, the classical phase space
in this case is known to be highly `mixed' \cite{Mossmann_Jung_PRA2006_semiclassical_BHtrimer,
  Hiller_Kottos_Geisel_BHtrimer_PRA2009, Viscondi_Furuya__JPhysA2011_BHtrimer,
  Vardi_DoronCohen_PRA2014_BHtrimer, Han_Wu_PRA2016_BHtrimer_ehrenfest,
  Buerkle_Anglin_4siteBH_PRA2019, Kolovsky_russians_AIPConf2020, Rautenberg_Gaertner_PRA2020}, and
the behavior of the finite-$N$ quantum system shows deviations from fully chaotic behavior.  For
larger site numbers, $k\gtrsim5$, the Bose-Hubbard systems tend to behave as quantum-chaotic systems
\cite{Kolovsky_Buchleitner_EPL2004_BHchaos, Kollath_Roux_Biroli_Laeuchli_JSM2010,
  Kolovsky_IntJModPhys2016, Hirsch_PRE2020_quantumchaos_BH, Buchleitner_arxiv2020_BHchaos}.

Assuming the eigenstates to be perfectly ergodic ``infinite-temperature'' states in the sense of
having Gaussian-distributed random coefficients, an exact formula for the EEV fluctuations is
available.  This relation [Eq.\ \eqref{eq:statistical_variance_2}] only depends on trace expressions
of operators, and holds for all operators, as long as the ``Gaussian eigenstates'' approximation is
valid.  It mathematically connects the Gaussianity of eigenstates to finite size scaling of ETH, and
provides an alternate derivation of ETH scaling.

The class of observables generally considered in ETH investigations (observables of physical
interest) are few-body operators or sums thereof, i.e., `local' operators.
However, having $k$ fixed and small means that the notion of `local' operators has to be
re-examined.  When considering $k\to\infty$, two-site operators are local in the sense that
$2\ll{k}$.  This is no longer true for our constant $k$; in particular it is strongly violated for
$k=3$.  Therefore we re-examine the expected scaling of EEV fluctuations, first for Gaussian
eigenstates.  We show that the expected behavior for $N\gg{k}$ is $\sim D^{-e_0}$ with $e_0 =
\frac{1}{2}-\frac{1}{k-1}$.  The exponent ranges from $e_0=0$ for the 3-site case ($k=3$) to
$e_0\to\frac{1}{2}$ for large $k$ ($N\gg{k}\gg1$).

The scaling exponent $e_0$ being different from 1/2 for moderate $k$ can be traced to the
unboundedness of the operators in the $N\to\infty$ limit.  The $D^{-1/2}$ scaling could be
retrieved by rescaling the operators $A$ and considering operators $\bar{A}=A/N$.  

For the actual mid-spectrum eigenstates of the Bose-Hubbard chains, we show numerically that the EEV
fluctuations match the behavior predicted by the Gaussian ansatz, once $k$ is large enough.  For
small $k$, the dependence on $D$ is seen to be a power law but with a \emph{different} exponent
$e\neq{e_0}$.  This is a remarkable manifestation of the `mixed' nature of the 3-site and 4-site
Bose-Hubbard systems.

At present, we do not have a predictive explanation for the observed exponents.  Small or subleading
deviations from Gaussian behavior (from random-matrix behavior) has been noted previously in
mid-spectrum eigenstates of chaotic many-body systems.  Such deviations can be seen in the
eigenstate coefficient distribution in the configuration basis \cite{Beugeling_coefficients_PRE2018,
  Luitz_Barlev_PRL2016, Baecker_Haque_Khaymovich_PRE2019,
  Luitz_Khaymovich_BarLev_multifrac_SciPost2020}, in the entanglement entropy of eigenstates
\cite{Beugeling_entanglement_JSM2015, Balents_SYK_entanglement_PRB2018,
  LeBlond_Mallaya_Vidmar_Rigol_PRE2019, Haque_McClarty_Khaymovich_entanglement_deviation}, and in
the fluctuations of off-diagonal matrix elements \cite{Khaymovich_Haque_McClarty_PRL2019}.  Since the present $k=3$
system has strong non-chaotic features, it is not surprising that there is even stronger deviation
from Gaussian-state behavior.  Perhaps unexpectedly, EEV fluctuations show a clear power law
dependence on the Hilbert space dimension, but the exponent is markedly different from the
random-matrix prediction.  This demonstrates that eigenstates of small-$k$ Bose-Hubbard chains have
non-random correlations.  We examine the most straightforward ways in which physical eigenstates
could be different from ideally ergodic eigenstates, e.g., non-Gaussian coefficient distributions
and two-point correlations between the eigenstate coefficients.  We demonstrate that none of these
effects explain our observed scaling.  The observed scaling exponent thus results from more subtle
structures in the eigenstate coefficients, which seems challenging to characterize or identify.

% The structure of this article is as follows.

In Section \ref{sec:preliminaries} we introduce the system, notations, and the numerical procedure
for extracting EEV fluctuations $\sigma$.  Section \ref{sec:scaling_results} is the heart of the
paper and reports the main results: analytically derived trace expressions and scaling laws for
$\sigma$ in the case of idealized (Gaussian) eigenstates (\ref{subsec:random_states_derivation}),
and results of extensive numerics showing where these predictions succeed and where they fail
(\ref{subsec:scaling_numerics}).  In Section \ref{sec:non_reasons} we examine possible reasons for
the few site systems (particularly $k=3$) deviating from the Gaussian prediction and displaying
anomalous exponents in the $D$-dependence of $\sigma$.  We conclude that the anomalous scaling is
due to subtle higher-order correlations among the eigenfunction coefficients which cannot be
captured by the 2-point correlations among eigenfunction components.
Section \ref{sec:conclusion} provides context and concluding discussion.

The appendices present further details, data and clarification.  We show the dependence on the
interaction parameter and justify the interaction value used in the main text (Appendices
\ref{sec:r_stat} and \ref{sec:lambda_dependence}).  We display the structure of covariance matrices
of the physical eigenstates (\ref{sec:visualizing_covariances}).  Appendices
\ref{sec:variance_calculation} and \ref{sec:operator_scaling} contain the derivations of the
analytic results announced in the main text.

\section{Preliminaries --- Systems and Observables}  \label{sec:preliminaries}

In this section, we introduce the system Hamiltonian (\ref{subsec:model}), discuss how to
define  the fluctuations of the eigenstate expectation values (EEVs) (\ref{subsec:EEVfluc_def}),
and introduce notation to be used in this paper.

\subsection{Hamiltonians}\label{subsec:model}

We will investigate Bose-Hubbard systems restricted to open-boundary chains of length $k$, with
nearest neighbor hoppings and on-site interactions.  The Hamiltonian is
\begin{equation}\label{eq:bose_hubbard}
H = -1/2 \sum_{\langle i,j\rangle} J_{i,j} a_i^\dagger a_j + \frac{U}{2}\sum_{j=1}^k n_j(n_j-1),
\end{equation}
where $\langle i,j\rangle$ denotes summation over adjacent sites ($j=i\pm1$), $a_j^\dagger$ and
$a_j$ are the bosonic creation and annihilation operators for the $j$-th site and
$n_j=a_j^{\dagger}a_j$ is the corresponding occupation number operator. $J_{i,j} = J_{j,i}$ is the
symmetric tunneling coefficient and $U$ is the two-particle on-site interaction strength.  We choose
$J_{1,2}=1.5$ and $J_{i,j}=1$ for $i,j\ge 2$ to avoid reflection symmetry.
The particle number $N$ is conserved by the Hamiltonian.  We introduce the tuning parameter
$\Lambda=UN$.

In the limits $\Lambda\to 0$ and $\Lambda\to\infty$ the system is integrable. If $\Lambda=0$, then
the free bosonic system is solvable in terms of single-particle eigenstates, while
$\Lambda\to\infty$ effectively means $J$ can be neglected, so that Eq.~\eqref{eq:bose_hubbard}
is diagonal already. For intermediate $\Lambda$ the systems behave chaotically, i.e., the 
spectrum approximately obeys Wigner-Dyson statistics.  For the smallest chain length $k=3$, there
are stronger deviations, but nevertheless the spectral statistics near the middle of the spectrum is
close to Wigner-Dyson form (Appendix \ref{sec:r_stat}).

The Hilbert space size is $D = \binom{N+k-1}{k-1} = \binom{N+k-1}{N}$, which scales for fixed $k$ in the large $N$ limit as $D\sim N^{k-1}$. Therefore, for fixed $k$ in the large $D$ limit, $N$ scales as $N \sim D^{1/(k-1)}$.

The operators we focus on are the tunnel operator $a_2^\dagger a_1$ from site 1 to site 2 and the
number operator $n_1$ on site 1.  We have checked that the overall scaling behaviors are the same
for $a_i^\dagger a_j$ with other $i$, $j$.  Note that $a_2^\dagger a_1$ is non-hermitian. If
hermiticity is required, we will instead investigate $(a_2^{\dagger}a_1+ a_1^{\dagger}a_2)$. This change
results in an additional factor of 2 in the corresponding EEV fluctuation.

\subsection{Defining EEV fluctuations}\label{subsec:EEVfluc_def}

%%%%%%%%% FIGURE %%%%%%%%%%%%% FIGURE %%%%%%%%%%%%% FIGURE %%%%%%%%%%%%% 
\begin{figure}%[htbp]
\begin{center}
%\the\columnwidth
\includegraphics[width=\columnwidth]{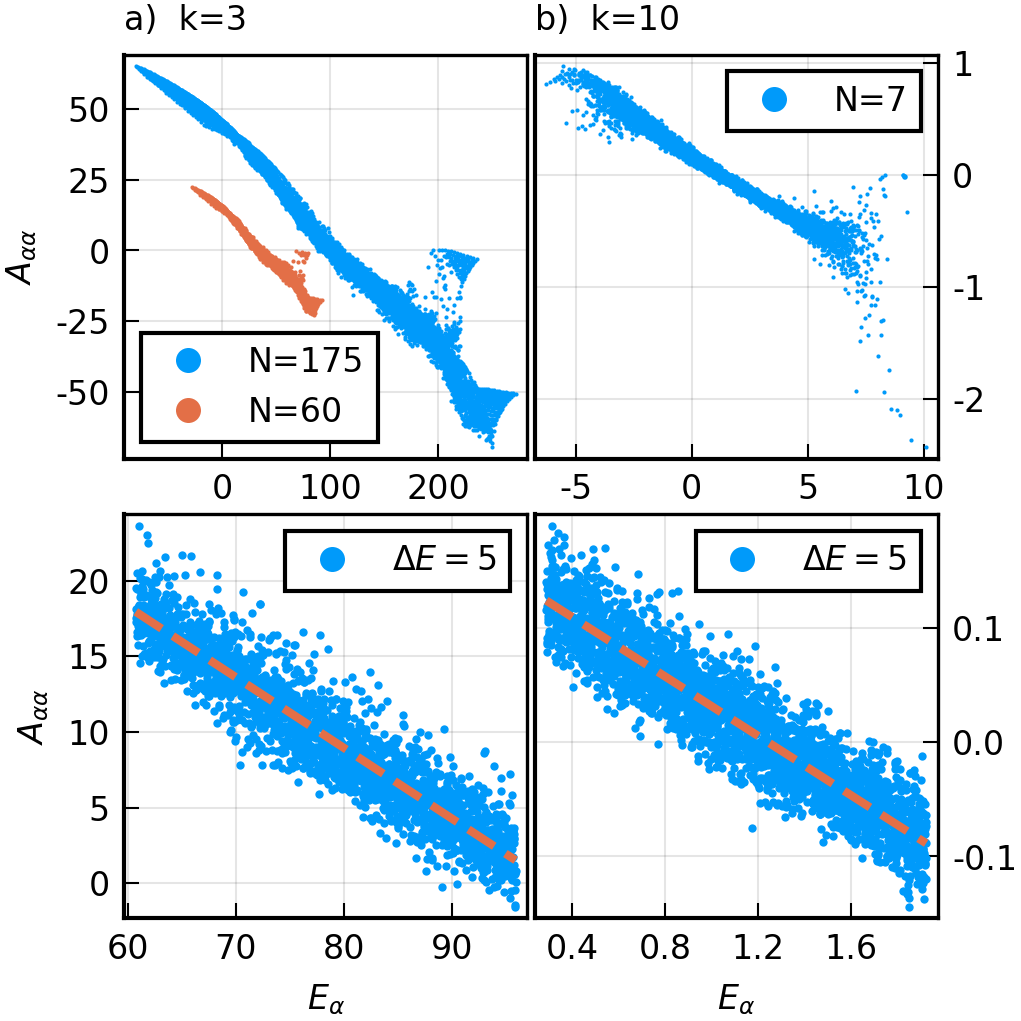}
\caption{ \label{fig:1} Eigenstate expectation values $A_{\alpha\alpha}$ of the tunnel operator
  $A=a_2^\dagger a_1$, plotted against eigenenergies, for a Bose-Hubbard chain with a) $k=3$ sites and
  b) $k=10$ sites.  The numbers of particles $N$ are listed in legends.  The interaction parameter
  is $\Lambda\approx2.477$ as explained in the text.  Top panels show full spectra.  Bottom panels
  zoom into the 5th of ten equal energy intervals, as indicated by the shorthand label
  ``$\Delta{E}=5$''.  Dotted lines are fitted linear functions.  }
\end{center}
\end{figure}
%%%%%%%%% FIGURE %%%%%%%%%%%%% FIGURE %%%%%%%%%%%%% FIGURE %%%%%%%%%%%%% 

We denote by $\ket{E_\alpha}$ the eigenstate of the Hamiltonian $H$ belonging to the eigenvalue
$E_\alpha$. If the ETH holds, $A_{\alpha\alpha}= \bra{E_\alpha} A \ket{E_\alpha}$ is constant on
small energy windows $[E_\alpha-\delta, E_\alpha+\delta]$ around $E_\alpha$ for suitable (small)
$\delta$, up to finite size fluctuations. Further, $A_{\alpha\alpha}$ coincides with the
microcanonical average
\begin{equation}
\langle A \rangle_{(E_\alpha, \delta)} 
=  \frac{1}{N_{E_\alpha, \delta E_\alpha}} \sum_{\beta:E_\beta\in [E_\alpha-\delta, E_\alpha +\delta]} A_{\beta\beta}.
\end{equation}
A quantitative measure of how well the EEVs coincide with the microcanonical average is the
magnitude of the finite size fluctuations, i.e., the statistical standard deviation of
$A_{\alpha\alpha}- \langle A \rangle_{(E_\alpha, \delta)}$,
\begin{equation}
	\left[ \frac{1}{N_{\Delta E}}\sum_{E_\alpha\text{ in }\Delta E}\left(A_{\alpha\alpha}-\langle A \rangle_{(E_\alpha, \delta)} \right)^2 \right]^{1/2}
\end{equation}
over some energy window $\Delta E$, where $N_{\Delta E}$ is the number of eigenvalues $E_\alpha$ in
$\Delta E$. Graphically this measures the width of the distribution of $A_{\alpha\alpha}$ values.

%% If the ETH holds, the quantity $A_{\alpha\alpha} = \bracket{E_\alpha\mid A\mid E_\alpha}$ will
%% vary smoothly as a function of $E_\alpha$, up to finite size fluctuations.

The above definition involves a choice of the window width $\delta$; this choice may depend on
various parameters of the physical model.  To avoid larger scale changes of $A_{\alpha\alpha}$ (e.g,
as visible in Figure \ref{fig:1}), the window should not be too wide.  The window should also not be
too small, so as to ensure that sufficiently many EEVs are within the window around $E_\alpha$ to
get a statistically significant estimate of the fluctuations.  Choosing $\delta$ can thus be a tricky
balancing act.

We can avoid these technicalities by noting that ETH implies that the quantity $A_{\alpha\alpha}$
will vary smoothly as a function of $E_\alpha$, up to finite size fluctuations.  If the ETH
holds in this sense, then, by definition of smoothness, $A_{\alpha\alpha}$ should be locally
linear. We divide the energy spectrum into 10 equal-length, disjoint intervals $\Delta E$. In these
intervals the large scale change of $A_{\alpha\alpha}$ in our case is
indeed linear to an excellent approximation, as seen in the 
lower panels of Figure \ref{fig:1}. We fit linear functions $E_\alpha \to b + m E_\alpha$, on each
interval to 
$A_{\alpha\alpha}$.  We then investigate the fluctuations around these
functions, i.e., 
\begin{equation}\label{eq:fluctuations}
\sigma^2(A, \Delta E) = \frac{1}{N_{\Delta E}} \sum_{E_{\alpha}\text{ in }\Delta E} |A_{\alpha\alpha} - b - mE_\alpha|^2.
\end{equation}

Since we are primarily interested in mid-spectrum eigenstates, we will show data from the 5th, 6th
and 7th energy intervals.  In shorthand these will be labeled as $\Delta{E}=5$, $\Delta{E}=6$,
$\Delta{E}=7$, with $\Delta{E}$ referring to the label and not the interval width.  In Figure
\ref{fig:1}, we display the EEV's for the full spectrum and for the $\Delta{E}=5$ interval, for two
different values of $k$.  Unless indicated otherwise, we present data for an intermediate value of
the interaction parameter $\Lambda = UN$ around which the systems are found to be significantly chaotic,
namely, $\Lambda = 10^{13/33} \approx 2.477$.  Appendix
\ref{sec:r_stat} provides further details on this choice of $\Lambda$.

\section{EEV fluctuation scaling}\label{sec:scaling_results}

In this section we first (\ref{subsec:random_states_derivation}) consider Gaussian-random states,
e.g. eigenstates of matrices drawn from the GOE ensemble.  We present expressions for the EEV
fluctuations for such idealized eigenstates.  For operators of the type $A=a_i^{\dagger}a_j$, we
show that the scaling of EEV fluctuations, $\sigma\sim{D}^{-e_0}$, occurs with exponent
$e_0=\frac{1}{2}-\frac{1}{k-1}$ in the classical limit ($N\to\infty$, fixed $k$).  For scaled
operators $\bar{A}=A/N$, the exponent is $e_0=\frac{1}{2}$.  In comparison, in the usual
thermodynamic limit ($N\to\infty$, fixed $k/N$), the exponent is $e_0=\frac{1}{2}$ for local
operators.

Next (\ref{subsec:scaling_numerics}), we present numerically calculated scaling results for
Bose-Hubbard chains.  We show that the EEV fluctuations have power-law dependence on the
Hilbert-space dimension, $\sim{D}^{-e}$.  The exponent  $e$ matches the Gaussian-state prediction
$e_0$ for larger chains, but  differs substantially for small $k$.

\subsection{Derivation for random Gaussian states} \label{subsec:random_states_derivation}

The EEV's calculated for random states do not have large-scale smooth variation as a function of
energy, in contrast to Figure \ref{fig:1}.  Thus, the statistical standard deviation of EEV's,
$\sigma$, can be directly compared with our measurement of EEV fluctuations in the physical
eigenstates.  The assumption of eigenstates being effectively random has been previously used to
derive scaling properties of EEV's in the thermodynamic limit \cite{Marquardt_PRE2012,
  Beugeling_scaling_PRE14, Khaymovich_Haque_McClarty_PRL2019}.  Here, we provide explicit expressions for $\sigma$ in
terms of trace properties of the operator matrix, and then specialize to both thermodynamic and
classical limits.  We present the important expressions here, and relegate derivation details to
Appendices \ref{sec:variance_calculation} and \ref{sec:operator_scaling}.

\subsubsection{Trace expressions}

Let $A$ be a $D\times D$ square matrix representing the operator of interest, and $\ket{Z}$ be a
$D$-dimensional, multivariate random state with identically and independently distributed (i.i.d.)
components $Z_i$, each with mean 0. Then the statistical variance of $\langle A\rangle = \langle
Z|A|Z\rangle$ can be expressed as
\begin{align}\label{eq:statistical_variance_1}
	\sigma^2(\langle A\rangle)
	&=\left(E[Z_1^4]-3E[Z_1^2]^2\right) \sum_i  A_{ii}^2 \nonumber\\
	&+ E[Z_1^2]^2\left[\tr(A^2) + \tr(AA^\dagger)\right].
\end{align}
Here $E[\cdot]$ represents the expectation value.  The manipulations leading to this expression can be
found in Appendix \ref{sec:variance_calculation}.

Eq.~\eqref{eq:statistical_variance_1} is valid irrespective of the distribution of the
i.i.d.\ variables $Z_i$.  Simplifications are achieved by specializing to the case of GOE
eigenstates, widely considered as reasonable models for the behavior of mid-spectrum eigenstates of
chaotic Hamiltonians.  Eigenstates of $D$-dimensional Gaussian matrices are uniformly distributed on
the $D$-dimensional unit sphere. In the large $D$ limit their components can be regarded as
independently normally distributed with mean 0 and variance $1/D$.  If we take $Z_i$ to be
normally distributed with variance $1/D$, in addition to the previous constraints, then the first
term in Eq.~\eqref{eq:statistical_variance_1} vanishes.  Thus
\begin{equation}\label{eq:statistical_variance_2}
	\sigma^2(\langle A\rangle)
	= \frac{1}{D^2}\left[\tr(A^2) + \tr(AA^\dagger)\right]
\end{equation}
for random states with Gaussian-distributed coefficients. 

The second of the two traces, $\tr(AA^\dagger)$, is the squared Hilbert-Schmidt norm or the Frobenius norm of the operator.  For
Hermitian $A$ the two trace terms are equal: $\tr(A^2)=\tr(AA^\dagger)$.
As $\tr(A^2)$ and $\tr(AA^\dagger)$ are invariant under a basis change, so is the variance
Eq.~\eqref{eq:statistical_variance_2}.  In contrast, Eq.~\eqref{eq:statistical_variance_1} is not
basis-invariant, due to the first term.
Expressions equivalent or analogous to Eq.\ \eqref{eq:statistical_variance_2} have appeared
previously in the literature, e.g., in Refs.\ \cite{Gemmer_book2009_QuantumThermo,
  Reimann_PRL2007_typicality, Reimann_JStatPhys2008_typicality, Lloyd_arxiv2013_thesischapter}.

For observables of the form $A=a_j^\dagger a_i$, the trace expression is shown in Appendix
\ref{sec:operator_scaling} to be given by
\begin{subequations} \label{eq:traces}
\begin{align}
\tr(A^2) + \tr(AA^\dagger) &= \frac{N(N+k)}{k(k+1)} D &i\neq j, 
\label{eq:traces_subeq1} \\
\tr(A^2) + \tr(AA^\dagger) &= \frac{2N(2N+k-1)}{k(k+1)} D &i=j.
\label{eq:traces_subeq2}
\end{align}
\end{subequations}
In the usual thermodynamic limit, the fraction multiplying $D$ has
$O(D^0)$ scaling in either case.  Thus one
obtains $\sigma^2\sim\frac{D}{D^2}$, i.e., the usual
$\sigma{\sim}D^{-1/2}$ scaling for EEV fluctuations in the thermodynamic limit.

\subsubsection{Scaling in the classical limit}

We now consider the classical limit, $k\ll N$.  If $A$ is a linear combination of terms like
$a_j^\dagger a_i$, the trace expression scales as
\begin{equation}\label{eq:traces_classical_limit}
	\tr(A^2) + \tr(AA^\dagger ) \sim N^2 D.
\end{equation}
Since $N\sim D^{1/(k-1)}$  in the classical limit, the variance scales as 
\begin{equation}\label{eq:statistical_variance_3}
\sigma^2(\langle A\rangle) \sim \frac{D\cdot D^{2/(k-1)}}{D^2} = D^{-2e_0},
\end{equation}
where
\begin{equation}\label{eq:exponent_e0}
  e_0=e_0(k)= \frac{1}{2}-\frac{1}{(k-1)}
\end{equation}
is the scaling exponent announced previously.  For $k\gg1$, but still $k\ll{N}$, the
second term becomes negligible and we obtain $\sigma\sim{D}^{-1/2}$ scaling similar to the 
thermodynamic limit.

For numbers of sites that are not too large, the EEV scaling of two-point operators (of the type
$a_j^\dagger a_i$ or their linear combinations) is different for the classical limit compared to
the thermodynamic limit.  Mathematically, this difference can be attributed to the operators $A$
scaling with $N$.  If we rescale $A$ to $\bar{A} = A/N$, then the traces
\eqref{eq:traces_classical_limit} scale as $D$ rather than as $N^2D$,
so that  the variance scales for all $k$ as 
\begin{equation}\label{eq:statistical_variance_4}
\sigma^2(\langle \bar{A}\rangle) \sim \frac{D}{D^2} = D^{-1}.
\end{equation}
Summarizing: in the classical limit, the EEV fluctuation scaling is ${\sim}D^{-1/2}$ for scaled operators $\bar{A} = A/N$
for all $k$, and also for un-scaled operators $A$ in the $k\gg1$ limit.  This is the same exponent
$e_0=\frac{1}{2}$ familiar from the thermodynamic limit \cite{Beugeling_scaling_PRE14,
  Khaymovich_Haque_McClarty_PRL2019}.  However, for moderate $k$ and for the operator $A$, the scaling is according to
the exponent $e_0= \frac{1}{2}-\frac{1}{(k-1)}$ of Eq.~\eqref{eq:exponent_e0}

\subsection{Numerical results: Bose-Hubbard eigenstates} \label{subsec:scaling_numerics}

%%%%%%%%% FIGURE %%%%%%%%%%%%% FIGURE %%%%%%%%%%%%% FIGURE %%%%%%%%%%%%%
\begin{figure}%[htbp]
	\begin{center}
\includegraphics[width=\columnwidth, height=\columnwidth]{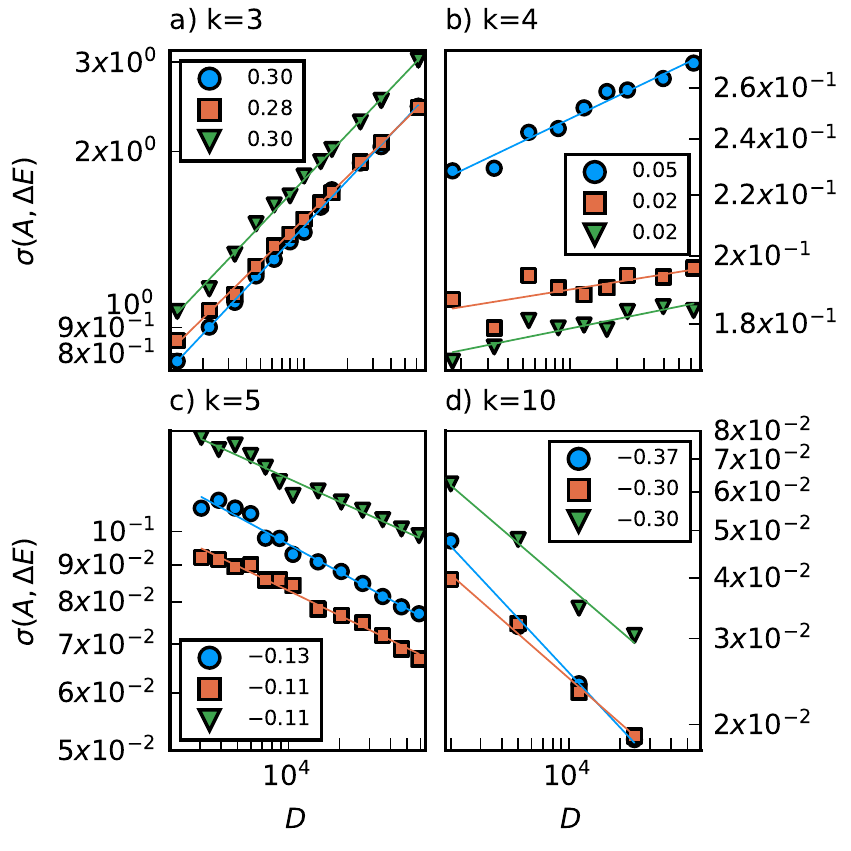}
\caption{\label{fig:2} EEV fluctuations $\sigma$ of the 
  operator $A=a_2^\dagger a_1$ against Hilbert space size $D$ for
  various chain lengths $k$.  Data shown for eigenstates in energy windows
  $\Delta E=$ 5 (blue circles), 6 (red squares), 7 (green inverted
  triangles). The $\sigma$ vs $D$ data sequences are arranged
  reasonably linearly in all cases in the log-log plots, suggesting
  $\sigma\sim{D}^{-e}$ behavior.  The slopes of fitted lines (i.e.,
  numerical estimates of $-e$) are given in the legends.  The Gaussian
  predictions $e_0$ for the exponents are $0$,
  $\frac{1}{6}\approx0.1667$, $0.25$, and $\frac{7}{18}\approx0.3889$
  respectively for $k=$ 3, 4, 5, and 10.  }
\end{center}
\end{figure}
%%%%%%%%% FIGURE %%%%%%%%%%%%% FIGURE %%%%%%%%%%%%% FIGURE %%%%%%%%%%%%%

In Figure \ref{fig:2} we show the fluctuations $\sigma$ of the EEVs for different energy windows
near the middle of the spectrum, plotted against $D$.  Each panel shows a different (fixed) number
of sites $k$; in each case the classical limit is approached by increasing $N$.  Generally, the
sequences follow clear power-law dependencies, $\sigma\sim{}D^{-e}$.  The power-law behavior sets in
at relatively small values of $D$ already.

It is clear from the $k=3$ data, panel (a), that the exponent $e$ does not match the value predicted
for Gaussian-random states, Eq.~\eqref{eq:exponent_e0}, which is $e_0=0$ for $k=3$.  The EEV
fluctuation for the system eigenstates \emph{increases} with a positive exponent ($e<0$) instead of
being flat as a function of $D$.  Similarly, for the $4$-site chain the exponent $e$ is seen to be
slightly negative --- $\sigma$ increases slowly with system size --- whereas the
predicted value is $e_0 = +1/6$.

%%%%%%%%% FIGURE %%%%%%%%%%%%% FIGURE %%%%%%%%%%%%% FIGURE %%%%%%%%%%%%%
\begin{figure}%[htbp]
\begin{center}
\includegraphics[width=\columnwidth]{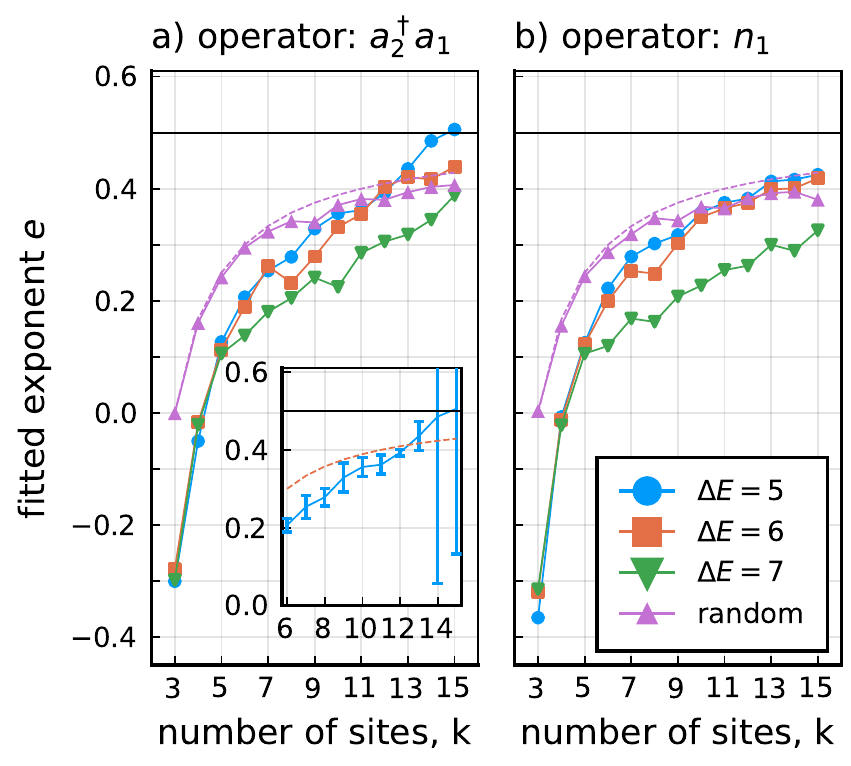}
\caption{\label{fig:3} The exponent $e$ versus the chain length $k$ for Bose-Hubbard eigenstates in
  different energy windows, and for Gaussian-random states.  The pink dashed curve is the
  predicted formula for exponents, $e_0$, which tends to $1/2$ (solid horizontal line) for large
  $k$.  Inset to left panel shows the error bars for the estimation of $e$ from $\Delta{E}=5$ data.
  The error bars are omitted elsewhere and will be omitted in later figures.}
\end{center}
\end{figure}
%%%%%%%%% FIGURE %%%%%%%%%%%%% FIGURE %%%%%%%%%%%%% FIGURE %%%%%%%%%%%%%

The calculations rely on full numerical diagonalization, and hence are limited by the Hilbert space
size $D$.  Our limit was $D\lesssim10^5$.  For each $k$, we increased the particle number $N$ as far
as possible such that $D$ did not exceed $100,000$.  For small $k$, this provides a satisfactory
number of available $N$ values, and extracting the exponent $e$ from a fit to  $\sigma\sim{}D^{-e}$
is quite reliable.  For large $k$, only a few $N$ values are available.  For the largest lattice
($k=15$), only three data points ($N=4$, $N=5$, and $N=6$) were used.  This means a large
uncertainty in the estimation of $e$ (Figure \ref{fig:3} inset).  It also means that the
regime $N\gg k$ is not reached.

In Figure \ref{fig:3} we present the exponents $e$ extracted from the numerical data.  In addition
to the exponents for the operator $a_2^\dagger a_1$ (corresponding to Figure \ref{fig:2}), we also
show the exponents for the operator $n_1$.  For small $k$, the numerically observed exponents $e$
fall significantly below the Gaussian-random case, for both operators.  For larger $k$ values, the
Bose-Hubbard systems show EEV fluctuations closer to the Gaussian-random case, at least for
$\Delta{E}=5,6$.  
(The $\Delta{E}=7$ window shows larger deviation, presumably becuase it is closer to the edges of
the spectrum.)
We interpret this as a signature of the large-$k$ Bose-Hubbard systems being
more chaotic, so that mid-spectrum eigenstates are better approximated by Gaussian-random states.
The deviation for small $k$ represents the `mixed' (chaotic+regular) nature of the few-site
Bose-Hubbard Hamiltonians. 

Figure \ref{fig:3} also shows numerically calculated exponents for EEV fluctuations in
Gaussian-random states (pink triangles), and compares with the $k\ll{N}$ prediction,
Eq.~\eqref{eq:exponent_e0} (pink dashed curve).  The agreement is good for all $k$ and excellent for small
$k$.  At larger $k$, computer memory limitations prevent our calculations from reaching particle
numbers $N\gg{k}$.  This explains the (minor) deviation of the numerical exponents from the
$N\gg{k}$ prediction.

%%%%%%%%% FIGURE %%%%%%%%%%%%% FIGURE %%%%%%%%%%%%% FIGURE %%%%%%%%%%%%%
\begin{figure}%[htbp]
\begin{center}
\includegraphics[width=\columnwidth]{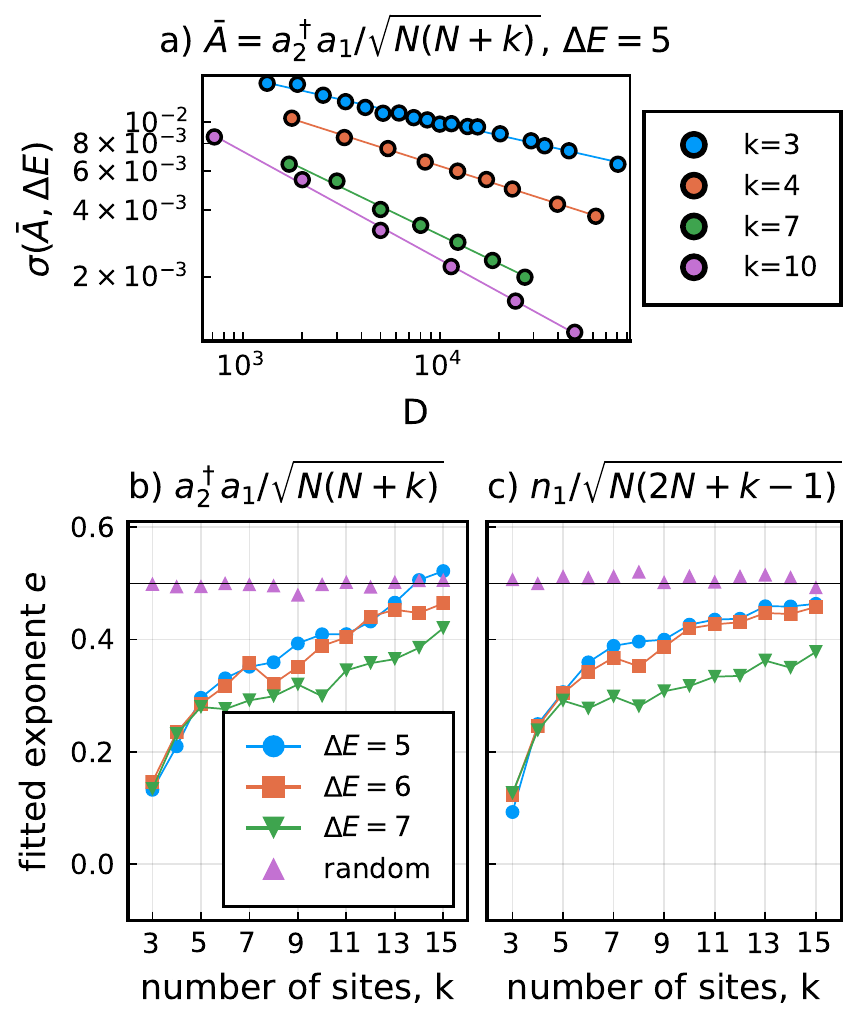}
\caption{\label{fig:scaled_operators} Similar data as in Figures \ref{fig:2} and \ref{fig:3} but
  with scaled operators $\bar{A}$.  The operators are scaled by factors $\sim{N}$; the precise
  factors are explained in the text.
}
\end{center}
\end{figure}
%%%%%%%%% FIGURE %%%%%%%%%%%%% FIGURE %%%%%%%%%%%%% FIGURE %%%%%%%%%%%%%

One can view the same effects through the fluctuations of the scaled
($\bar{A}$) operators $a_2^{\dagger}a_1/N$ and $n_1/N$.  For these
operators, the predicted exponent is $1/2$ for all $k$.  We present
the numerical exponents for such operators in Figure
\ref{fig:scaled_operators}; however we scale by factors slightly
different from $N$.  The prediction $e_0=1/2$ was obtained in the
previous subsection by assuming $N\gg k$. In particular in the trace
expressions of Eq.~\eqref{eq:traces}, this led to $N(N+k)\approx{N^2}$
and $2N(2N+k-1)\approx{4N^2}$.  In our numerical calculations, for
larger $k$ we do not have access to $N$ values in this regime.
Therefore we scale the operators by the factor $\sqrt{N(N+k)}$ for
$A=a_i^{\dagger}a_j$ with $i\neq{j}$ and by the factor
$\sqrt{N(2N+k-1)}$ for $i=j$.  With this modification, the numerically
calculated exponents using Gaussian-random states (pink triangles) do not deviate
systematically from $1/2$ at large $k$, even though the $N\gg{k}$
regime is not reached.  The observed physical exponents (for
Bose-Hubbard eigenstates) are significantly different from the
predicted $e_0=1/2$ for small $k$, but approach this value as $k$ is
increased.

Summarizing our numerical findings: power-law dependence of EEV fluctuations on $D$ is seen for all
$k$.  The exponent at larger $k$ (more fully chaotic systems) matches well the Gaussian prediction.
For smaller $k$ (mixed systems), there are significant deviations from the Gaussian prediction.
Perhaps the most remarkable aspect of these numeric results is that, for small $k$, the departure of
the eigenstates from Gaussian-random or ergodic behavior does not destroy the power-law dependence
of EEV fluctuations with $D$ but changes the exponent substantially.  In the following section we
will examine possible explanations for this phenomenon.

\section{Non-reasons for anomalous scaling} \label{sec:non_reasons}

The mismatch with predicted scaling for small $k$ must be due to the eigenstates of few-site
Bose-Hubbard systems deviating from idealized Gaussian-random states, which we refer to as `ergodic'
states.  In this section, we examine various types of deviation from the idealized case, and rule
out several plausible explanations for the anomalous scaling.

One could imagine that the eigenstates effectively occupy a smaller fraction of the Hilbert space
than a random state, and that this fraction has sublinear scaling with $D$.  This can be quantified
through analysis of the participation ratio or multifractal dimensions of the eigenstates.
Therefore, in Subsection \ref{subsec:participation_ratios}, we discuss the participation ratio
$P_{\alpha}$.  We show how $P_{\alpha}\sim D^1$ scaling implies the expected EEV fluctuation
scalings that we have derived previously in  \ref{subsec:random_states_derivation}.  We
also show that Hilbert space occupancy is not the explanation for our anomalous EEV scaling
exponents, because eigenstates of the $k=3$ system have $P_{\alpha}\sim D^1$ scaling.

One could also imagine that the anomaly of scaling exponents is due to the eigenstate components not
being identically distributed.  In \ref{subsec:coeff_dist} we present data showing that
this is not the reason for the anomalous scaling of EEV fluctuations.

These results show that the non-ergodic scaling must be due to correlations present in the
eigenstates.  In \ref{subsec:eigenstate_correlations} we examine the simplest (and most
prominent) type of correlations between eigenstate coefficients, namely, those captured in the
covariance matrix.  Perhaps surprisingly, we find that these correlations do not explain the
anomalous scaling either.  The anomalous scaling exponent for $k=3$ EEV fluctuations is thus caused
by more subtle (higher-order) correlations.

\subsection{Participation Ratios}\label{subsec:participation_ratios}

For this Subsection it will be convenient to consider our operator $A$ to be Hermitian.
We expand the eigenstates $\ket{E_\alpha}$ of the Hamiltonian in the eigenstate basis
$\ket{\phi_\gamma}$ of the operator $A$ 
\begin{equation}
\ket{E_\alpha} = \sum_{\gamma} c_\gamma^{(\alpha)} \ket{\phi_\gamma},
\end{equation}
where $c_\gamma^{(\alpha)} = \bracket{\phi_\gamma | E_\alpha}$.  If we denote the eigenvalues of $A$
as $a_\gamma$, then the EEVs can be written as
\begin{equation}
A_{\alpha\alpha} = \sum_{\gamma=1}^D  |c_\gamma^{(\alpha)}|^2 a_\gamma.
\end{equation}
We regard the coefficients $c_\gamma^{(\alpha)}$ to be random variables, with each eigenstate index
$(\alpha)$ denoting a different sample from the same underlying distribution.  The distribution is
assumed to be the same for every $\gamma$.  As usual, this framework will not capture large-scale
dependences of the EEVs $A_{\alpha\alpha}$ on the energies $E_\alpha$.
This is acceptable because we are interested in the fluctuations only.

The variance of the EEVs is 
\begin{multline}\label{eq:participation_ratio_variance}
\var\left(\sum_{\gamma=1}^D |c_\gamma|^2 a_\gamma\right) 
= \sum_{\gamma=1}^D \var\left(|c_\gamma|^2\right) a_\gamma^2
\\
=  \var\left(|c_\gamma|^2\right) \tr(A^2) .
\end{multline}
The variance of $|c_\gamma|^2$ can be written as
\begin{equation}
\var\left(|c_\gamma|^2\right) = E\left[|c_\gamma|^4\right]  -
\left(E\left[|c_\gamma|^2\right]\right)^2
= \frac{1}{DP} - \frac{1}{D^2}
\end{equation}
where we have used the definition of the participation ratio to be 
\begin{equation}
P_\alpha = \left(\sum_{\gamma=1}^D |c_\gamma^{(\alpha)}|^4\right)^{-1} = \left(\sum_{\gamma=1}^D
D\times E\left[|c_\gamma^{(\alpha)}|^4\right]\right)^{-1} \ . 
\end{equation}
We thus have the  prediction for the EEV variance
\begin{equation}\label{eq:participation_ratio_variance_2}
  \sigma^2 = \left(\frac{1}{DP} - \frac{1}{D^2}\right) \tr(A^2)
  =  \left(\frac{1}{DP} - \frac{1}{D^2}\right) \tr(AA^\dagger)
\end{equation}
for Hermitian operators.  For \emph{Gaussian} states, $P= D/3$, so that this expression for
$\sigma^2$ reduces to Eq.\ \eqref{eq:statistical_variance_2}.  More generally, as long as the
participation ratio scales linearly with $D$, the factor in brackets $\sim1/D^2$, so that we obtain
the same scaling as for Gaussian states.  For non-ergodic states $P\sim D^K$ with $K<1$, the first
term in brackets would dominate and one would obtain different scaling.

%%%%%%%%% FIGURE %%%%%%%%%%%%% FIGURE %%%%%%%%%%%%% FIGURE %%%%%%%%%%%%% 
\begin{figure}%[htbp]
\begin{center}
\includegraphics[width=\columnwidth]{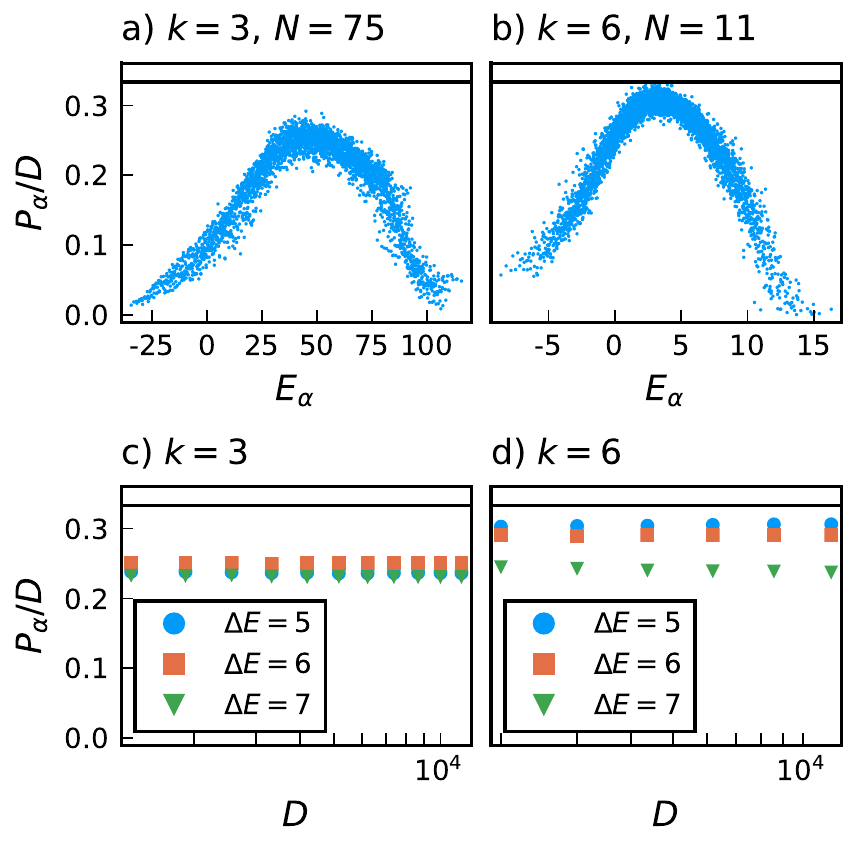}
\caption{ \label{fig:4} Participation ratio in the basis of eigenstates of the operator
  $a_2^{\dagger}a_1+a_1^{\dagger}a_2$.  Horizontal lines indicate  the Gaussian
  expectation ($=\frac{1}{3}$).  Top panels: Scaled participation ratio $P_\alpha/D$ of
  energy eigenstates, versus corresponding energy eigenvalues.  Lower panels: average scaled
  participation ratio for different energy intervals plotted against Hilbert space dimension
  $D$, for fixed chain length $k$ and increasing particle number $N$.  }
\end{center}
\end{figure}
%%%%%%%%% FIGURE %%%%%%%%%%%%% FIGURE %%%%%%%%%%%%% FIGURE %%%%%%%%%%%%% 

In Figure \ref{fig:4} we show the behavior of the participation ratio in the basis of eigenstates of
$(a_2^\dagger a_1 + a_1^\dagger a_2)$.  The mid-spectrum $P_{\alpha}$ is close to the Gaussian
expectation for highly chaotic (larger $k$) systems. For $k=3$ the deviation from Gaussianity
($P=D/3$)  is
strong.  However, in both cases the scaling of $P_{\alpha}$ with Hilbert space dimension $D$ is very
linear (lower panels).  

Thus, Hilbert space occupancy does not explain the observed anomalous exponent of EEV fluctuations.

\subsection{Non-identical distribution of eigenstate coefficients}\label{subsec:coeff_dist}

The analysis in Subsections \ref{subsec:participation_ratios} and
\ref{subsec:random_states_derivation} are based on eigenstate coefficients being \emph{identically
  distributed} and \emph{independent}.  Could it be that the failure to capture the EEV fluctuation
scaling at small $k$ results from the eigenstate components having non-identical distributions?

\begin{figure}%[htbp]
\begin{center}
\includegraphics[width=\columnwidth]{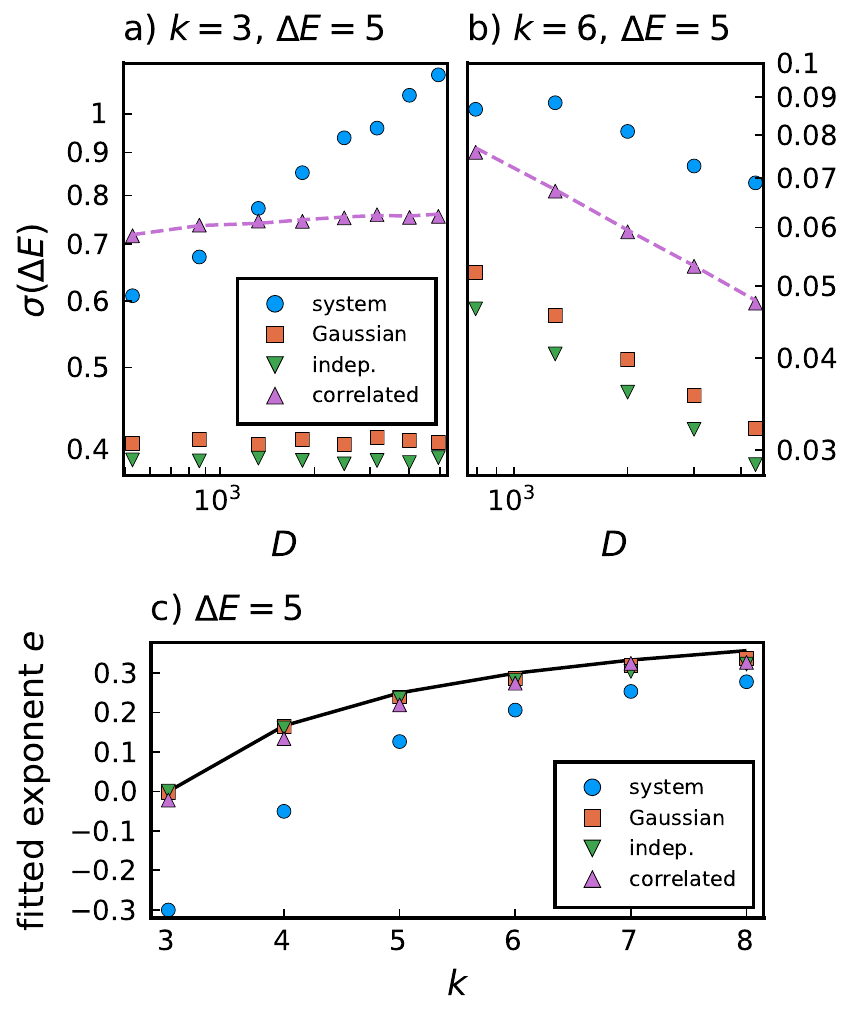}
\caption{ The operator in all plots is $a_2^\dagger a_1$.  Blue dots: eigenstates of physical
  Bose-Hubbard system.  Red squares: Gaussian states with i.i.d.\ components.  Green down triangles:
  vectors with independent but non-identically distributed components, each component sampled from
  system eigenstates.  Purple up triangles: multivariate Gaussian states with covariance matrix
  estimated from system eigenstates.
  a)-b) EEV fluctuations.  The dashed purple line is the prediction by
  Eq.~\eqref{eq:statistical_variance_correlations}.  c) The exponent $e$, such that $\sigma\sim{D}^{-e}$,
  versus the number of sites $k$ for the same distributions as in a)-b).  }
\label{fig:5}
\end{center}
\end{figure}

To check, we drop the condition that the eigenstates have identically distributed components and only
assume that the components are independent. In Figure \ref{fig:5} we show the EEV fluctuations
obtained from an estimation of the underlying distributions of the eigenstate components of the
Bose-Hubbard systems.  We sampled uniformly from components of all eigenstates of the Bose-Hubbard
model in a specific energy interval.  Effectively, for each component of the state, we randomly
picked an eigenstate in the energy interval and used the entry of the corresponding component.  The
EEV fluctuations are then calculated from such sampled states.  The resulting data  are marked `independent' in
Figure \ref{fig:5}.
The results match well with the EEV fluctuations obtained from Gaussian states and do not match the EEV fluctuations of the actual
physical systems for small $k$.

We conclude that that any model for the wavefunctions with independent components does not explain
the observed anomalous scaling at small $k$.  In other words, the root cause of the phenomenon is
not the eigenstate coefficients having non-identical or non-Gaussian distributions, or insufficient
occupancy of the Hilbert space.  Rather, we have traced the cause to the fact that the eigenstate
coefficients are not independent.

%%%%%%%%% FIGURE %%%%%%%%%%%%% FIGURE %%%%%%%%%%%%% FIGURE %%%%%%%%%%%%% 
%%%%%%%%% FIGURE %%%%%%%%%%%%% FIGURE %%%%%%%%%%%%% FIGURE %%%%%%%%%%%%% 

\subsection{Eigenstate Correlations}\label{subsec:eigenstate_correlations}

Continuing our effort to identify what feature of small-$k$ eigenstates is responsible for the
anomalous ETH scaling, we relax the constraints on the model states even further.  Now, we assume
that eigenstates are drawn independently from a multivariate distribution $Z$, but the $D$
components of $Z$ are allowed to be correlated.  Given a multivariate distribution $Z$, the
statistical correlations between components are quantified by the covariance matrix $\Sigma$.  The
covariance matrix in the case of eigenstates can be estimated by regarding the eigenstates within a
specific energy interval as different samples of $Z$.

It is reasonable to assume that the mean of all components of $Z$ is zero, as components of
mid-spectrum eigenstates generally have zero mean.  Then an estimate of $\Sigma$ is given by
\begin{equation}\label{eq:covariance_matrix}
	\Sigma 
	= N_{\Delta E}^{-1} \sum_{E_\alpha \text{ in } \Delta E} \ket{E_\alpha} \bra{E_\alpha}
	%- N_{\Delta E}^{-2} \sum_{E_\alpha, E_\beta \text{ in } \Delta E} \ket{E_\alpha}\bra{E_\beta}
	,
\end{equation}
where $E_\alpha$ and $\ket{E_\alpha}$ denote eigenvalues and eigenstates, respectively, and
$N_{\Delta{E}}$ is the number of eigenstates in the energy window $\Delta{E}$. Eq.~\eqref{eq:covariance_matrix} follows directly from the definition
of the sample covariance matrix of $Z$, where the samples are the eigenstates $\ket{E_\alpha}$.

In Appendix \ref{sec:visualizing_covariances}, Figure \ref{fig:appendix_covariance}, we have provided
visualizations of estimated covariance matrices for $k=3$ and $k=6$.  The $k=3$ case is seen to have
significant off-diagonal elements (arranged in intriguing patterns), indicating non-negligible
correlations between the eigenstate components.

Once we have constructed the correlation matrix $\Sigma$ fom the actual eigenstates, we can sample
$D$-component vectors whose components are distributed \emph{and} correlated according to $\Sigma$.
In Figure \ref{fig:5} the EEV fluctuations obtained from such sampled states are marked as
`correlated'.  The values of the fluctuations thus obtained are larger than those obtained from the
independent-component random states, and more comparable to the fluctuations obtained from the
physical eigenstates.  In the chaotic cases (large $k$), all of these cases have the same scaling.
However, in the $k=3$ case, the scaling exponent is close to the ergodic (Gaussian) case and does not
reproduce the anomalous scaling at all.  This is seen in panel a) of Figure \ref{fig:5}, and also in
panel c) where the fitted exponents are plotted.  The fitted exponent is slightly off the Gaussian
value for small $k$, but far from the anomalous values of the physical system.

These results show that the deviation of the physical system from expected `ergodic' behavior is
only partially captured by the two-point correlations between the eigenstate components.  This
suggests that the small-$k$ eigenstates deviate from randomness in some more drastic manner, which
does not seem easy to quantify.

In addition to the direct numerical verification discussed above, one can use analytic results to
argue (non-rigorously) that inclusion of reasonable two-point correlations in the model of random
states should not change the EEV fluctuation scaling exponent.  Given a multivariate Gaussian
distribution with mean zero and covariance matrix $\Sigma$, we can show (details in Appendix
\ref{sec:variance_calculation}) that the variance of EEV fluctuations is given by
\begin{equation}\label{eq:statistical_variance_correlations}
\var(\langle A \rangle) 
= \frac{1}{D^2}\left[
\tr((D\Sigma A)^2) + \tr(D\Sigma A D\Sigma A^t) 
\right]. 
\end{equation}
This is Eq.~\eqref{eq:statistical_variance_2} with the change $A\to D\Sigma A$. The variance of
components of the wavefunction is not fixed by normalization of the eigenstates any more.  But it is
reasonable to assume that it still scales as $\sim 1/D$, and we have checked numerically that this
scaling holds for the mid-spectrum eigenstates of all our physical systems, including $k=3$.  Since
the variances of the wavefunction components are the diagonal entries of $\Sigma$, the diagonals of
$D\Sigma$ scale (at most) as constant in $D$.  By the Cauchy-Schwartz theorem the off-diagonal terms of
$\Sigma$ are bounded by the diagonal, so \textit{every} component of $D\Sigma$ is (at most) constant
in $D$.

Without making assumptions about the detailed structure of $\Sigma$, we cannot derive rigorously the
scaling of the traces in Eq.~\eqref{eq:statistical_variance_correlations}, which was possible for
Eq.\ \eqref{eq:traces} or \eqref{eq:traces_classical_limit}.  However, since $D\Sigma$ is
elementwise at most $\sim{D^0}$, and assuming $\Sigma$ is not too exotic, one can argue that the
derivation of Eq.\ \eqref{sec:operator_scaling} should hold for this case as well.  In other words,
for `reasonable' $\Sigma$, one expects the same scaling as in the case of independent Gaussian
eigenstates.  This is consistent with Figure \ref{fig:5}, where the $\Sigma$ is estimated
numerically from the physical eigenstates.

\section{Discussion and Context}\label{sec:conclusion}

Motivated by questions relating quantum dynamics to statistical mechanics, we have undertaken a
study of the eigenstate thermalization hypothesis in the scaling sense, but considering increasing
Hilbert space dimensions along the (semi)classical limit rather than the usual thermodynamic limit.
This has led to a characterization of the distinctive  properties of few-site 
Bose-Hubbard systems in terms of anomalous scaling exponents.  

\textbf{Summary of analytic results.}
For GOE eigenstates, i.e., states whose components can be approximated by Gaussian-distributed
random variables, we have used trace expressions for the EEV fluctuation $\sigma$,
Eq.\ \eqref{eq:statistical_variance_2}.  For operators of the type $A= a_j^{\dagger}a_i$, the trace
operators can be expressed as Eq.\ \eqref{eq:traces}.  Based on these main expressions, we are able to predict ideal scaling behaviors of EEV fluctuations
in the classical limit, for both un-scaled operators of the type $A$ and scaled operators
$\bar{A}=A/N$.  Of course, the usual ETH scaling of the thermodynamic limit also follows from these
expressions.

In the classical limit $N\gg{k}$, the EEV fluctuations are found for such idealized eigenstates to behave
as $\sigma\sim{D}^{-e_0}$, with $e_0= \frac{1}{2}-\frac{1}{(k-1)}$ for un-scaled $A$ operators and
$e_0=1/2$ for scaled $\bar{A}$ operators.

In addition, we have presented expressions for $\sigma$ for a number of related cases, e.g., for
i.i.d.\ distributed eigenstate components with the distribution not assumed to be Gaussian,
Eq.\ \eqref{eq:statistical_variance_1}, in terms of the participation ratio,
Eq.\ \eqref{eq:participation_ratio_variance_2}, and for the more general case where the eigenstate
components are allowed to be correlated according to a covariance matrix,
Eq.\ \eqref{eq:statistical_variance_correlations}.

\textbf{Summary of numerical results.}
We have explored the scaling exponent for various lattice lengths $k$, increasing the boson number
$N$ with fixed $k$ to approach the classical limit.  At larger $k$, the exponent matches the
random-eigenstate prediction. At small $k$, the fluctuation appears to have power-law dependence
$\sigma\sim{D}^{-e}$ on the Hilbert space dimension, i.e., $e$ is well-defined, but the value of $e$
differs markedly from the ergodic prediction.  Through a series of additional numerical tests, we
have shown that this anomalous scaling is not explained by 2-point correlations between eigenstate
components.

The small-size Bose-Hubbard systems thus have mid-spectrum eigenstates which violate the usual
randomness approximation in some subtle higher-order manner.

\textbf{Deviation from Ergodicity.}
The deviation of quantum many-body systems from ergodicity has been the subject of interest from
multiple viewpoints in recent years.  Other than the strong violations of ETH/ergodicity due to
integrability or many-body localization, more subtle departures have also been addressed or
observed.  In many-body systems that are nominally chaotic, mid-spectrum states are largely
well-modeled by random states, but small or subleading deviations have been observed in various properties
\cite{Beugeling_coefficients_PRE2018, Luitz_Barlev_PRL2016, Baecker_Haque_Khaymovich_PRE2019,
  Luitz_Khaymovich_BarLev_multifrac_SciPost2020, Beugeling_entanglement_JSM2015,
  Balents_SYK_entanglement_PRB2018, LeBlond_Mallaya_Vidmar_Rigol_PRE2019,
  Haque_McClarty_Khaymovich_entanglement_deviation, Khaymovich_Haque_McClarty_PRL2019}.  However,
\emph{scaling} properties in these systems generally follow random-state predictions.  In the
small-$k$ Bose-Hubbard systems, we have shown a striking exception: a system which is not integrable
or many-body localized, but nevertheless violates the usual scaling behavior expected in chaotic
systems.  

The three-site and four-site Bose-Hubbard systems are widely known to be imperfectly chaotic, in
particular because their classical phase space has been explored and is known to have both chaotic
and regular regions \cite{Mossmann_Jung_PRA2006_semiclassical_BHtrimer,
  Hiller_Kottos_Geisel_BHtrimer_PRA2009, Viscondi_Furuya__JPhysA2011_BHtrimer,
  Vardi_DoronCohen_PRA2014_BHtrimer, Han_Wu_PRA2016_BHtrimer_ehrenfest,
  Buerkle_Anglin_4siteBH_PRA2019, Kolovsky_russians_AIPConf2020, Rautenberg_Gaertner_PRA2020}.  With
the current interest in chaos in many-body eigenstates, it is important to characterize such
deviations from chaoticity.  In this work, ETH scaling (i.e., the scaling of EEV fluctuations) has
turned out to be fruitful approach to characterize these special  systems.

%% \textbf{Recent interest.}
%% %

\textbf{Open questions and ideas arising from this work.}
The present work opens up a number of new questions deserving investigation: 

(1) We have found that the small-$k$ Bose-Hubbard systems display EEV fluctuations scaling with
exponents that appear numerically well-defined but very clearly different from the random-state
prediction.  An analytic explanation for these observed new exponents is currently not available,
and remains an open question.  The anomalous scaling is related to the insufficient chaoticity of
few-site systems, which is closely connected to the mixed phase space of the corresponding classical
system.  Hence a tempting conjecture is that some property measuring the degree of chaos in the
classical limit might explain the exponents.

(2) Our analytic results have focused on essentially infinite-temperature states.  It would be
interesting to develop trace expressions for finite temperatures.  This is likely not possible
to do in complete generality without making assumptions on the system Hamiltonian, but perhaps some
results can be derived with minimal assumptions, such as locality of the Hamiltonian. 

(3) For the thermodynamic limit, the EEV fluctuations of \emph{integrable} systems generally show
power-law scaling in $N$ or $k$ \cite{Ziraldo_Santoro_relaxation_PRB2013,
  Ikeda_Ueda_PRE2013_LiebLiniger, Beugeling_scaling_PRE14, Alba_PRB2015, ArnabSenArnabDas_PRB16,
  Magan_randomfreefermions_PRL2016, HaqueMcClarty_SYKETH_PRB2019, Mierzejewski_Vidmar_PRL2020},
which translates into $\sim\ln{D}$ scaling.  For the classical limit, however, power-law scaling in
$N$ would mean power-law scaling in $D$ as well.  Our numerics (Appendix
\ref{sec:lambda_dependence}) shows that fitting $\sigma\sim{D}^{-e}$ in the integrable regime yields
$e=-1/(k-1)$ for $A$-type operators and $e=0$ for $\bar{A}$-type operators.  A detailed
understanding and explanation of these exponents remains a task for future research.

(4) Bose-Hubbard systems are, of course, not the only quantum systems with a classical limit.  It
remains to be discovered how generic our findings are.  Comparing chaos-related properties between
quantum systems and the corresponding classical systems has also been of interest in few-spin
systems \cite{Feingold_Peres_PhysicaD1983_coupled_rotators, Feingold_Moiseyev_Peres_PRA1984,
  Reichl_PRE1998_twospin, Emerson_Ballentine_PRA2001_twospin, Ballentine_PRA2002_twospins,
  Fan_Gnutzmann_Liang_PRE2017_FeingoldPeres, Pappalardi_Silva_Fazio_PRB2018_longrange,
Lewenstein_Quach_PRE2019_entanglement_kickedrotor, 
  Pappalardi_PRA2020_semiclassicalsystems, Scaffidi_Cao_PRL2020_scrambling,
  Pappalardi_Polkovnikov_Silva_SciPost2020_SherringtonKirkpatrick, Mondal_Sinha_PRE2020_twospins}
and in spin-boson systems \cite{Hirsch_PRA2014_Dicke_comparative2, Hirsch_PhysicaScrpta2015_Dicke,
  Ghosh_Sinha_PRE2016_kicked_Dicke,  Hirsch_PRE2016_Dicke,
  Bollinger_AMRey_NatureComm2019_Dicke,  LSantos_Hirsch_PRL2019_Dicke_Lyapunovs,
  Pappalardi_PRA2020_semiclassicalsystems, 
  Scaffidi_Cao_PRL2020_scrambling,  Robnik_PRE2020_Dicke, LSantos_Hirsch_NJP2020_Dicke,
  LSantos_Hirsch_PRE2020, Sinha_PRL2020_BJJ_bosonicmode}.  (The classical
system is obtained in the limit of large spin quantum number.)
Studying the behavior of EEV fluctuations in such systems when approaching the classical limit would
provide interesting characterizations of ergodicity, e.g., of how well randomness approximations
work.

(5) As part of our effort to address the anomalous scaling at small $k$, we have briefly examined
the covariance matrix of eigenstates, treating each eigenstate as a sample drawn from the
distribution of eigenstates, according to Eq.\ \eqref{eq:covariance_matrix}.  Studying the
thus-defined covariance matrix might be fruitful for various quantum systems, as the departure of
this matrix from the identity matrix tells us how different the eigenstates are from
infinite-temperature states.  A further significance of this covariance matrix is that the same
object is the microcanonical density matrix, and hence its structure should provide insights into
the connection betwen quantum eigenstates and thermodynamics.

\begin{acknowledgments} 
MH thanks P.~Ribeiro and R.~Mondaini for enlightening discussions.  We thank J.~D.~Urbina and
K.~Richter for useful correspondence.  Some computations were performed on the Irish Center for
High-End Computing. GN is supported by the Irish Research Council Government of Ireland Postgraduate Scholarship Scheme (GOIPG/2019/58).
\end{acknowledgments}

\appendix

\section{Level spacing statistics}\label{sec:r_stat}

%%%%%%%%% FIGURE %%%%%%%%%%%%% FIGURE %%%%%%%%%%%%% FIGURE %%%%%%%%%%%%%
\begin{figure}%[htbp]
\begin{center}
\includegraphics[width=\columnwidth]{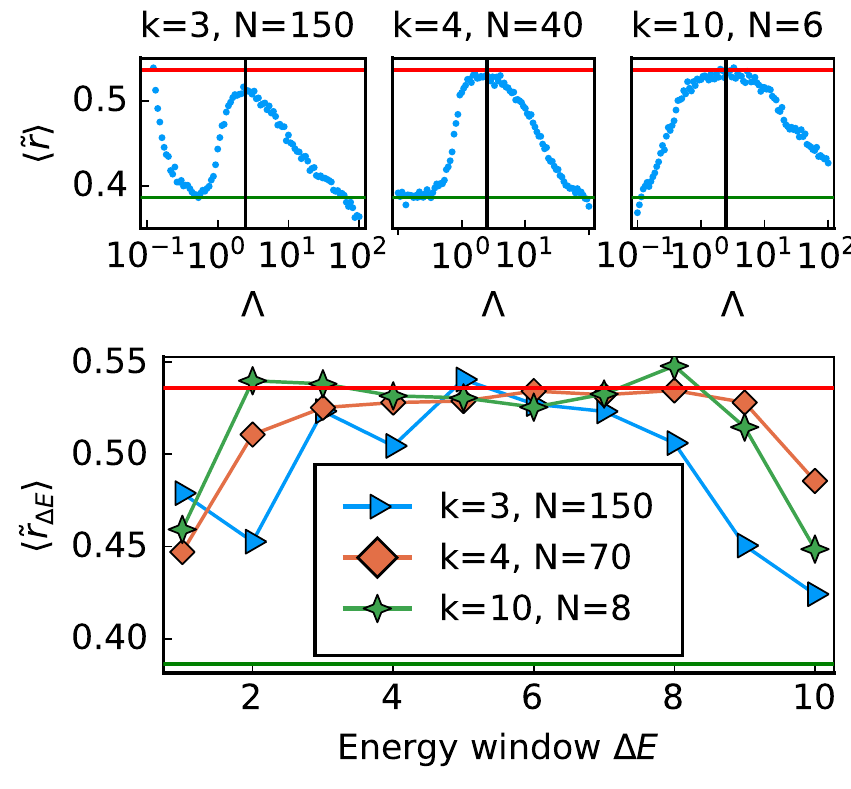}
\caption{\label{fig:rstat} Top: Level ratios $\tilde{r}$ averaged over the whole spectrum, as a
  function of interaction strength $\Lambda=UN$, shown in the range $\Lambda\in(0.1,100)$.  The
  horizontal lines are $\langle\tilde{r}\rangle_{\rm integr}$ and $\langle\tilde{r}\rangle_{\rm
    GOE}$. The vertical line indicates $\Lambda=10^{13/33}\approx 2.477$.  Bottom: Mean level ratio
  against ten evenly spaced energy windows $\Delta E$, labeled 1 through 10 and plotted against
  these labels, for $\Lambda\approx 2.477$, various system sizes.  The
  match to the GOE value is better away from spectral edges.  }
\end{center}
\end{figure}
%%%%%%%%% FIGURE %%%%%%%%%%%%% FIGURE %%%%%%%%%%%%% FIGURE %%%%%%%%%%%%%

The distribution of energy eigenvalue spacings, $s_\alpha = E_{\alpha+1} - E_\alpha$ for ordered
$E_\alpha < E_{\alpha+1}$, is often used as an indicator for quantum chaos.  Measuring this
distribution involves unfolding the spectrum. This is bypassed by investigating the distribution of
spacing ratios \cite{Oganesyan_Huse_PRB2007, Atas_Bogomolny_Roux_PRL2013}
\[
r_\alpha = \frac{s_{\alpha+1}}{s_\alpha} \quad \text{and} \quad \tilde{r}_\alpha =
\min\left(r_\alpha, \frac{1}{r_\alpha}\right) . 
\]
The latter has the advantage that it is bounded.  Quantum Hamiltonians are considered chaotic when
the distributions follow that of the relevant Wigner-Dyson ensemble, which in our case is the GOE
ensemble. The mean $\tilde{r}$ for GOE is $\langle \tilde{r} \rangle_{\rm GOE} \approx 4-2\sqrt{3}
\approx 0.53590$.  For an integrable Hamiltonian, the level spacing distribution is Poissonian
(exponential); this leads to $\langle \tilde{r} \rangle_{\rm integr} = 2\log 2 -1 \approx 0.38629$.

Figure \ref{fig:rstat} (top) shows that, with increasing interaction strength $\Lambda=UN$, the
Bose-Hubbard systems turn from integrable to chaotic and back to integrable.  Some anomalies are
visible for short chains (small $k$).  First, the distributions of the level ratios for very small
$\Lambda$ or very large $\Lambda$ match neither GOE nor Poisson statistics.  Also, for
small $k$, in the most chaotic regime, the peak of $\langle \tilde{r} \rangle$ is still noticeably
lower than the GOE value $\approx0.534$.  As $k$ increases, the peak gets closer and closer to the
GOE value, consistent with the intuition that  larger-$k$ chains are more chaotic.

Our grid of $\Lambda$ values contains 100 values spaced logarithmically from 0.1 to 100.  Among
these values, $\Lambda=10^{13/33}\approx 2.477$ is found to be close to the location of the chaotic
peak, for all $k$.  Therefore, in this work we have shown data for this value of the interaction
parameter.

Of course, we do not expect that the full spectrum obeys GOE statistics --- the spectral edges are
expected to deviate.  This can be observed in Eq.~\eqref{fig:rstat} (bottom panel), where we plot $\langle
\tilde{r} \rangle$ against different energy windows.

\section{Dependence of EEV Scaling on interaction strength $\Lambda$} \label{sec:lambda_dependence}

%%%%%%%%% FIGURE %%%%%%%%%%%%% FIGURE %%%%%%%%%%%%% FIGURE %%%%%%%%%%%%%
\begin{figure}%[htbp]
\begin{center}
\includegraphics[width=\columnwidth]{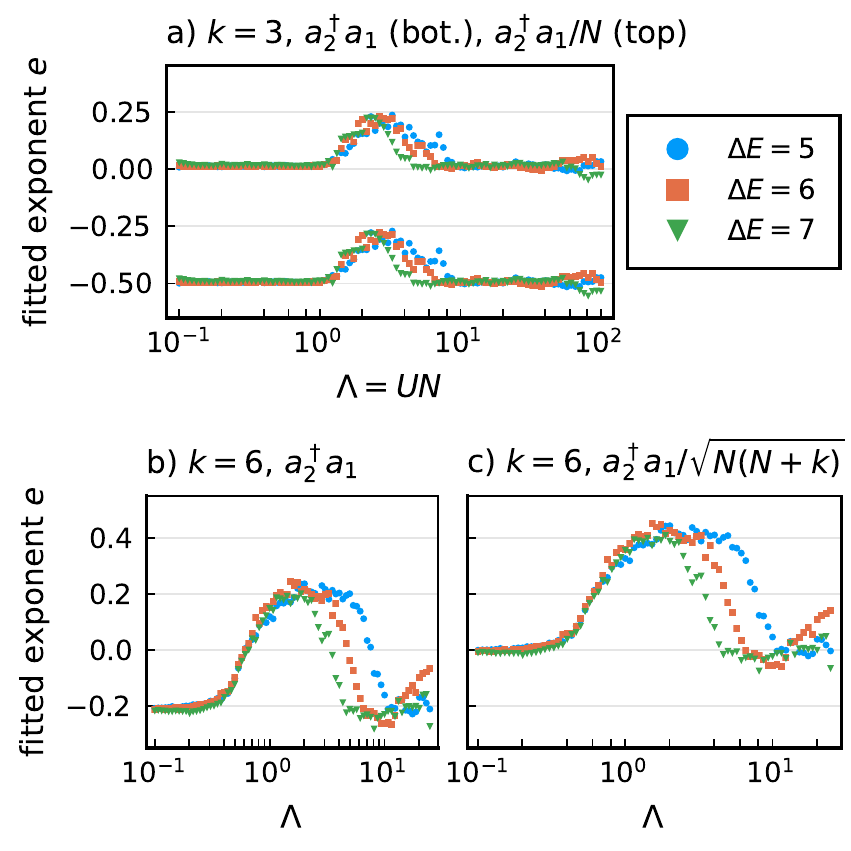}
\caption{\label{fig:scaling_lambda} Exponent $e$ obtained from $\sigma\sim{D}^{-e}$ fits, as a
  function of the interaction parameter $\Lambda = UN$.  Results
  are shown for both un-scaled operator  $A=a_2^\dagger a_1$ and its scaled version $\bar{A}$, with
  $\bar{A}=A/N$ for the $k=3$ system (top) and $\bar{A}=A/\sqrt{N(N+k)}$ for the $k=6$ system (bottom).
}
\end{center}
\end{figure}
%%%%%%%%% FIGURE %%%%%%%%%%%%% FIGURE %%%%%%%%%%%%% FIGURE %%%%%%%%%%%%%

In the main text we focused on the value of the interaction parameter $\Lambda=UN$ at which our
systems are most chaotic.  Here, we show how changing $\Lambda$ affects the scaling of EEV
fluctuations.  Fitting the fluctuation data to $\sigma\sim D^{-e}$, we extract exponents $e$ which
we show in Figure \ref{fig:scaling_lambda} as a function of $\Lambda$.  We compare the smallest
Bose-Hubbard chain with $k=3$ sites with a larger, thus more chaotic chain with $k=6$.

We show data both for the operator $A=a_2^\dagger a_1$ and the scaled operator $\bar{A}\approx
a_2^\dagger a_1/N$.  For $k=3$, we show results for $\bar{A}=a_2^\dagger a_1/N$.  (As the data for
small $k$ extends into the $N\gg{k}$ regime, it is reasonable to approximate $\sqrt{N(N+k)}\approx
N$.)  For larger $k$ values, the regime $N\gg{k}$ is difficult to reach numerically.  Therefore it
is appropriate to scale with the factor $\sqrt{N(N+k)}$ instead of the factor $N$, as the
approximation $\sqrt{N(N+k)}\approx N$ made for  Eq.~\eqref{eq:traces}
or Eq.~\eqref{eq:a_j_a_i_exact} may not be accurate.

Since $N\sim D^{1/(k-1)}$, the exponent $e$ for $\sigma(\bar{A})$ is shifted upwards compared to
that for $\sigma(A)$ by $1/(k-1)$.  This is seen in Figure \ref{fig:scaling_lambda} for both $k=3$ (shift of
$\frac{1}{3-1}=0.5$) and for $k=6$ (shift of $\frac{1}{6-1}=0.2$).

At intermediate values of $\Lambda$, when the system is (partially) chaotic, the value of $e$ is
near that discussed in the main text.  If we had fully chaotic (Gaussian) eigenstates, we would
expect $e=0.5$ for the scaled operators $\bar{A}$ and $e=0.5-0.5=0$ ($k=3$) or $e=0.5-0.2=0.3$
($k=6$) for the un-scaled operators $A$.  Compared to the Gaussian prediction, the observed
exponents are significantly smaller for the $k=3$ system, and somewhat smaller for the $k=6$ system,
as discussed in detail in the main text.

For small or large $\Lambda$, the system is (near-)integrable.  In these limits, the fluctuations
for scaled operators $\bar{A}$ do not have power-law dependence on $D$, and the fit to $D^{-e}$
results in $e=0$.  Thus eigenstate fluctuation scaling of \emph{scaled} operators $\bar{A}=A/N$ in
the classical limit is analogous to that of \emph{local} operators in the thermodynamic limit:
$e\to0$ in the integrable limits and $e\approx0.5$ in the chaotic regime
\cite{Beugeling_scaling_PRE14}.  For local operators in the thermodynamic limit, the fluctuations
have power-law dependence on system size \cite{Ziraldo_Santoro_relaxation_PRB2013,
  Ikeda_Ueda_PRE2013_LiebLiniger, Beugeling_scaling_PRE14, Alba_PRB2015, ArnabSenArnabDas_PRB16,
  Magan_randomfreefermions_PRL2016, HaqueMcClarty_SYKETH_PRB2019, Mierzejewski_Vidmar_PRL2020},
i.e., logarithmic dependence on $D$.  A detailed study of the (near-)integrable models in the
classical limit remains an interesting task for future studies.

Because $e$ for $\bar{A}$ settles to zero for $\Lambda\to0$ and $\Lambda\to\infty$, the exponent for
the unscaled operator $A=a_2^\dagger a_1$ settles to $-1/(k-1)$ in these limits, i.e., to $-0.5$ for
the trimer and to $-0.2$ for $k=6$.  This is clearly seen in Figure \ref{fig:scaling_lambda}.

\section{Estimated covariance matrices from physical eigenstates} \label{sec:visualizing_covariances}

%%%%%%%%% FIGURE %%%%%%%%%%%%% FIGURE %%%%%%%%%%%%% FIGURE %%%%%%%%%%%%%
\begin{figure}%[htbp]
\begin{center}
\includegraphics[width=\columnwidth]{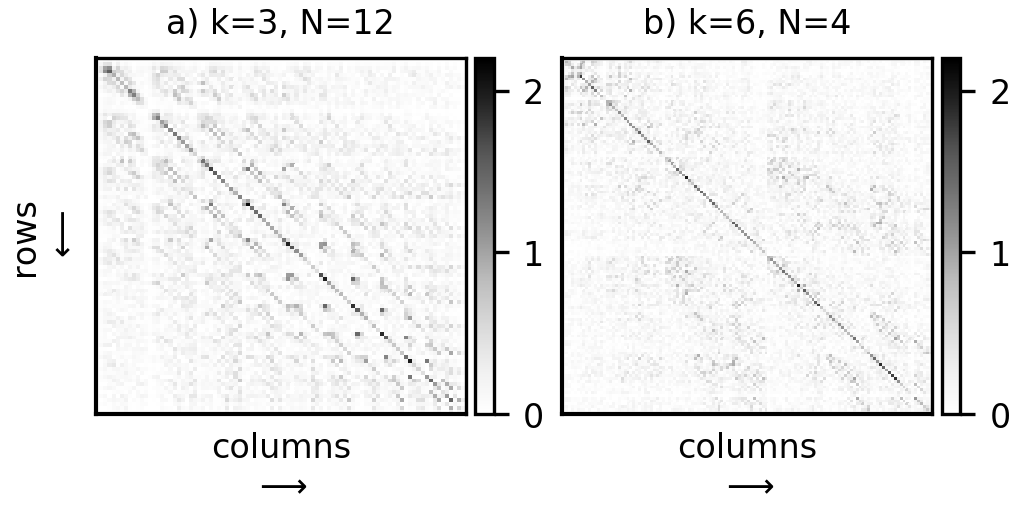}
\caption{\label{fig:appendix_covariance} Scaled estimated covariance matrix $D\Sigma$ of mid-spectrum eigenstates   (energy window $\Delta E=5$) for Bose-Hubbard chains with $k$ sites
  and $N$ particles.  The estimate 
  $\Sigma$ is defined in Eq.~\eqref{eq:covariance_matrix},  The
  absolute values of matrix entries, $|D\cdot \Sigma_{ij}|$, corresponding to basis $\mathcal{B}$
  described in text, are shown. The arrows indicate the ordering of the basis, e.g. in the top left
  is the entry $|D\cdot \Sigma_{11}|$. $N$ had to be chosen significantly smaller than in the rest
  of the paper to visualize the patterns without zooming. The patterns are stable for increasing
  $N$.}
\end{center}
\end{figure}
%%%%%%%%% FIGURE %%%%%%%%%%%%% FIGURE %%%%%%%%%%%%% FIGURE %%%%%%%%%%%%%

In Subection \ref{subsec:eigenstate_correlations} we discussed results incorporating the 2-point
correlations between eigenstate components, as embodied in the covariance matrix.  The covariance
matrix for physical mid-spectrum eigenstates is calculated according to 
Eq.~\eqref{eq:covariance_matrix}, i.e., as an equal-weight mixture of the density matrices
$\ket{E_{\alpha}}\bra{E_{\alpha}}$ corresponding to each eigenstate in the energy window.  Here we
show the structure of the covariance matrices obtained in this way, i.e., we visualize the 2-point
correlations present in the physical eigenstates.    

To visualize the structure of covariance matrices one has to fix a basis of the Hilbert space. In
Eq.~\eqref{fig:appendix_covariance} we show covariance matrices with respect to the defining basis
$\mathcal{B}$. The elements of $\mathcal{B}$ are the mutual eigenstates of all number operators
$n_j=a_j^\dagger a_j$, denoted by $\ket{n_1,\dots,n_k}$ where $n_1+\dots + n_k = N$. For two states
$\ket{\psi}=\ket{n_1,\dots,n_k}$ and $\ket{\phi}=\ket{m_1,\dots,m_k}$ we say that $\ket{\psi} <
\ket{\phi}$ if and only if the states interpreted as $(N+1)$-adic numbers satisfy $(\psi)_{N+1} < (\phi)_{N+1}$,
i.e.  
\begin{equation}\label{eq:basis_ordering}
	\ket{\psi} < \ket{\phi} \iff \sum_{j=1}^k n_j (N+1)^j < \sum_{j=1}^k m_j (N+1)^j.
\end{equation}
The basis states in $\mathcal{B}$ are in descendent order with respect to the ordering given by
Eq.~\eqref{eq:basis_ordering}. For example, for $k=3$ and $N=2$, $\mathcal{B}$ is ordered as 
\begin{equation}
	\mathcal{B} = \left(  \ket{0,0,2}, \ket{0,1,1}, \ket{0,2,0}, \ket{1,0,1}, \ket{1,1,0}, \ket{2,0,0} \right).
\end{equation}

If the components of eigenstates of Bose-Hubbard chains would be independent, their estimated
covariance matrix would be close to the identity matrix.  Figure \ref{fig:appendix_covariance} shows
that, even for $k=6$, there are significant off-diagonal terms, indicating that the eigenstate
components are not independent.  The deviation from identity matrix suggests that the mid-spectrum
states deviate from infinite-temperature states, even for significantly chaotic states, consistent
with Refs.\ \cite{Beugeling_coefficients_PRE2018, Luitz_Barlev_PRL2016,
  Baecker_Haque_Khaymovich_PRE2019, Luitz_Khaymovich_BarLev_multifrac_SciPost2020,
  Beugeling_entanglement_JSM2015, Balents_SYK_entanglement_PRB2018,
  LeBlond_Mallaya_Vidmar_Rigol_PRE2019, Haque_McClarty_Khaymovich_entanglement_deviation,
  Khaymovich_Haque_McClarty_PRL2019}.  For small $k$, nonzero off-diagonal elements are even more pronounced, and
these appear in intricate patterns (Figure \ref{fig:appendix_covariance}(a)).

%% ---------- 

%% As the estimated covariance matrix is effectively a microcanonical density ensemble of the energy
%% interval $\Delta E$, see \cref{eq:covariance_matrix}, the former implies that an eventually assigned
%% temperature to $\Delta E$ is not infinite. More specifically, for a suitable norm $\|\cdot\|$ and
%% normalization constants $M_1,M_2$ we look for a $\beta$ minimizing
%% \begin{equation}\label{eq:beta}
%% %N_{\Delta E}^{-1} 
%% \left\| M_1 \sum_{E_\alpha \text{ in } \Delta E} \ket{E_\alpha}\bra{E_\alpha} - M_2 e^{-\beta H} \right\|
%% \end{equation}
%% and assign a temperature $T=1/\beta$ to $\Delta E$. As the first term in \cref{eq:beta} is not
%% close to the identity, see \cref{fig:appendix_covariance}, the optimal $\beta$ will not be 0,
%% thus $T < \infty$. Making this statement rigorous and exploring, if this is just a side effect of
%% finite Hilbert space dimension $D$, falls out of the scope of this paper and we leave it open for
%% future research.  

%% Nevertheless, we showed in \cref{subsec:eigenstate_correlations} numerically and gave analytical
%% arguments that correlation between eigenstate components do not change the scaling in the Hilbert
%% space dimension $D$ of EEV fluctuations. So a finite temperature $T$ does not explain the
%% deviation of small Bose-Hubbard EEV fluctuation scaling from Gaussian scaling.  

\section{Variance and Covariance of EEVs of Random Eigenstates}\label{sec:variance_calculation}

In this section, we provide derivations of some of the equations that were announced and used in the
main text without proof.  First, we derive the results announced in Section
\ref{sec:scaling_results} for random eigenstates with independent and identically distributed
(i.i.d.) components (Subsection \ref{subsec:calculation_iid}).  Next, we prove the result
Eq.\ \eqref{eq:statistical_variance_correlations}, announced in Subsection
\ref{subsec:eigenstate_correlations}, for random vectors whose components are not independent, with
2-point correlations described by a covariance matrix (Subsection \ref{subsec:calculation_correlated}).

\subsection{Random vectors with i.i.d.\ components} \label{subsec:calculation_iid}

Let $A$ be a $D\times D$ square matrix and $Z$
be a $D$-dimensional, multivariate random state with identically and independently distributed
components $Z_i$, each with mean 0. The statistical variance of $\langle A \rangle = \langle Z |A| Z
\rangle $ is given by
\begin{align}
\var(\langle A \rangle) &= \sum_{i,j,i',j'=1}^D A_{i,j} A_{i',j'} \cov( Z_i Z_j, Z_{i'} Z_{j'}),
\end{align}
where
\begin{align}
\cov( Z_i Z_j, Z_{i'} Z_{j'})
&= E[Z_i Z_j Z_{i'} Z_{j'}] - E[Z_iZ_j]\cdot E[ Z_{i'} Z_{j'}] \nonumber \\
&= E[Z_i Z_j Z_{i'} Z_{j'}] - \delta_{ij}\delta_{i'j'}E[Z_i^2]^2.
\end{align}
By the independence of $Z_i,Z_j$ for $i\neq j$, $E[Z_i Z_j Z_{i'} Z_{j'}]$ is only non-zero if there is no index $i,j,i',j'$ different to the other three. The only possibilities for this are
\begin{align}
&i=j=i'=j' \nonumber\\
&i=j \text{ and } i'=j'\nonumber\\
&i=i' \text{ and } j=j' \nonumber\\
&i=j' \text{ and } j=i',
\end{align}
so
\begin{align}
E[Z_i &Z_j Z_{i'} Z_{j'}]
= \delta_{ij}\delta_{ii'}\delta_{ij'} E[Z_i^4] \nonumber \\
&+ (1-\delta_{ij}\delta_{ii'}\delta_{ij'}) \nonumber \\
&\times
\left[
\delta_{ij}\delta_{i'j'}E[Z_i^2]^2
+ \delta_{ii'}\delta_{jj'} E[Z_i^2]^2
+ \delta_{ij'}\delta_{ji'} E[Z_i^2]^2
\right].
\end{align}
Then we get
\begin{align}
\var(\langle A \rangle)
&= E[Z_1^4] \sum_i  A_{ii}^2 + E[Z_1^2]^2 \nonumber \\
&\times \left[
\sum_{i\neq i'} A_{ii}A_{i'i'} + \sum_{i\neq j}  A_{ij}^2 + \sum_{i\neq j}A_{ij} A_{ji} 
\right] \nonumber \\
&- E[Z_1^2]^2 \sum_{i,i'} A_{ii}A_{i'i'}
\end{align}
and
\begin{align}
\var(\langle A \rangle)
&= \left(E[Z_1^4]-E[Z_1^2]^2\right) \sum_i  A_{ii}^2 \nonumber \\
& + E[Z_1^2]^2\left[\sum_{i\neq j}  A_{ij}^2 + \sum_{i\neq j}A_{ij} A_{ji}
\right] \nonumber \\
&= \left(E[Z_1^4]-3E[Z_1^2]^2\right) \sum_i  A_{ii}^2 \nonumber \\
&+ E[Z_1^2]^2\left[\tr(A^2) + \tr(AA^\dagger)\right]. 
\end{align}
This is Eq.~\eqref{eq:statistical_variance_1} of the main text.  It is a rather general result, not
assuming a particular distribution of the components, only that they should be independent and
identically distributed and have mean 0.

Now, we specialize to the case that the components $Z_i$ are each normally distributed with mean 0
and variance $1/D$. Then
\begin{equation}
E[Z_i^2] = \frac{1}{D} E[(\sqrt{D}Z_i)^2] = \frac{1}{D}
\end{equation}
and
\begin{equation}
E[Z_i^4] = \frac{1}{D^2} E[(\sqrt{D}Z_i)^4] = \frac{3}{D^2},
\end{equation}
because $\sqrt{DZ_i}$ is normally distributed with mean 0 and variance 1. Plugging this into the above formula, we get
\begin{align}
\var(\langle A \rangle) 
&= \left(\frac{3}{D^2} - \frac{3}{D^2}\right) \sum_i  A_{ii}^2 \\
&+ \frac{1}{D^2}\left[\tr(A^2) + \tr(AA^\dagger)\right] \nonumber \\
&= \frac{1}{D^2}\left[\tr(A^2) + \tr(AA^\dagger)\right].
\end{align}
This concludes the proof of Eq.~\eqref{eq:statistical_variance_2}.

\subsection{Including 2-point correlations} \label{subsec:calculation_correlated}

We now relax the assumption of independence and allow the wavefunction components to have 2-point
correlations.  The trace expression in this case is
Eq.~\eqref{eq:statistical_variance_correlations}, which we will now prove.

Assume $Z=LX$, where $X$ is a vector whose independent components are Gaussian-distributed with mean
zero and variance $1$, and $L$ is the Cholesky root of a non-random $D\times D$ (covariance-) matrix
$\Sigma$, i.e $\Sigma = LL^t$. Then $Z$ has mean zero and the covariances between components of $Z$
are given by $\Sigma$.  We compute the statistical variance of the EEVs as
\begin{align}
\var(\langle A \rangle) 
&=\var((LX)^t A LX) \nonumber \\
&= \var(X^t L^t A LX) \nonumber \\
&= \var(1/\sqrt{D}X^t DL^t A L 1/\sqrt{D}X)
\end{align}
Because the components of $1/\sqrt{D} X$ are normally distributed with mean 0 and variance $1/D$, we
can use Eq.~\eqref{eq:statistical_variance_2} with $A$ replaced by $L^t A L$.  This leads to
\begin{align}
\var(\langle A \rangle) 
&= \frac{1}{D^2} (\tr(D^2 (L^tAL)^2) + \tr(D^2 L^tAL(L^tAL)^t) \nonumber \\
&= \frac{1}{D^2}(\tr((D\Sigma A)^2) + \tr(D\Sigma A D\Sigma A^t),
\end{align}
which concludes the proof of Eq.~\eqref{eq:statistical_variance_correlations}.

\section{Scaling of operators}\label{sec:operator_scaling}

In this section we derive results announced in the main text concerning the scaling of the traces.
For $A=a_j^\dagger a_i$ we prove the trace expression of Eq.~\eqref{eq:traces}.  For the classical
limit, we prove the scaling relation, Eq.~\eqref{eq:traces_classical_limit}, for $A$ being a
(finite) linear combination of terms of the type $a_j^\dagger a_i$.

Considering $A$ to be such a linear combination, $A^2$ and $A^\dagger A$ can be written as a sum of
terms $a_j^\dagger a_i a_{j'}^\dagger a_{i'}$.  Consider a defining basis state $|\psi\rangle
=|n_1,\dots,n_k\rangle$, which has $n_j$ particles on site $j$. Then
\begin{align}\label{eq:a_j_a_i_psi}
	a_j^\dagger a_i |\psi\rangle
	&= \delta_{ij} n_i |\psi\rangle \\
	&+ (1-\delta_{ij}) \sqrt{n_j+1}\sqrt{n_i}|\dots,n_i-1, n_j+1,\dots\rangle \nonumber.
\end{align}
%

%So for another defining basis state $|\varphi\rangle$
%%
%\begin{equation}
%|\langle \varphi|a_j^\dagger a_i |\psi\rangle |
%\le |\langle \varphi|\psi\rangle| N,
%\end{equation}
%which implies that $a_j^\dagger a_i/N$ is bounded in $k$, $N$ and $D$.
 
Using Eq.~\eqref{eq:a_j_a_i_psi} twice we get
\begin{align}
	\langle \psi | a_j^\dagger a_i &a_{j'}^\dagger a_{i'} |\psi\rangle
	= \delta_{i'j'}\delta_{ij} n_{i'}n_i \nonumber\\
	&+ (1-\delta_{i'j'})(1-\delta_{ij}) \delta_{ij'}\delta_{ji'}(n_{j'}+1)n_{i'}.
\end{align}

Now we will calculate $\tr(a_j^\dagger a_i a_{j'}^\dagger a_{i'})$. First let $i=j=i'=j'$. Using that there are $\binom{N-l+k-2}{k-2}$ states with $l$ particles on site $i$, we get
\begin{equation}
	\tr(n_i^2)
	= \sum_{l=0}^N l^2 \binom{N-l+k-2}{k-2}.
\end{equation}
Writing $l^2$ in terms of binomial coefficients and exploiting an upper index Vandermonde identity, namely
\begin{equation}\label{eq:vandermonde}
	\sum_{l=0}^n \binom{l}{c_1} \binom{n-l}{c_2} = \binom{n+1}{c_1+c_2+1},
\end{equation}
for constants $n$, $c_1$ and $c_2$, we get 
\begin{equation}\label{eq:n_i_exact}
 	\tr(n_i^2) 
 	= \frac{N(2N+k-1)}{k(k+1)} D.
\end{equation}

Now let $i=j$ and $i'=j'$ but $i\neq i'$. There are $\binom{N-l-s+k-3}{k-3}$ many states with $l$ particles on site $i$ and $s$ particles on site $i'$, so
\begin{equation}
\tr(n_i n_{i'}) = \sum_{l=0}^N \sum_{s=0}^{N-l} ls \binom{N-l-s+k-3}{k-3}.
\end{equation}
Invoking the Vandermonde identity Eq.~\eqref{eq:vandermonde} twice gives us
\begin{equation}\label{eq:n_i_n_i_prime_exact}
	\tr(n_i n_{i'})
	= \frac{N(N-1)}{(k+1)k} D.
\end{equation}

The case $i\neq j$ and $i'\neq j'$, but $j=i'$ and $i=j'$, works exactly the same as $i=j$ and $i'=j'$ but $i\neq i'$. Using Eq.~\eqref{eq:vandermonde} twice on
\begin{equation}
	\tr(a_j^\dagger a_i a_{i}^\dagger a_{j}) = \sum_{l=0}^{N} \sum_{s=0}^{N-l}(l+1) s \binom{N-l-s+k-3}{k-3}
\end{equation}
results in
\begin{equation}\label{eq:a_j_a_i_exact}
	\tr(a_j^\dagger a_i a_{i}^\dagger a_{j})
	= \frac{N(N+k)}{k(k+1)} D.
\end{equation}

In the classical limit, $k\ll N$, Eqs.\ \eqref{eq:n_i_exact}, \eqref{eq:n_i_n_i_prime_exact} and \eqref{eq:a_j_a_i_exact} all scale as $\sim N^2 D$, so the corresponding trace expressions of all typical observables scale as
\begin{equation}
	\tr(A^2) + \tr(A^\dagger A) 
	\sim N^2 D.
\end{equation}

%For intermediate regimes of $N$ and $k$ the exact scaling in these regimes depends on the linear combinations of $a_j^\dagger a_i$. For the observables investigated in this paper, $A=a_j^\dagger a_i$ being a single linear combination, the scaling for all $N$ and $k$, up to a constant not depending on $N$ nor on $k$, is retrieved by combining \cref{eq:n_i_exact} and \cref{eq:a_j_a_i_exact}
%\begin{equation}
%	\tr(A^2) + \tr(A^\dagger A) 
%	\sim \frac{N(N+k)}{k(k+1)} D.
%\end{equation}
%Fixing $k$ and varying $N$, the trace expressions scale, up to a constant only depending on $k$, as
%\begin{equation}
%	\tr(A^2) + \tr(A^\dagger A) 
%	\sim N(N+k) D.
%\end{equation}

In the thermodynamic limit, $k=\rho N$ for a density $\rho$ not depending on $N$ nor $k$ and $k,N\to\infty$, the trace expressions scale as
\begin{equation}
	\tr(A^2) + \tr(A^\dagger A) 
	\sim D,
\end{equation}
because the leading order of $k$ and $N$ is quadratic in both the numerators and the denumerators in
Eqs.\ \eqref{eq:n_i_exact}, \eqref{eq:n_i_n_i_prime_exact} and \eqref{eq:a_j_a_i_exact}. By combining this
with the results in Eq.~\eqref{sec:variance_calculation} we rediscover the $D^{-1/2}$ dependence of the
EEV fluctuations in the thermodynamic limit for Gaussian states and observables $A$, where
$\tr(A^2)$ and $\tr(AA^\dagger)$ are bounded in $D$.

\bibliography{document}

\begin{thebibliography}{118}%
\makeatletter
\providecommand \@ifxundefined [1]{%
 \@ifx{#1\undefined}
}%
\providecommand \@ifnum [1]{%
 \ifnum #1\expandafter \@firstoftwo
 \else \expandafter \@secondoftwo
 \fi
}%
\providecommand \@ifx [1]{%
 \ifx #1\expandafter \@firstoftwo
 \else \expandafter \@secondoftwo
 \fi
}%
\providecommand \natexlab [1]{#1}%
\providecommand \enquote  [1]{``#1''}%
\providecommand \bibnamefont  [1]{#1}%
\providecommand \bibfnamefont [1]{#1}%
\providecommand \citenamefont [1]{#1}%
\providecommand \href@noop [0]{\@secondoftwo}%
\providecommand \href [0]{\begingroup \@sanitize@url \@href}%
\providecommand \@href[1]{\@@startlink{#1}\@@href}%
\providecommand \@@href[1]{\endgroup#1\@@endlink}%
\providecommand \@sanitize@url [0]{\catcode `\\12\catcode `\$12\catcode
  `\&12\catcode `\#12\catcode `\^12\catcode `\_12\catcode `\%12\relax}%
\providecommand \@@startlink[1]{}%
\providecommand \@@endlink[0]{}%
\providecommand \url  [0]{\begingroup\@sanitize@url \@url }%
\providecommand \@url [1]{\endgroup\@href {#1}{\urlprefix }}%
\providecommand \urlprefix  [0]{URL }%
\providecommand \Eprint [0]{\href }%
\providecommand \doibase [0]{http://dx.doi.org/}%
\providecommand \selectlanguage [0]{\@gobble}%
\providecommand \bibinfo  [0]{\@secondoftwo}%
\providecommand \bibfield  [0]{\@secondoftwo}%
\providecommand \translation [1]{[#1]}%
\providecommand \BibitemOpen [0]{}%
\providecommand \bibitemStop [0]{}%
\providecommand \bibitemNoStop [0]{.\EOS\space}%
\providecommand \EOS [0]{\spacefactor3000\relax}%
\providecommand \BibitemShut  [1]{\csname bibitem#1\endcsname}%
\let\auto@bib@innerbib\@empty
%</preamble>
\bibitem [{\citenamefont {Deutsch}(1991)}]{Deutsch1991}%
  \BibitemOpen
  \bibfield  {author} {\bibinfo {author} {\bibfnamefont {J.~M.}\ \bibnamefont
  {Deutsch}},\ }\bibfield  {title} {\enquote {\bibinfo {title} {Quantum
  statistical mechanics in a closed system},}\ }\href {\doibase
  10.1103/PhysRevA.43.2046} {\bibfield  {journal} {\bibinfo  {journal} {Phys.
  Rev. A}\ }\textbf {\bibinfo {volume} {43}},\ \bibinfo {pages} {2046--2049}
  (\bibinfo {year} {1991})}\BibitemShut {NoStop}%
\bibitem [{\citenamefont {Srednicki}(1994)}]{Srednicki1994}%
  \BibitemOpen
  \bibfield  {author} {\bibinfo {author} {\bibfnamefont {Mark}\ \bibnamefont
  {Srednicki}},\ }\bibfield  {title} {\enquote {\bibinfo {title} {Chaos and
  quantum thermalization},}\ }\href {\doibase 10.1103/PhysRevE.50.888}
  {\bibfield  {journal} {\bibinfo  {journal} {Phys. Rev. E}\ }\textbf {\bibinfo
  {volume} {50}},\ \bibinfo {pages} {888--901} (\bibinfo {year}
  {1994})}\BibitemShut {NoStop}%
\bibitem [{\citenamefont {Srednicki}(1996)}]{srednicki1996thermal}%
  \BibitemOpen
  \bibfield  {author} {\bibinfo {author} {\bibfnamefont {Mark}\ \bibnamefont
  {Srednicki}},\ }\bibfield  {title} {\enquote {\bibinfo {title} {Thermal
  fluctuations in quantized chaotic systems},}\ }\href {\doibase
  10.1088/0305-4470/10/12/016} {\bibfield  {journal} {\bibinfo  {journal}
  {Journal of Physics A: Mathematical and General}\ }\textbf {\bibinfo {volume}
  {29}},\ \bibinfo {pages} {L75} (\bibinfo {year} {1996})}\BibitemShut
  {NoStop}%
\bibitem [{\citenamefont {Srednicki}(1999)}]{srednicki1999approach}%
  \BibitemOpen
  \bibfield  {author} {\bibinfo {author} {\bibfnamefont {Mark}\ \bibnamefont
  {Srednicki}},\ }\bibfield  {title} {\enquote {\bibinfo {title} {The approach
  to thermal equilibrium in quantized chaotic systems},}\ }\href {\doibase
  10.1088/0305-4470/32/7/007} {\bibfield  {journal} {\bibinfo  {journal}
  {Journal of Physics A: Mathematical and General}\ }\textbf {\bibinfo {volume}
  {32}},\ \bibinfo {pages} {1163} (\bibinfo {year} {1999})}\BibitemShut
  {NoStop}%
\bibitem [{\citenamefont {Rigol}\ \emph {et~al.}(2008)\citenamefont {Rigol},
  \citenamefont {Dunjko},\ and\ \citenamefont {Olshanii}}]{Rigol_Nature2008}%
  \BibitemOpen
  \bibfield  {author} {\bibinfo {author} {\bibfnamefont {Marcos}\ \bibnamefont
  {Rigol}}, \bibinfo {author} {\bibfnamefont {Vanja}\ \bibnamefont {Dunjko}}, \
  and\ \bibinfo {author} {\bibfnamefont {Maxim}\ \bibnamefont {Olshanii}},\
  }\bibfield  {title} {\enquote {\bibinfo {title} {{Thermalization and its
  mechanism for generic isolated quantum systems}},}\ }\href {\doibase
  10.1038/nature06838} {\bibfield  {journal} {\bibinfo  {journal} {Nature}\
  }\textbf {\bibinfo {volume} {452}},\ \bibinfo {pages} {854--858} (\bibinfo
  {year} {2008})}\BibitemShut {NoStop}%
\bibitem [{\citenamefont {Polkovnikov}\ \emph {et~al.}(2011)\citenamefont
  {Polkovnikov}, \citenamefont {Sengupta}, \citenamefont {Silva},\ and\
  \citenamefont {Vengalattore}}]{Polkovnikov_RMP2011}%
  \BibitemOpen
  \bibfield  {author} {\bibinfo {author} {\bibfnamefont {Anatoli}\ \bibnamefont
  {Polkovnikov}}, \bibinfo {author} {\bibfnamefont {Krishnendu}\ \bibnamefont
  {Sengupta}}, \bibinfo {author} {\bibfnamefont {Alessandro}\ \bibnamefont
  {Silva}}, \ and\ \bibinfo {author} {\bibfnamefont {Mukund}\ \bibnamefont
  {Vengalattore}},\ }\bibfield  {title} {\enquote {\bibinfo {title}
  {Colloquium: Nonequilibrium dynamics of closed interacting quantum
  systems},}\ }\href {\doibase 10.1103/RevModPhys.83.863} {\bibfield  {journal}
  {\bibinfo  {journal} {Rev. Mod. Phys.}\ }\textbf {\bibinfo {volume} {83}},\
  \bibinfo {pages} {863--883} (\bibinfo {year} {2011})}\BibitemShut {NoStop}%
\bibitem [{\citenamefont {Rigol}\ and\ \citenamefont
  {Srednicki}(2012)}]{Rigol_Srednicki_PRL2012_alternatives}%
  \BibitemOpen
  \bibfield  {author} {\bibinfo {author} {\bibfnamefont {Marcos}\ \bibnamefont
  {Rigol}}\ and\ \bibinfo {author} {\bibfnamefont {Mark}\ \bibnamefont
  {Srednicki}},\ }\bibfield  {title} {\enquote {\bibinfo {title} {Alternatives
  to eigenstate thermalization},}\ }\href {\doibase
  10.1103/PhysRevLett.108.110601} {\bibfield  {journal} {\bibinfo  {journal}
  {Phys. Rev. Lett.}\ }\textbf {\bibinfo {volume} {108}},\ \bibinfo {pages}
  {110601} (\bibinfo {year} {2012})}\BibitemShut {NoStop}%
\bibitem [{\citenamefont {Eisert}\ \emph {et~al.}(2015)\citenamefont {Eisert},
  \citenamefont {Friesdorf},\ and\ \citenamefont {Gogolin}}]{Eisert2015}%
  \BibitemOpen
  \bibfield  {author} {\bibinfo {author} {\bibfnamefont {Jens}\ \bibnamefont
  {Eisert}}, \bibinfo {author} {\bibfnamefont {Mathis}\ \bibnamefont
  {Friesdorf}}, \ and\ \bibinfo {author} {\bibfnamefont {Christian}\
  \bibnamefont {Gogolin}},\ }\bibfield  {title} {\enquote {\bibinfo {title}
  {Quantum many-body systems out of equilibrium},}\ }\href {\doibase
  10.1038/nphys3215} {\bibfield  {journal} {\bibinfo  {journal} {Nature
  Physics}\ }\textbf {\bibinfo {volume} {11}},\ \bibinfo {pages} {124--130}
  (\bibinfo {year} {2015})}\BibitemShut {NoStop}%
\bibitem [{\citenamefont {D'Alessio}\ \emph {et~al.}(2016)\citenamefont
  {D'Alessio}, \citenamefont {Kafri}, \citenamefont {Polkovnikov},\ and\
  \citenamefont {Rigol}}]{Dalessio2016}%
  \BibitemOpen
  \bibfield  {author} {\bibinfo {author} {\bibfnamefont {Luca}\ \bibnamefont
  {D'Alessio}}, \bibinfo {author} {\bibfnamefont {Yariv}\ \bibnamefont
  {Kafri}}, \bibinfo {author} {\bibfnamefont {Anatoli}\ \bibnamefont
  {Polkovnikov}}, \ and\ \bibinfo {author} {\bibfnamefont {Marcos}\
  \bibnamefont {Rigol}},\ }\bibfield  {title} {\enquote {\bibinfo {title} {From
  quantum chaos and eigenstate thermalization to statistical mechanics and
  thermodynamics},}\ }\href {\doibase 10.1080/00018732.2016.1198134} {\bibfield
   {journal} {\bibinfo  {journal} {Advances in Physics}\ }\textbf {\bibinfo
  {volume} {65}},\ \bibinfo {pages} {239} (\bibinfo {year} {2016})}\BibitemShut
  {NoStop}%
\bibitem [{\citenamefont {Neuenhahn}\ and\ \citenamefont
  {Marquardt}(2012)}]{Marquardt_PRE2012}%
  \BibitemOpen
  \bibfield  {author} {\bibinfo {author} {\bibfnamefont {Clemens}\ \bibnamefont
  {Neuenhahn}}\ and\ \bibinfo {author} {\bibfnamefont {Florian}\ \bibnamefont
  {Marquardt}},\ }\bibfield  {title} {\enquote {\bibinfo {title}
  {Thermalization of interacting fermions and delocalization in fock space},}\
  }\href {\doibase 10.1103/PhysRevE.85.060101} {\bibfield  {journal} {\bibinfo
  {journal} {Phys. Rev. E}\ }\textbf {\bibinfo {volume} {85}},\ \bibinfo
  {pages} {060101} (\bibinfo {year} {2012})}\BibitemShut {NoStop}%
\bibitem [{\citenamefont {Beugeling}\ \emph {et~al.}(2014)\citenamefont
  {Beugeling}, \citenamefont {Moessner},\ and\ \citenamefont
  {Haque}}]{Beugeling_scaling_PRE14}%
  \BibitemOpen
  \bibfield  {author} {\bibinfo {author} {\bibfnamefont {W.}~\bibnamefont
  {Beugeling}}, \bibinfo {author} {\bibfnamefont {R.}~\bibnamefont {Moessner}},
  \ and\ \bibinfo {author} {\bibfnamefont {Masudul}\ \bibnamefont {Haque}},\
  }\bibfield  {title} {\enquote {\bibinfo {title} {Finite-size scaling of
  eigenstate thermalization},}\ }\href {\doibase 10.1103/PhysRevE.89.042112}
  {\bibfield  {journal} {\bibinfo  {journal} {Phys. Rev. E}\ }\textbf {\bibinfo
  {volume} {89}},\ \bibinfo {pages} {042112} (\bibinfo {year}
  {2014})}\BibitemShut {NoStop}%
\bibitem [{\citenamefont {Beugeling}\ \emph
  {et~al.}(2015{\natexlab{a}})\citenamefont {Beugeling}, \citenamefont
  {Moessner},\ and\ \citenamefont {Haque}}]{Beugeling_offdiag_PRE2015}%
  \BibitemOpen
  \bibfield  {author} {\bibinfo {author} {\bibfnamefont {Wouter}\ \bibnamefont
  {Beugeling}}, \bibinfo {author} {\bibfnamefont {Roderich}\ \bibnamefont
  {Moessner}}, \ and\ \bibinfo {author} {\bibfnamefont {Masudul}\ \bibnamefont
  {Haque}},\ }\bibfield  {title} {\enquote {\bibinfo {title} {Off-diagonal
  matrix elements of local operators in many-body quantum systems},}\ }\href
  {\doibase 10.1103/PhysRevE.91.012144} {\bibfield  {journal} {\bibinfo
  {journal} {Phys. Rev. E}\ }\textbf {\bibinfo {volume} {91}},\ \bibinfo
  {pages} {012144} (\bibinfo {year} {2015}{\natexlab{a}})}\BibitemShut
  {NoStop}%
\bibitem [{\citenamefont {Ziraldo}\ and\ \citenamefont
  {Santoro}(2013)}]{Ziraldo_Santoro_relaxation_PRB2013}%
  \BibitemOpen
  \bibfield  {author} {\bibinfo {author} {\bibfnamefont {Simone}\ \bibnamefont
  {Ziraldo}}\ and\ \bibinfo {author} {\bibfnamefont {Giuseppe~E.}\ \bibnamefont
  {Santoro}},\ }\bibfield  {title} {\enquote {\bibinfo {title} {Relaxation and
  thermalization after a quantum quench: Why localization is important},}\
  }\href {\doibase 10.1103/PhysRevB.87.064201} {\bibfield  {journal} {\bibinfo
  {journal} {Phys. Rev. B}\ }\textbf {\bibinfo {volume} {87}},\ \bibinfo
  {pages} {064201} (\bibinfo {year} {2013})}\BibitemShut {NoStop}%
\bibitem [{\citenamefont {Ikeda}\ \emph {et~al.}(2013)\citenamefont {Ikeda},
  \citenamefont {Watanabe},\ and\ \citenamefont
  {Ueda}}]{Ikeda_Ueda_PRE2013_LiebLiniger}%
  \BibitemOpen
  \bibfield  {author} {\bibinfo {author} {\bibfnamefont {Tatsuhiko~N.}\
  \bibnamefont {Ikeda}}, \bibinfo {author} {\bibfnamefont {Yu}~\bibnamefont
  {Watanabe}}, \ and\ \bibinfo {author} {\bibfnamefont {Masahito}\ \bibnamefont
  {Ueda}},\ }\bibfield  {title} {\enquote {\bibinfo {title} {Finite-size
  scaling analysis of the eigenstate thermalization hypothesis in a
  one-dimensional interacting bose gas},}\ }\href {\doibase
  10.1103/PhysRevE.87.012125} {\bibfield  {journal} {\bibinfo  {journal} {Phys.
  Rev. E}\ }\textbf {\bibinfo {volume} {87}},\ \bibinfo {pages} {012125}
  (\bibinfo {year} {2013})}\BibitemShut {NoStop}%
\bibitem [{\citenamefont {Alba}(2015)}]{Alba_PRB2015}%
  \BibitemOpen
  \bibfield  {author} {\bibinfo {author} {\bibfnamefont {Vincenzo}\
  \bibnamefont {Alba}},\ }\bibfield  {title} {\enquote {\bibinfo {title}
  {Eigenstate thermalization hypothesis and integrability in quantum spin
  chains},}\ }\href {\doibase 10.1103/PhysRevB.91.155123} {\bibfield  {journal}
  {\bibinfo  {journal} {Phys. Rev. B}\ }\textbf {\bibinfo {volume} {91}},\
  \bibinfo {pages} {155123} (\bibinfo {year} {2015})}\BibitemShut {NoStop}%
\bibitem [{\citenamefont {Nandy}\ \emph {et~al.}(2016)\citenamefont {Nandy},
  \citenamefont {Sen}, \citenamefont {Das},\ and\ \citenamefont
  {Dhar}}]{ArnabSenArnabDas_PRB16}%
  \BibitemOpen
  \bibfield  {author} {\bibinfo {author} {\bibfnamefont {Sourav}\ \bibnamefont
  {Nandy}}, \bibinfo {author} {\bibfnamefont {Arnab}\ \bibnamefont {Sen}},
  \bibinfo {author} {\bibfnamefont {Arnab}\ \bibnamefont {Das}}, \ and\
  \bibinfo {author} {\bibfnamefont {Abhishek}\ \bibnamefont {Dhar}},\
  }\bibfield  {title} {\enquote {\bibinfo {title} {Eigenstate gibbs ensemble in
  integrable quantum systems},}\ }\href {\doibase 10.1103/PhysRevB.94.245131}
  {\bibfield  {journal} {\bibinfo  {journal} {Phys. Rev. B}\ }\textbf {\bibinfo
  {volume} {94}},\ \bibinfo {pages} {245131} (\bibinfo {year}
  {2016})}\BibitemShut {NoStop}%
\bibitem [{\citenamefont {Mag\'an}(2016)}]{Magan_randomfreefermions_PRL2016}%
  \BibitemOpen
  \bibfield  {author} {\bibinfo {author} {\bibfnamefont {Javier~M.}\
  \bibnamefont {Mag\'an}},\ }\bibfield  {title} {\enquote {\bibinfo {title}
  {Random free fermions: An analytical example of eigenstate thermalization},}\
  }\href {\doibase 10.1103/PhysRevLett.116.030401} {\bibfield  {journal}
  {\bibinfo  {journal} {Phys. Rev. Lett.}\ }\textbf {\bibinfo {volume} {116}},\
  \bibinfo {pages} {030401} (\bibinfo {year} {2016})}\BibitemShut {NoStop}%
\bibitem [{\citenamefont {Haque}\ and\ \citenamefont
  {McClarty}(2019)}]{HaqueMcClarty_SYKETH_PRB2019}%
  \BibitemOpen
  \bibfield  {author} {\bibinfo {author} {\bibfnamefont {M.}~\bibnamefont
  {Haque}}\ and\ \bibinfo {author} {\bibfnamefont {P.~A.}\ \bibnamefont
  {McClarty}},\ }\bibfield  {title} {\enquote {\bibinfo {title} {Eigenstate
  thermalization scaling in majorana clusters: From chaotic to integrable
  sachdev-ye-kitaev models},}\ }\href {\doibase 10.1103/PhysRevB.100.115122}
  {\bibfield  {journal} {\bibinfo  {journal} {Phys. Rev. B}\ }\textbf {\bibinfo
  {volume} {100}},\ \bibinfo {pages} {115122} (\bibinfo {year}
  {2019})}\BibitemShut {NoStop}%
\bibitem [{\citenamefont {Mierzejewski}\ and\ \citenamefont
  {Vidmar}(2020)}]{Mierzejewski_Vidmar_PRL2020}%
  \BibitemOpen
  \bibfield  {author} {\bibinfo {author} {\bibfnamefont {Marcin}\ \bibnamefont
  {Mierzejewski}}\ and\ \bibinfo {author} {\bibfnamefont {Lev}\ \bibnamefont
  {Vidmar}},\ }\bibfield  {title} {\enquote {\bibinfo {title} {Quantitative
  impact of integrals of motion on the eigenstate thermalization hypothesis},}\
  }\href {\doibase 10.1103/PhysRevLett.124.040603} {\bibfield  {journal}
  {\bibinfo  {journal} {Phys. Rev. Lett.}\ }\textbf {\bibinfo {volume} {124}},\
  \bibinfo {pages} {040603} (\bibinfo {year} {2020})}\BibitemShut {NoStop}%
\bibitem [{\citenamefont {LeBlond}\ \emph {et~al.}(2019)\citenamefont
  {LeBlond}, \citenamefont {Mallayya}, \citenamefont {Vidmar},\ and\
  \citenamefont {Rigol}}]{LeBlond_Mallaya_Vidmar_Rigol_PRE2019}%
  \BibitemOpen
  \bibfield  {author} {\bibinfo {author} {\bibfnamefont {Tyler}\ \bibnamefont
  {LeBlond}}, \bibinfo {author} {\bibfnamefont {Krishnanand}\ \bibnamefont
  {Mallayya}}, \bibinfo {author} {\bibfnamefont {Lev}\ \bibnamefont {Vidmar}},
  \ and\ \bibinfo {author} {\bibfnamefont {Marcos}\ \bibnamefont {Rigol}},\
  }\bibfield  {title} {\enquote {\bibinfo {title} {Entanglement and matrix
  elements of observables in interacting integrable systems},}\ }\href
  {\doibase 10.1103/PhysRevE.100.062134} {\bibfield  {journal} {\bibinfo
  {journal} {Phys. Rev. E}\ }\textbf {\bibinfo {volume} {100}},\ \bibinfo
  {pages} {062134} (\bibinfo {year} {2019})}\BibitemShut {NoStop}%
\bibitem [{\citenamefont {Rigol}(2009)}]{Rigol_PRL2009}%
  \BibitemOpen
  \bibfield  {author} {\bibinfo {author} {\bibfnamefont {Marcos}\ \bibnamefont
  {Rigol}},\ }\bibfield  {title} {\enquote {\bibinfo {title} {Breakdown of
  thermalization in finite one-dimensional systems},}\ }\href {\doibase
  10.1103/PhysRevLett.103.100403} {\bibfield  {journal} {\bibinfo  {journal}
  {Phys. Rev. Lett.}\ }\textbf {\bibinfo {volume} {103}},\ \bibinfo {pages}
  {100403} (\bibinfo {year} {2009})}\BibitemShut {NoStop}%
\bibitem [{\citenamefont {Biroli}\ \emph {et~al.}(2010)\citenamefont {Biroli},
  \citenamefont {Kollath},\ and\ \citenamefont
  {L\"auchli}}]{BiroliKollathLauchli_PRL10}%
  \BibitemOpen
  \bibfield  {author} {\bibinfo {author} {\bibfnamefont {Giulio}\ \bibnamefont
  {Biroli}}, \bibinfo {author} {\bibfnamefont {Corinna}\ \bibnamefont
  {Kollath}}, \ and\ \bibinfo {author} {\bibfnamefont {Andreas~M.}\
  \bibnamefont {L\"auchli}},\ }\bibfield  {title} {\enquote {\bibinfo {title}
  {Effect of rare fluctuations on the thermalization of isolated quantum
  systems},}\ }\href {\doibase 10.1103/PhysRevLett.105.250401} {\bibfield
  {journal} {\bibinfo  {journal} {Phys. Rev. Lett.}\ }\textbf {\bibinfo
  {volume} {105}},\ \bibinfo {pages} {250401} (\bibinfo {year}
  {2010})}\BibitemShut {NoStop}%
\bibitem [{\citenamefont {Rigol}\ and\ \citenamefont
  {Santos}(2010)}]{RigolSantos_PRA10}%
  \BibitemOpen
  \bibfield  {author} {\bibinfo {author} {\bibfnamefont {Marcos}\ \bibnamefont
  {Rigol}}\ and\ \bibinfo {author} {\bibfnamefont {Lea~F.}\ \bibnamefont
  {Santos}},\ }\bibfield  {title} {\enquote {\bibinfo {title} {Quantum chaos
  and thermalization in gapped systems},}\ }\href {\doibase
  10.1103/PhysRevA.82.011604} {\bibfield  {journal} {\bibinfo  {journal} {Phys.
  Rev. A}\ }\textbf {\bibinfo {volume} {82}},\ \bibinfo {pages} {011604}
  (\bibinfo {year} {2010})}\BibitemShut {NoStop}%
\bibitem [{\citenamefont {Santos}\ and\ \citenamefont
  {Rigol}(2010)}]{SantosRigol_PRE10}%
  \BibitemOpen
  \bibfield  {author} {\bibinfo {author} {\bibfnamefont {Lea~F.}\ \bibnamefont
  {Santos}}\ and\ \bibinfo {author} {\bibfnamefont {Marcos}\ \bibnamefont
  {Rigol}},\ }\bibfield  {title} {\enquote {\bibinfo {title} {Localization and
  the effects of symmetries in the thermalization properties of one-dimensional
  quantum systems},}\ }\href {\doibase 10.1103/PhysRevE.82.031130} {\bibfield
  {journal} {\bibinfo  {journal} {Phys. Rev. E}\ }\textbf {\bibinfo {volume}
  {82}},\ \bibinfo {pages} {031130} (\bibinfo {year} {2010})}\BibitemShut
  {NoStop}%
\bibitem [{\citenamefont {Roux}(2010)}]{Roux_PRA2010}%
  \BibitemOpen
  \bibfield  {author} {\bibinfo {author} {\bibfnamefont {Guillaume}\
  \bibnamefont {Roux}},\ }\bibfield  {title} {\enquote {\bibinfo {title}
  {Finite-size effects in global quantum quenches: {E}xamples from free bosons
  in an harmonic trap and the one-dimensional {B}ose-{H}ubbard model},}\ }\href
  {\doibase 10.1103/PhysRevA.81.053604} {\bibfield  {journal} {\bibinfo
  {journal} {Phys. Rev. A}\ }\textbf {\bibinfo {volume} {81}},\ \bibinfo
  {pages} {053604} (\bibinfo {year} {2010})}\BibitemShut {NoStop}%
\bibitem [{\citenamefont {Motohashi}(2011)}]{Motohashi2011}%
  \BibitemOpen
  \bibfield  {author} {\bibinfo {author} {\bibfnamefont {Atsushi}\ \bibnamefont
  {Motohashi}},\ }\bibfield  {title} {\enquote {\bibinfo {title}
  {Thermalization of atom-molecule bose gases in a double-well potential},}\
  }\href {\doibase 10.1103/PhysRevA.84.063631} {\bibfield  {journal} {\bibinfo
  {journal} {Phys. Rev. A}\ }\textbf {\bibinfo {volume} {84}},\ \bibinfo
  {pages} {063631} (\bibinfo {year} {2011})}\BibitemShut {NoStop}%
\bibitem [{\citenamefont {Brandino}\ \emph {et~al.}(2012)\citenamefont
  {Brandino}, \citenamefont {De~Luca}, \citenamefont {Konik},\ and\
  \citenamefont {Mussardo}}]{BrandinoKonikMussardo_PRB12}%
  \BibitemOpen
  \bibfield  {author} {\bibinfo {author} {\bibfnamefont {G.~P.}\ \bibnamefont
  {Brandino}}, \bibinfo {author} {\bibfnamefont {A.}~\bibnamefont {De~Luca}},
  \bibinfo {author} {\bibfnamefont {R.~M.}\ \bibnamefont {Konik}}, \ and\
  \bibinfo {author} {\bibfnamefont {G.}~\bibnamefont {Mussardo}},\ }\bibfield
  {title} {\enquote {\bibinfo {title} {Quench dynamics in randomly generated
  extended quantum models},}\ }\href {\doibase 10.1103/PhysRevB.85.214435}
  {\bibfield  {journal} {\bibinfo  {journal} {Phys. Rev. B}\ }\textbf {\bibinfo
  {volume} {85}},\ \bibinfo {pages} {214435} (\bibinfo {year}
  {2012})}\BibitemShut {NoStop}%
\bibitem [{\citenamefont {Steinigeweg}\ \emph {et~al.}(2013)\citenamefont
  {Steinigeweg}, \citenamefont {Herbrych},\ and\ \citenamefont
  {Prelov\ifmmode~\check{s}\else \v{s}\fi{}ek}}]{Steinigeweg_Prelovsek_PRE13}%
  \BibitemOpen
  \bibfield  {author} {\bibinfo {author} {\bibfnamefont {R.}~\bibnamefont
  {Steinigeweg}}, \bibinfo {author} {\bibfnamefont {J.}~\bibnamefont
  {Herbrych}}, \ and\ \bibinfo {author} {\bibfnamefont {P.}~\bibnamefont
  {Prelov\ifmmode~\check{s}\else \v{s}\fi{}ek}},\ }\bibfield  {title} {\enquote
  {\bibinfo {title} {Eigenstate thermalization within isolated spin-chain
  systems},}\ }\href {\doibase 10.1103/PhysRevE.87.012118} {\bibfield
  {journal} {\bibinfo  {journal} {Phys. Rev. E}\ }\textbf {\bibinfo {volume}
  {87}},\ \bibinfo {pages} {012118} (\bibinfo {year} {2013})}\BibitemShut
  {NoStop}%
\bibitem [{\citenamefont {Kim}\ \emph {et~al.}(2014)\citenamefont {Kim},
  \citenamefont {Ikeda},\ and\ \citenamefont {Huse}}]{Kim_Ikeda_Huse_PRE2014}%
  \BibitemOpen
  \bibfield  {author} {\bibinfo {author} {\bibfnamefont {Hyungwon}\
  \bibnamefont {Kim}}, \bibinfo {author} {\bibfnamefont {Tatsuhiko~N.}\
  \bibnamefont {Ikeda}}, \ and\ \bibinfo {author} {\bibfnamefont {David~A.}\
  \bibnamefont {Huse}},\ }\bibfield  {title} {\enquote {\bibinfo {title}
  {Testing whether all eigenstates obey the eigenstate thermalization
  hypothesis},}\ }\href {\doibase 10.1103/PhysRevE.90.052105} {\bibfield
  {journal} {\bibinfo  {journal} {Phys. Rev. E}\ }\textbf {\bibinfo {volume}
  {90}},\ \bibinfo {pages} {052105} (\bibinfo {year} {2014})}\BibitemShut
  {NoStop}%
\bibitem [{\citenamefont {Sorg}\ \emph {et~al.}(2014)\citenamefont {Sorg},
  \citenamefont {Vidmar}, \citenamefont {Pollet},\ and\ \citenamefont
  {Heidrich-Meisner}}]{Sorg_Vidmar_Pollet_HeidrichMeisner_PRA2014}%
  \BibitemOpen
  \bibfield  {author} {\bibinfo {author} {\bibfnamefont {S.}~\bibnamefont
  {Sorg}}, \bibinfo {author} {\bibfnamefont {L.}~\bibnamefont {Vidmar}},
  \bibinfo {author} {\bibfnamefont {L.}~\bibnamefont {Pollet}}, \ and\ \bibinfo
  {author} {\bibfnamefont {F.}~\bibnamefont {Heidrich-Meisner}},\ }\bibfield
  {title} {\enquote {\bibinfo {title} {Relaxation and thermalization in the
  one-dimensional bose-hubbard model: A case study for the interaction quantum
  quench from the atomic limit},}\ }\href {\doibase 10.1103/PhysRevA.90.033606}
  {\bibfield  {journal} {\bibinfo  {journal} {Phys. Rev. A}\ }\textbf {\bibinfo
  {volume} {90}},\ \bibinfo {pages} {033606} (\bibinfo {year}
  {2014})}\BibitemShut {NoStop}%
\bibitem [{\citenamefont {Steinigeweg}\ \emph {et~al.}(2014)\citenamefont
  {Steinigeweg}, \citenamefont {Khodja}, \citenamefont {Niemeyer},
  \citenamefont {Gogolin},\ and\ \citenamefont
  {Gemmer}}]{Steinigeweg_Gogolin_Gemmer_PRL2014}%
  \BibitemOpen
  \bibfield  {author} {\bibinfo {author} {\bibfnamefont {R.}~\bibnamefont
  {Steinigeweg}}, \bibinfo {author} {\bibfnamefont {A.}~\bibnamefont {Khodja}},
  \bibinfo {author} {\bibfnamefont {H.}~\bibnamefont {Niemeyer}}, \bibinfo
  {author} {\bibfnamefont {C.}~\bibnamefont {Gogolin}}, \ and\ \bibinfo
  {author} {\bibfnamefont {J.}~\bibnamefont {Gemmer}},\ }\bibfield  {title}
  {\enquote {\bibinfo {title} {Pushing the limits of the eigenstate
  thermalization hypothesis towards mesoscopic quantum systems},}\ }\href
  {\doibase 10.1103/PhysRevLett.112.130403} {\bibfield  {journal} {\bibinfo
  {journal} {Phys. Rev. Lett.}\ }\textbf {\bibinfo {volume} {112}},\ \bibinfo
  {pages} {130403} (\bibinfo {year} {2014})}\BibitemShut {NoStop}%
\bibitem [{\citenamefont {Fratus}\ and\ \citenamefont
  {Srednicki}(2015)}]{Fratus_Srednicki_PRE2015}%
  \BibitemOpen
  \bibfield  {author} {\bibinfo {author} {\bibfnamefont {Keith~R.}\
  \bibnamefont {Fratus}}\ and\ \bibinfo {author} {\bibfnamefont {Mark}\
  \bibnamefont {Srednicki}},\ }\bibfield  {title} {\enquote {\bibinfo {title}
  {Eigenstate thermalization in systems with spontaneously broken symmetry},}\
  }\href {\doibase 10.1103/PhysRevE.92.040103} {\bibfield  {journal} {\bibinfo
  {journal} {Phys. Rev. E}\ }\textbf {\bibinfo {volume} {92}},\ \bibinfo
  {pages} {040103} (\bibinfo {year} {2015})}\BibitemShut {NoStop}%
\bibitem [{\citenamefont {Nandkishore}\ and\ \citenamefont
  {Huse}(2015)}]{Nandkishore_Huse_AnnuRev2015}%
  \BibitemOpen
  \bibfield  {author} {\bibinfo {author} {\bibfnamefont {Rahul}\ \bibnamefont
  {Nandkishore}}\ and\ \bibinfo {author} {\bibfnamefont {David~A.}\
  \bibnamefont {Huse}},\ }\bibfield  {title} {\enquote {\bibinfo {title}
  {{Many-Body Localization and Thermalization in Quantum Statistical
  Mechanics}},}\ }\href {\doibase 10.1146/annurev-conmatphys-031214-014726}
  {\bibfield  {journal} {\bibinfo  {journal} {Annual Review of Condensed Matter
  Physics}\ }\textbf {\bibinfo {volume} {6}},\ \bibinfo {pages} {15--38}
  (\bibinfo {year} {2015})}\BibitemShut {NoStop}%
\bibitem [{\citenamefont {Khodja}\ \emph {et~al.}(2015)\citenamefont {Khodja},
  \citenamefont {Steinigeweg},\ and\ \citenamefont
  {Gemmer}}]{Khodja_Steinigeweg_Gemmer_PRE2015}%
  \BibitemOpen
  \bibfield  {author} {\bibinfo {author} {\bibfnamefont {Abdellah}\
  \bibnamefont {Khodja}}, \bibinfo {author} {\bibfnamefont {Robin}\
  \bibnamefont {Steinigeweg}}, \ and\ \bibinfo {author} {\bibfnamefont
  {Jochen}\ \bibnamefont {Gemmer}},\ }\bibfield  {title} {\enquote {\bibinfo
  {title} {Relevance of the eigenstate thermalization hypothesis for thermal
  relaxation},}\ }\href {\doibase 10.1103/PhysRevE.91.012120} {\bibfield
  {journal} {\bibinfo  {journal} {Phys. Rev. E}\ }\textbf {\bibinfo {volume}
  {91}},\ \bibinfo {pages} {012120} (\bibinfo {year} {2015})}\BibitemShut
  {NoStop}%
\bibitem [{\citenamefont {Mondaini}\ \emph {et~al.}(2016)\citenamefont
  {Mondaini}, \citenamefont {Fratus}, \citenamefont {Srednicki},\ and\
  \citenamefont {Rigol}}]{Mondaini_Srednicki_Rigol_PRE2016}%
  \BibitemOpen
  \bibfield  {author} {\bibinfo {author} {\bibfnamefont {Rubem}\ \bibnamefont
  {Mondaini}}, \bibinfo {author} {\bibfnamefont {Keith~R.}\ \bibnamefont
  {Fratus}}, \bibinfo {author} {\bibfnamefont {Mark}\ \bibnamefont
  {Srednicki}}, \ and\ \bibinfo {author} {\bibfnamefont {Marcos}\ \bibnamefont
  {Rigol}},\ }\bibfield  {title} {\enquote {\bibinfo {title} {Eigenstate
  thermalization in the two-dimensional transverse field ising model},}\ }\href
  {\doibase 10.1103/PhysRevE.93.032104} {\bibfield  {journal} {\bibinfo
  {journal} {Phys. Rev. E}\ }\textbf {\bibinfo {volume} {93}},\ \bibinfo
  {pages} {032104} (\bibinfo {year} {2016})}\BibitemShut {NoStop}%
\bibitem [{\citenamefont {Chandran}\ \emph {et~al.}(2016)\citenamefont
  {Chandran}, \citenamefont {Schulz},\ and\ \citenamefont
  {Burnell}}]{Chandran_Burnell_PRB2016}%
  \BibitemOpen
  \bibfield  {author} {\bibinfo {author} {\bibfnamefont {A.}~\bibnamefont
  {Chandran}}, \bibinfo {author} {\bibfnamefont {Marc~D.}\ \bibnamefont
  {Schulz}}, \ and\ \bibinfo {author} {\bibfnamefont {F.~J.}\ \bibnamefont
  {Burnell}},\ }\bibfield  {title} {\enquote {\bibinfo {title} {The eigenstate
  thermalization hypothesis in constrained hilbert spaces: A case study in
  non-abelian anyo n chains},}\ }\href {\doibase 10.1103/PhysRevB.94.235122}
  {\bibfield  {journal} {\bibinfo  {journal} {Phys. Rev. B}\ }\textbf {\bibinfo
  {volume} {94}},\ \bibinfo {pages} {235122} (\bibinfo {year}
  {2016})}\BibitemShut {NoStop}%
\bibitem [{\citenamefont {Luitz}\ and\ \citenamefont
  {Bar~Lev}(2016)}]{Luitz_Barlev_PRL2016}%
  \BibitemOpen
  \bibfield  {author} {\bibinfo {author} {\bibfnamefont {David~J.}\
  \bibnamefont {Luitz}}\ and\ \bibinfo {author} {\bibfnamefont {Yevgeny}\
  \bibnamefont {Bar~Lev}},\ }\bibfield  {title} {\enquote {\bibinfo {title}
  {Anomalous thermalization in ergodic systems},}\ }\href {\doibase
  10.1103/PhysRevLett.117.170404} {\bibfield  {journal} {\bibinfo  {journal}
  {Phys. Rev. Lett.}\ }\textbf {\bibinfo {volume} {117}},\ \bibinfo {pages}
  {170404} (\bibinfo {year} {2016})}\BibitemShut {NoStop}%
\bibitem [{\citenamefont {Mondaini}\ and\ \citenamefont
  {Rigol}(2017)}]{Mondaini_Rigol_PRE2017_tfIsing}%
  \BibitemOpen
  \bibfield  {author} {\bibinfo {author} {\bibfnamefont {Rubem}\ \bibnamefont
  {Mondaini}}\ and\ \bibinfo {author} {\bibfnamefont {Marcos}\ \bibnamefont
  {Rigol}},\ }\bibfield  {title} {\enquote {\bibinfo {title} {Eigenstate
  thermalization in the two-dimensional transverse field ising model. ii.
  off-diagonal matrix elements of observables},}\ }\href {\doibase
  10.1103/PhysRevE.96.012157} {\bibfield  {journal} {\bibinfo  {journal} {Phys.
  Rev. E}\ }\textbf {\bibinfo {volume} {96}},\ \bibinfo {pages} {012157}
  (\bibinfo {year} {2017})}\BibitemShut {NoStop}%
\bibitem [{\citenamefont {Sonner}\ and\ \citenamefont
  {Vielma}(2017)}]{Sonner_JHEP2017_eth_syk}%
  \BibitemOpen
  \bibfield  {author} {\bibinfo {author} {\bibfnamefont {Julian}\ \bibnamefont
  {Sonner}}\ and\ \bibinfo {author} {\bibfnamefont {Manuel}\ \bibnamefont
  {Vielma}},\ }\bibfield  {title} {\enquote {\bibinfo {title} {Eigenstate
  thermalization in the sachdev-ye-kitaev model},}\ }\href
  {https://doi.org/10.1007/JHEP11(2017)149} {\bibfield  {journal} {\bibinfo
  {journal} {Journal of High Energy Physics}\ }\textbf {\bibinfo {volume}
  {2017}},\ \bibinfo {pages} {149} (\bibinfo {year} {2017})}\BibitemShut
  {NoStop}%
\bibitem [{\citenamefont {Lan}\ and\ \citenamefont
  {Powell}(2017)}]{Powell_PRB2017_dimermodelETH}%
  \BibitemOpen
  \bibfield  {author} {\bibinfo {author} {\bibfnamefont {Zhihao}\ \bibnamefont
  {Lan}}\ and\ \bibinfo {author} {\bibfnamefont {Stephen}\ \bibnamefont
  {Powell}},\ }\bibfield  {title} {\enquote {\bibinfo {title} {Eigenstate
  thermalization hypothesis in quantum dimer models},}\ }\href {\doibase
  10.1103/PhysRevB.96.115140} {\bibfield  {journal} {\bibinfo  {journal} {Phys.
  Rev. B}\ }\textbf {\bibinfo {volume} {96}},\ \bibinfo {pages} {115140}
  (\bibinfo {year} {2017})}\BibitemShut {NoStop}%
\bibitem [{\citenamefont {Yoshizawa}\ \emph {et~al.}(2018)\citenamefont
  {Yoshizawa}, \citenamefont {Iyoda},\ and\ \citenamefont
  {Sagawa}}]{Sagawa_PRL2018_LargeDeviationETH}%
  \BibitemOpen
  \bibfield  {author} {\bibinfo {author} {\bibfnamefont {Toru}\ \bibnamefont
  {Yoshizawa}}, \bibinfo {author} {\bibfnamefont {Eiki}\ \bibnamefont {Iyoda}},
  \ and\ \bibinfo {author} {\bibfnamefont {Takahiro}\ \bibnamefont {Sagawa}},\
  }\bibfield  {title} {\enquote {\bibinfo {title} {Numerical large deviation
  analysis of the eigenstate thermalization hypothesis},}\ }\href {\doibase
  10.1103/PhysRevLett.120.200604} {\bibfield  {journal} {\bibinfo  {journal}
  {Phys. Rev. Lett.}\ }\textbf {\bibinfo {volume} {120}},\ \bibinfo {pages}
  {200604} (\bibinfo {year} {2018})}\BibitemShut {NoStop}%
\bibitem [{\citenamefont {Dymarsky}\ \emph {et~al.}(2018)\citenamefont
  {Dymarsky}, \citenamefont {Lashkari},\ and\ \citenamefont
  {Liu}}]{Dymarsky_Lashkari_Liu_PRE2018}%
  \BibitemOpen
  \bibfield  {author} {\bibinfo {author} {\bibfnamefont {Anatoly}\ \bibnamefont
  {Dymarsky}}, \bibinfo {author} {\bibfnamefont {Nima}\ \bibnamefont
  {Lashkari}}, \ and\ \bibinfo {author} {\bibfnamefont {Hong}\ \bibnamefont
  {Liu}},\ }\bibfield  {title} {\enquote {\bibinfo {title} {Subsystem
  eigenstate thermalization hypothesis},}\ }\href {\doibase
  10.1103/PhysRevE.97.012140} {\bibfield  {journal} {\bibinfo  {journal} {Phys.
  Rev. E}\ }\textbf {\bibinfo {volume} {97}},\ \bibinfo {pages} {012140}
  (\bibinfo {year} {2018})}\BibitemShut {NoStop}%
\bibitem [{\citenamefont {Hunter-Jones}\ \emph {et~al.}(2018)\citenamefont
  {Hunter-Jones}, \citenamefont {Liu},\ and\ \citenamefont
  {Zhou}}]{HunterJones_Zhou_JHEP2018_eth_syk}%
  \BibitemOpen
  \bibfield  {author} {\bibinfo {author} {\bibfnamefont {Nicholas}\
  \bibnamefont {Hunter-Jones}}, \bibinfo {author} {\bibfnamefont {Junyu}\
  \bibnamefont {Liu}}, \ and\ \bibinfo {author} {\bibfnamefont {Yehao}\
  \bibnamefont {Zhou}},\ }\bibfield  {title} {\enquote {\bibinfo {title} {On
  thermalization in the syk and supersymmetric syk models},}\ }\href
  {https://doi.org/10.1007/JHEP02(2018)142} {\bibfield  {journal} {\bibinfo
  {journal} {Journal of High Energy Physics}\ }\textbf {\bibinfo {volume}
  {2018}},\ \bibinfo {pages} {142} (\bibinfo {year} {2018})}\BibitemShut
  {NoStop}%
\bibitem [{\citenamefont {Jansen}\ \emph {et~al.}(2019)\citenamefont {Jansen},
  \citenamefont {Stolpp}, \citenamefont {Vidmar},\ and\ \citenamefont
  {Heidrich-Meisner}}]{Vidmar_HeidrichM_PRB2019_HolsteinpolaronETH}%
  \BibitemOpen
  \bibfield  {author} {\bibinfo {author} {\bibfnamefont {David}\ \bibnamefont
  {Jansen}}, \bibinfo {author} {\bibfnamefont {Jan}\ \bibnamefont {Stolpp}},
  \bibinfo {author} {\bibfnamefont {Lev}\ \bibnamefont {Vidmar}}, \ and\
  \bibinfo {author} {\bibfnamefont {Fabian}\ \bibnamefont {Heidrich-Meisner}},\
  }\bibfield  {title} {\enquote {\bibinfo {title} {Eigenstate thermalization
  and quantum chaos in the holstein polaron model},}\ }\href {\doibase
  10.1103/PhysRevB.99.155130} {\bibfield  {journal} {\bibinfo  {journal} {Phys.
  Rev. B}\ }\textbf {\bibinfo {volume} {99}},\ \bibinfo {pages} {155130}
  (\bibinfo {year} {2019})}\BibitemShut {NoStop}%
\bibitem [{\citenamefont {Khaymovich}\ \emph {et~al.}(2019)\citenamefont
  {Khaymovich}, \citenamefont {Haque},\ and\ \citenamefont
  {McClarty}}]{Khaymovich_Haque_McClarty_PRL2019}%
  \BibitemOpen
  \bibfield  {author} {\bibinfo {author} {\bibfnamefont {Ivan~M.}\ \bibnamefont
  {Khaymovich}}, \bibinfo {author} {\bibfnamefont {Masudul}\ \bibnamefont
  {Haque}}, \ and\ \bibinfo {author} {\bibfnamefont {Paul~A.}\ \bibnamefont
  {McClarty}},\ }\bibfield  {title} {\enquote {\bibinfo {title} {Eigenstate
  thermalization, random matrix theory, and behemoths},}\ }\href {\doibase
  10.1103/PhysRevLett.122.070601} {\bibfield  {journal} {\bibinfo  {journal}
  {Phys. Rev. Lett.}\ }\textbf {\bibinfo {volume} {122}},\ \bibinfo {pages}
  {070601} (\bibinfo {year} {2019})}\BibitemShut {NoStop}%
\bibitem [{\citenamefont {Hamazaki}\ and\ \citenamefont
  {Ueda}(2018)}]{Ramazaki_Ueda_PRL2018_mostfewbody}%
  \BibitemOpen
  \bibfield  {author} {\bibinfo {author} {\bibfnamefont {Ryusuke}\ \bibnamefont
  {Hamazaki}}\ and\ \bibinfo {author} {\bibfnamefont {Masahito}\ \bibnamefont
  {Ueda}},\ }\bibfield  {title} {\enquote {\bibinfo {title} {Atypicality of
  most few-body observables},}\ }\href {\doibase
  10.1103/PhysRevLett.120.080603} {\bibfield  {journal} {\bibinfo  {journal}
  {Phys. Rev. Lett.}\ }\textbf {\bibinfo {volume} {120}},\ \bibinfo {pages}
  {080603} (\bibinfo {year} {2018})}\BibitemShut {NoStop}%
\bibitem [{\citenamefont {Khemani}\ \emph {et~al.}(2019)\citenamefont
  {Khemani}, \citenamefont {Laumann},\ and\ \citenamefont
  {Chandran}}]{Khemani_Laumann_Chandran_PRB2019_rydberg}%
  \BibitemOpen
  \bibfield  {author} {\bibinfo {author} {\bibfnamefont {Vedika}\ \bibnamefont
  {Khemani}}, \bibinfo {author} {\bibfnamefont {Chris~R.}\ \bibnamefont
  {Laumann}}, \ and\ \bibinfo {author} {\bibfnamefont {Anushya}\ \bibnamefont
  {Chandran}},\ }\bibfield  {title} {\enquote {\bibinfo {title} {Signatures of
  integrability in the dynamics of rydberg-blockaded chains},}\ }\href
  {\doibase 10.1103/PhysRevB.99.161101} {\bibfield  {journal} {\bibinfo
  {journal} {Phys. Rev. B}\ }\textbf {\bibinfo {volume} {99}},\ \bibinfo
  {pages} {161101} (\bibinfo {year} {2019})}\BibitemShut {NoStop}%
\bibitem [{\citenamefont {Brenes}\ \emph {et~al.}(2020)\citenamefont {Brenes},
  \citenamefont {LeBlond}, \citenamefont {Goold},\ and\ \citenamefont
  {Rigol}}]{Goold_Rigol_PRL2020}%
  \BibitemOpen
  \bibfield  {author} {\bibinfo {author} {\bibfnamefont {Marlon}\ \bibnamefont
  {Brenes}}, \bibinfo {author} {\bibfnamefont {Tyler}\ \bibnamefont {LeBlond}},
  \bibinfo {author} {\bibfnamefont {John}\ \bibnamefont {Goold}}, \ and\
  \bibinfo {author} {\bibfnamefont {Marcos}\ \bibnamefont {Rigol}},\ }\bibfield
   {title} {\enquote {\bibinfo {title} {Eigenstate thermalization in a locally
  perturbed integrable system},}\ }\href {\doibase
  10.1103/PhysRevLett.125.070605} {\bibfield  {journal} {\bibinfo  {journal}
  {Phys. Rev. Lett.}\ }\textbf {\bibinfo {volume} {125}},\ \bibinfo {pages}
  {070605} (\bibinfo {year} {2020})}\BibitemShut {NoStop}%
\bibitem [{\citenamefont
  {Polkovnikov}(2003)}]{Polkovnikov_semiclassics_TWA_PRA2003}%
  \BibitemOpen
  \bibfield  {author} {\bibinfo {author} {\bibfnamefont {Anatoli}\ \bibnamefont
  {Polkovnikov}},\ }\bibfield  {title} {\enquote {\bibinfo {title} {Quantum
  corrections to the dynamics of interacting bosons: Beyond the truncated
  wigner approximation},}\ }\href {\doibase 10.1103/PhysRevA.68.053604}
  {\bibfield  {journal} {\bibinfo  {journal} {Phys. Rev. A}\ }\textbf {\bibinfo
  {volume} {68}},\ \bibinfo {pages} {053604} (\bibinfo {year}
  {2003})}\BibitemShut {NoStop}%
\bibitem [{\citenamefont {Mahmud}\ \emph {et~al.}(2005)\citenamefont {Mahmud},
  \citenamefont {Perry},\ and\ \citenamefont
  {Reinhardt}}]{Mahmud_Reinhardt_BHdimer_semiclassical_PRA2005}%
  \BibitemOpen
  \bibfield  {author} {\bibinfo {author} {\bibfnamefont {Khan~W.}\ \bibnamefont
  {Mahmud}}, \bibinfo {author} {\bibfnamefont {Heidi}\ \bibnamefont {Perry}}, \
  and\ \bibinfo {author} {\bibfnamefont {William~P.}\ \bibnamefont
  {Reinhardt}},\ }\bibfield  {title} {\enquote {\bibinfo {title} {Quantum
  phase-space picture of bose-einstein condensates in a double well},}\ }\href
  {\doibase 10.1103/PhysRevA.71.023615} {\bibfield  {journal} {\bibinfo
  {journal} {Phys. Rev. A}\ }\textbf {\bibinfo {volume} {71}},\ \bibinfo
  {pages} {023615} (\bibinfo {year} {2005})}\BibitemShut {NoStop}%
\bibitem [{\citenamefont {Hiller}\ \emph {et~al.}(2006)\citenamefont {Hiller},
  \citenamefont {Kottos},\ and\ \citenamefont
  {Geisel}}]{Hiller_Kottos_Geisel_PRA2006}%
  \BibitemOpen
  \bibfield  {author} {\bibinfo {author} {\bibfnamefont {Moritz}\ \bibnamefont
  {Hiller}}, \bibinfo {author} {\bibfnamefont {Tsampikos}\ \bibnamefont
  {Kottos}}, \ and\ \bibinfo {author} {\bibfnamefont {T.}~\bibnamefont
  {Geisel}},\ }\bibfield  {title} {\enquote {\bibinfo {title} {Complexity in
  parametric bose-hubbard hamiltonians and structural analysis of
  eigenstates},}\ }\href {\doibase 10.1103/PhysRevA.73.061604} {\bibfield
  {journal} {\bibinfo  {journal} {Phys. Rev. A}\ }\textbf {\bibinfo {volume}
  {73}},\ \bibinfo {pages} {061604} (\bibinfo {year} {2006})}\BibitemShut
  {NoStop}%
\bibitem [{\citenamefont {Mossmann}\ and\ \citenamefont
  {Jung}(2006)}]{Mossmann_Jung_PRA2006_semiclassical_BHtrimer}%
  \BibitemOpen
  \bibfield  {author} {\bibinfo {author} {\bibfnamefont {S.}~\bibnamefont
  {Mossmann}}\ and\ \bibinfo {author} {\bibfnamefont {C.}~\bibnamefont
  {Jung}},\ }\bibfield  {title} {\enquote {\bibinfo {title} {Semiclassical
  approach to bose-einstein condensates in a triple well potential},}\ }\href
  {\doibase 10.1103/PhysRevA.74.033601} {\bibfield  {journal} {\bibinfo
  {journal} {Phys. Rev. A}\ }\textbf {\bibinfo {volume} {74}},\ \bibinfo
  {pages} {033601} (\bibinfo {year} {2006})}\BibitemShut {NoStop}%
\bibitem [{\citenamefont {Graefe}\ and\ \citenamefont
  {Korsch}(2007)}]{Graefe_Korsch_semiclassical_PRA2007}%
  \BibitemOpen
  \bibfield  {author} {\bibinfo {author} {\bibfnamefont {E.~M.}\ \bibnamefont
  {Graefe}}\ and\ \bibinfo {author} {\bibfnamefont {H.~J.}\ \bibnamefont
  {Korsch}},\ }\bibfield  {title} {\enquote {\bibinfo {title} {Semiclassical
  quantization of an $n$-particle bose-hubbard model},}\ }\href {\doibase
  10.1103/PhysRevA.76.032116} {\bibfield  {journal} {\bibinfo  {journal} {Phys.
  Rev. A}\ }\textbf {\bibinfo {volume} {76}},\ \bibinfo {pages} {032116}
  (\bibinfo {year} {2007})}\BibitemShut {NoStop}%
\bibitem [{\citenamefont {Trimborn}\ \emph {et~al.}(2009)\citenamefont
  {Trimborn}, \citenamefont {Witthaut},\ and\ \citenamefont
  {Korsch}}]{Trimborn_Witthaut_Korsch_PRA2009}%
  \BibitemOpen
  \bibfield  {author} {\bibinfo {author} {\bibfnamefont {F.}~\bibnamefont
  {Trimborn}}, \bibinfo {author} {\bibfnamefont {D.}~\bibnamefont {Witthaut}},
  \ and\ \bibinfo {author} {\bibfnamefont {H.~J.}\ \bibnamefont {Korsch}},\
  }\bibfield  {title} {\enquote {\bibinfo {title} {Beyond mean-field dynamics
  of small bose-hubbard systems based on the number-conserving phase-space
  approach},}\ }\href {\doibase 10.1103/PhysRevA.79.013608} {\bibfield
  {journal} {\bibinfo  {journal} {Phys. Rev. A}\ }\textbf {\bibinfo {volume}
  {79}},\ \bibinfo {pages} {013608} (\bibinfo {year} {2009})}\BibitemShut
  {NoStop}%
\bibitem [{\citenamefont
  {Polkovnikov}(2010)}]{Polkovnikov_AnnPhys2010_PhaseSpace}%
  \BibitemOpen
  \bibfield  {author} {\bibinfo {author} {\bibfnamefont {Anatoli}\ \bibnamefont
  {Polkovnikov}},\ }\bibfield  {title} {\enquote {\bibinfo {title} {Phase space
  representation of quantum dynamics},}\ }\href {\doibase
  10.1016/j.aop.2010.02.006} {\bibfield  {journal} {\bibinfo  {journal} {Annals
  of Physics}\ }\textbf {\bibinfo {volume} {325}},\ \bibinfo {pages} {1790}
  (\bibinfo {year} {2010})}\BibitemShut {NoStop}%
\bibitem [{\citenamefont {Paw\l{}owski}\ \emph {et~al.}(2011)\citenamefont
  {Paw\l{}owski}, \citenamefont {Zi\ifmmode~\acute{n}\else \'{n}\fi{}},
  \citenamefont {Rz{a}\ifmmode~\dot{z}\else \.{z}\fi{}ewski},\ and\
  \citenamefont {Trippenbach}}]{Trippenbach_BHdimer_PRA2011}%
  \BibitemOpen
  \bibfield  {author} {\bibinfo {author} {\bibfnamefont {Krzysztof}\
  \bibnamefont {Paw\l{}owski}}, \bibinfo {author} {\bibfnamefont {Pawe\l{}}\
  \bibnamefont {Zi\ifmmode~\acute{n}\else \'{n}\fi{}}}, \bibinfo {author}
  {\bibfnamefont {Kazimierz}\ \bibnamefont {Rz{a}\ifmmode~\dot{z}\else
  \.{z}\fi{}ewski}}, \ and\ \bibinfo {author} {\bibfnamefont {Marek}\
  \bibnamefont {Trippenbach}},\ }\bibfield  {title} {\enquote {\bibinfo {title}
  {Revivals in an attractive bose-einstein condensate in a double-well
  potential and their decoherence},}\ }\href {\doibase
  10.1103/PhysRevA.83.033606} {\bibfield  {journal} {\bibinfo  {journal} {Phys.
  Rev. A}\ }\textbf {\bibinfo {volume} {83}},\ \bibinfo {pages} {033606}
  (\bibinfo {year} {2011})}\BibitemShut {NoStop}%
\bibitem [{\citenamefont {Simon}\ and\ \citenamefont
  {Strunz}(2012)}]{Simon_Strunz_PRA2012}%
  \BibitemOpen
  \bibfield  {author} {\bibinfo {author} {\bibfnamefont {Lena}\ \bibnamefont
  {Simon}}\ and\ \bibinfo {author} {\bibfnamefont {Walter~T.}\ \bibnamefont
  {Strunz}},\ }\bibfield  {title} {\enquote {\bibinfo {title} {Analytical
  results for josephson dynamics of ultracold bosons},}\ }\href {\doibase
  10.1103/PhysRevA.86.053625} {\bibfield  {journal} {\bibinfo  {journal} {Phys.
  Rev. A}\ }\textbf {\bibinfo {volume} {86}},\ \bibinfo {pages} {053625}
  (\bibinfo {year} {2012})}\BibitemShut {NoStop}%
\bibitem [{\citenamefont {Simon}\ and\ \citenamefont
  {Strunz}(2014)}]{Simon_Strunz_PRA2014}%
  \BibitemOpen
  \bibfield  {author} {\bibinfo {author} {\bibfnamefont {Lena}\ \bibnamefont
  {Simon}}\ and\ \bibinfo {author} {\bibfnamefont {Walter~T.}\ \bibnamefont
  {Strunz}},\ }\bibfield  {title} {\enquote {\bibinfo {title} {Time-dependent
  semiclassics for ultracold bosons},}\ }\href {\doibase
  10.1103/PhysRevA.89.052112} {\bibfield  {journal} {\bibinfo  {journal} {Phys.
  Rev. A}\ }\textbf {\bibinfo {volume} {89}},\ \bibinfo {pages} {052112}
  (\bibinfo {year} {2014})}\BibitemShut {NoStop}%
\bibitem [{\citenamefont {Veksler}\ and\ \citenamefont
  {Fishman}(2015)}]{Veksler_Fishman_semiclassical_NJP2015}%
  \BibitemOpen
  \bibfield  {author} {\bibinfo {author} {\bibfnamefont {Hagar}\ \bibnamefont
  {Veksler}}\ and\ \bibinfo {author} {\bibfnamefont {Shmuel}\ \bibnamefont
  {Fishman}},\ }\bibfield  {title} {\enquote {\bibinfo {title} {Semiclassical
  analysis of bose{\textendash}hubbard dynamics},}\ }\href {\doibase
  10.1088/1367-2630/17/5/053030} {\bibfield  {journal} {\bibinfo  {journal}
  {New Journal of Physics}\ }\textbf {\bibinfo {volume} {17}},\ \bibinfo
  {pages} {053030} (\bibinfo {year} {2015})}\BibitemShut {NoStop}%
\bibitem [{\citenamefont {Engl}\ \emph {et~al.}(2015)\citenamefont {Engl},
  \citenamefont {Urbina},\ and\ \citenamefont
  {Richter}}]{Engl_Urbina_Richter_PRE2015}%
  \BibitemOpen
  \bibfield  {author} {\bibinfo {author} {\bibfnamefont {Thomas}\ \bibnamefont
  {Engl}}, \bibinfo {author} {\bibfnamefont {Juan~Diego}\ \bibnamefont
  {Urbina}}, \ and\ \bibinfo {author} {\bibfnamefont {Klaus}\ \bibnamefont
  {Richter}},\ }\bibfield  {title} {\enquote {\bibinfo {title} {Periodic
  mean-field solutions and the spectra of discrete bosonic fields: Trace
  formula for bose-hubbard models},}\ }\href {\doibase
  10.1103/PhysRevE.92.062907} {\bibfield  {journal} {\bibinfo  {journal} {Phys.
  Rev. E}\ }\textbf {\bibinfo {volume} {92}},\ \bibinfo {pages} {062907}
  (\bibinfo {year} {2015})}\BibitemShut {NoStop}%
\bibitem [{\citenamefont {Engl}\ \emph {et~al.}(2016)\citenamefont {Engl},
  \citenamefont {Urbina},\ and\ \citenamefont
  {Richter}}]{Engl_Urbina_KRichter_PhilTrans2016}%
  \BibitemOpen
  \bibfield  {author} {\bibinfo {author} {\bibfnamefont {Thomas}\ \bibnamefont
  {Engl}}, \bibinfo {author} {\bibfnamefont {Juan~Diego}\ \bibnamefont
  {Urbina}}, \ and\ \bibinfo {author} {\bibfnamefont {Klaus}\ \bibnamefont
  {Richter}},\ }\bibfield  {title} {\enquote {\bibinfo {title} {The
  semiclassical propagator in fock space: dynamical echo and many-body
  interference},}\ }\href {\doibase 10.1098/rsta.2015.0159} {\bibfield
  {journal} {\bibinfo  {journal} {Philosophical Transactions of the Royal
  Society A: Mathematical, Physical and Engineering Sciences}\ }\textbf
  {\bibinfo {volume} {374}},\ \bibinfo {pages} {20150159} (\bibinfo {year}
  {2016})}\BibitemShut {NoStop}%
\bibitem [{\citenamefont {Ray}\ \emph {et~al.}(2016{\natexlab{a}})\citenamefont
  {Ray}, \citenamefont {Ostmann}, \citenamefont {Simon}, \citenamefont
  {Grossmann},\ and\ \citenamefont
  {Strunz}}]{Grossmann_Strunz_semiclassical_JPA2016}%
  \BibitemOpen
  \bibfield  {author} {\bibinfo {author} {\bibfnamefont {Shouryya}\
  \bibnamefont {Ray}}, \bibinfo {author} {\bibfnamefont {Paula}\ \bibnamefont
  {Ostmann}}, \bibinfo {author} {\bibfnamefont {Lena}\ \bibnamefont {Simon}},
  \bibinfo {author} {\bibfnamefont {Frank}\ \bibnamefont {Grossmann}}, \ and\
  \bibinfo {author} {\bibfnamefont {Walter~T.}\ \bibnamefont {Strunz}},\
  }\bibfield  {title} {\enquote {\bibinfo {title} {Dynamics of interacting
  bosons using the herman-kluk semiclassical initial value representation},}\
  }\href {https://iopscience.iop.org/article/10.1088/1751-8113/49/16/165303}
  {\bibfield  {journal} {\bibinfo  {journal} {Journal of Physics A:
  Mathematical and Theoretical}\ }\textbf {\bibinfo {volume} {49}},\ \bibinfo
  {pages} {165303} (\bibinfo {year} {2016}{\natexlab{a}})}\BibitemShut
  {NoStop}%
\bibitem [{\citenamefont {Dubertrand}\ and\ \citenamefont
  {M\"uller}(2016)}]{Dubertrand_Mueller_NJP2016_semiclassics_spectralstat}%
  \BibitemOpen
  \bibfield  {author} {\bibinfo {author} {\bibfnamefont {R{\'{e}}my}\
  \bibnamefont {Dubertrand}}\ and\ \bibinfo {author} {\bibfnamefont
  {Sebastian}\ \bibnamefont {M\"uller}},\ }\bibfield  {title} {\enquote
  {\bibinfo {title} {Spectral statistics of chaotic many-body systems},}\
  }\href {\doibase 10.1088/1367-2630/18/3/033009} {\bibfield  {journal}
  {\bibinfo  {journal} {New Journal of Physics}\ }\textbf {\bibinfo {volume}
  {18}},\ \bibinfo {pages} {033009} (\bibinfo {year} {2016})}\BibitemShut
  {NoStop}%
\bibitem [{\citenamefont {Tomsovic}\ \emph {et~al.}(2018)\citenamefont
  {Tomsovic}, \citenamefont {Schlagheck}, \citenamefont {Ullmo}, \citenamefont
  {Urbina},\ and\ \citenamefont
  {Richter}}]{Tomsovic_Schlagheck_Richter_PRA2018_postEhrenfest}%
  \BibitemOpen
  \bibfield  {author} {\bibinfo {author} {\bibfnamefont {Steven}\ \bibnamefont
  {Tomsovic}}, \bibinfo {author} {\bibfnamefont {Peter}\ \bibnamefont
  {Schlagheck}}, \bibinfo {author} {\bibfnamefont {Denis}\ \bibnamefont
  {Ullmo}}, \bibinfo {author} {\bibfnamefont {Juan-Diego}\ \bibnamefont
  {Urbina}}, \ and\ \bibinfo {author} {\bibfnamefont {Klaus}\ \bibnamefont
  {Richter}},\ }\bibfield  {title} {\enquote {\bibinfo {title} {Post-ehrenfest
  many-body quantum interferences in ultracold atoms far out of equilibrium},}\
  }\href {\doibase 10.1103/PhysRevA.97.061606} {\bibfield  {journal} {\bibinfo
  {journal} {Phys. Rev. A}\ }\textbf {\bibinfo {volume} {97}},\ \bibinfo
  {pages} {061606} (\bibinfo {year} {2018})}\BibitemShut {NoStop}%
\bibitem [{\citenamefont {Tomsovic}(2018)}]{Tomsovic_PRE2018_saddle_BH}%
  \BibitemOpen
  \bibfield  {author} {\bibinfo {author} {\bibfnamefont {Steven}\ \bibnamefont
  {Tomsovic}},\ }\bibfield  {title} {\enquote {\bibinfo {title} {Complex saddle
  trajectories for multidimensional quantum wave packet and coherent state
  propagation: Application to a many-body system},}\ }\href {\doibase
  10.1103/PhysRevE.98.023301} {\bibfield  {journal} {\bibinfo  {journal} {Phys.
  Rev. E}\ }\textbf {\bibinfo {volume} {98}},\ \bibinfo {pages} {023301}
  (\bibinfo {year} {2018})}\BibitemShut {NoStop}%
\bibitem [{\citenamefont {Rammensee}\ \emph {et~al.}(2018)\citenamefont
  {Rammensee}, \citenamefont {Urbina},\ and\ \citenamefont
  {Richter}}]{Rammensee_PRL2018}%
  \BibitemOpen
  \bibfield  {author} {\bibinfo {author} {\bibfnamefont {Josef}\ \bibnamefont
  {Rammensee}}, \bibinfo {author} {\bibfnamefont {Juan~Diego}\ \bibnamefont
  {Urbina}}, \ and\ \bibinfo {author} {\bibfnamefont {Klaus}\ \bibnamefont
  {Richter}},\ }\bibfield  {title} {\enquote {\bibinfo {title} {{Many-Body
  Quantum Interference and the Saturation of Out-of-Time-Order Correlators}},}\
  }\href {\doibase 10.1103/PhysRevLett.121.124101} {\bibfield  {journal}
  {\bibinfo  {journal} {Physical Review Letters}\ }\textbf {\bibinfo {volume}
  {121}},\ \bibinfo {pages} {124101} (\bibinfo {year} {2018})}\BibitemShut
  {NoStop}%
\bibitem [{\citenamefont {Kidd}\ \emph {et~al.}(2019)\citenamefont {Kidd},
  \citenamefont {Olsen},\ and\ \citenamefont
  {Corney}}]{Corney_PRA2019_BHdimer}%
  \BibitemOpen
  \bibfield  {author} {\bibinfo {author} {\bibfnamefont {R.~A.}\ \bibnamefont
  {Kidd}}, \bibinfo {author} {\bibfnamefont {M.~K.}\ \bibnamefont {Olsen}}, \
  and\ \bibinfo {author} {\bibfnamefont {J.~F.}\ \bibnamefont {Corney}},\
  }\bibfield  {title} {\enquote {\bibinfo {title} {Quantum chaos in a
  bose-hubbard dimer with modulated tunneling},}\ }\href {\doibase
  10.1103/PhysRevA.100.013625} {\bibfield  {journal} {\bibinfo  {journal}
  {Phys. Rev. A}\ }\textbf {\bibinfo {volume} {100}},\ \bibinfo {pages}
  {013625} (\bibinfo {year} {2019})}\BibitemShut {NoStop}%
\bibitem [{\citenamefont {Schlagheck}\ \emph {et~al.}(2019)\citenamefont
  {Schlagheck}, \citenamefont {Ullmo}, \citenamefont {Urbina}, \citenamefont
  {Richter},\ and\ \citenamefont {Tomsovic}}]{Schlagheck_PRL2019}%
  \BibitemOpen
  \bibfield  {author} {\bibinfo {author} {\bibfnamefont {Peter}\ \bibnamefont
  {Schlagheck}}, \bibinfo {author} {\bibfnamefont {Denis}\ \bibnamefont
  {Ullmo}}, \bibinfo {author} {\bibfnamefont {Juan~Diego}\ \bibnamefont
  {Urbina}}, \bibinfo {author} {\bibfnamefont {Klaus}\ \bibnamefont {Richter}},
  \ and\ \bibinfo {author} {\bibfnamefont {Steven}\ \bibnamefont {Tomsovic}},\
  }\bibfield  {title} {\enquote {\bibinfo {title} {Enhancement of many-body
  quantum interference in chaotic bosonic systems: The role of symmetry and
  dynamics},}\ }\href {\doibase 10.1103/PhysRevLett.123.215302} {\bibfield
  {journal} {\bibinfo  {journal} {Phys. Rev. Lett.}\ }\textbf {\bibinfo
  {volume} {123}},\ \bibinfo {pages} {215302} (\bibinfo {year}
  {2019})}\BibitemShut {NoStop}%
\bibitem [{\citenamefont {Nemoto}\ \emph {et~al.}(2000)\citenamefont {Nemoto},
  \citenamefont {Holmes}, \citenamefont {Milburn},\ and\ \citenamefont
  {Munro}}]{Milburn_Munro_PRA2000_BHtrimer}%
  \BibitemOpen
  \bibfield  {author} {\bibinfo {author} {\bibfnamefont {K.}~\bibnamefont
  {Nemoto}}, \bibinfo {author} {\bibfnamefont {C.~A.}\ \bibnamefont {Holmes}},
  \bibinfo {author} {\bibfnamefont {G.~J.}\ \bibnamefont {Milburn}}, \ and\
  \bibinfo {author} {\bibfnamefont {W.~J.}\ \bibnamefont {Munro}},\ }\bibfield
  {title} {\enquote {\bibinfo {title} {Quantum dynamics of three coupled atomic
  bose-einstein condensates},}\ }\href {\doibase 10.1103/PhysRevA.63.013604}
  {\bibfield  {journal} {\bibinfo  {journal} {Phys. Rev. A}\ }\textbf {\bibinfo
  {volume} {63}},\ \bibinfo {pages} {013604} (\bibinfo {year}
  {2000})}\BibitemShut {NoStop}%
\bibitem [{\citenamefont {Weiss}\ and\ \citenamefont
  {Teichmann}(2008)}]{Weiss_Teichmann_PRL2008_BHdimer_quantumclassical}%
  \BibitemOpen
  \bibfield  {author} {\bibinfo {author} {\bibfnamefont {Christoph}\
  \bibnamefont {Weiss}}\ and\ \bibinfo {author} {\bibfnamefont {Niklas}\
  \bibnamefont {Teichmann}},\ }\bibfield  {title} {\enquote {\bibinfo {title}
  {Differences between mean-field dynamics and $n$-particle quantum dynamics as
  a signature of entanglement},}\ }\href {\doibase
  10.1103/PhysRevLett.100.140408} {\bibfield  {journal} {\bibinfo  {journal}
  {Phys. Rev. Lett.}\ }\textbf {\bibinfo {volume} {100}},\ \bibinfo {pages}
  {140408} (\bibinfo {year} {2008})}\BibitemShut {NoStop}%
\bibitem [{\citenamefont {Viscondi}\ and\ \citenamefont
  {Furuya}(2011)}]{Viscondi_Furuya__JPhysA2011_BHtrimer}%
  \BibitemOpen
  \bibfield  {author} {\bibinfo {author} {\bibfnamefont {Thiago~F}\
  \bibnamefont {Viscondi}}\ and\ \bibinfo {author} {\bibfnamefont
  {K}~\bibnamefont {Furuya}},\ }\bibfield  {title} {\enquote {\bibinfo {title}
  {Dynamics of a bose{\textendash}einstein condensate in a symmetric
  triple-well trap},}\ }\href {\doibase 10.1088/1751-8113/44/17/175301}
  {\bibfield  {journal} {\bibinfo  {journal} {Journal of Physics A:
  Mathematical and Theoretical}\ }\textbf {\bibinfo {volume} {44}},\ \bibinfo
  {pages} {175301} (\bibinfo {year} {2011})}\BibitemShut {NoStop}%
\bibitem [{\citenamefont {Gertjerenken}\ and\ \citenamefont
  {Weiss}(2013)}]{Gertjerenken_Weiss_BHdimer_quantumclassical_PRA2013}%
  \BibitemOpen
  \bibfield  {author} {\bibinfo {author} {\bibfnamefont {Bettina}\ \bibnamefont
  {Gertjerenken}}\ and\ \bibinfo {author} {\bibfnamefont {Christoph}\
  \bibnamefont {Weiss}},\ }\bibfield  {title} {\enquote {\bibinfo {title}
  {Beyond-mean-field behavior of large bose-einstein condensates in double-well
  potentials},}\ }\href {\doibase 10.1103/PhysRevA.88.033608} {\bibfield
  {journal} {\bibinfo  {journal} {Phys. Rev. A}\ }\textbf {\bibinfo {volume}
  {88}},\ \bibinfo {pages} {033608} (\bibinfo {year} {2013})}\BibitemShut
  {NoStop}%
\bibitem [{\citenamefont {Kolovsky}(2016)}]{Kolovsky_IntJModPhys2016}%
  \BibitemOpen
  \bibfield  {author} {\bibinfo {author} {\bibfnamefont {Andrey~R.}\
  \bibnamefont {Kolovsky}},\ }\bibfield  {title} {\enquote {\bibinfo {title}
  {Bose-hubbard hamiltonian: Quantum chaos approach},}\ }\href {\doibase
  10.1142/S0217979216300097} {\bibfield  {journal} {\bibinfo  {journal} {Int.\
  J.~Mod.\ Phys.\ B}\ }\textbf {\bibinfo {volume} {30}},\ \bibinfo {pages}
  {1630009} (\bibinfo {year} {2016})}\BibitemShut {NoStop}%
\bibitem [{\citenamefont {Heinisch}\ and\ \citenamefont
  {Holthaus}(2016)}]{Heinisch_Holthaus_ZNforschungA2016}%
  \BibitemOpen
  \bibfield  {author} {\bibinfo {author} {\bibfnamefont {Christoph}\
  \bibnamefont {Heinisch}}\ and\ \bibinfo {author} {\bibfnamefont {Martin}\
  \bibnamefont {Holthaus}},\ }\bibfield  {title} {\enquote {\bibinfo {title}
  {Entropy production within a pulsed bose–einstein condensate},}\ }\href
  {\doibase 10.1515/zna-2016-0073} {\bibfield  {journal} {\bibinfo  {journal}
  {Zeitschrift für Naturforschung A}\ }\textbf {\bibinfo {volume} {71}},\
  \bibinfo {pages} {875} (\bibinfo {year} {2016})}\BibitemShut {NoStop}%
\bibitem [{\citenamefont {Rautenberg}\ and\ \citenamefont
  {G\"arttner}(2020)}]{Rautenberg_Gaertner_PRA2020}%
  \BibitemOpen
  \bibfield  {author} {\bibinfo {author} {\bibfnamefont {Michael}\ \bibnamefont
  {Rautenberg}}\ and\ \bibinfo {author} {\bibfnamefont {Martin}\ \bibnamefont
  {G\"arttner}},\ }\bibfield  {title} {\enquote {\bibinfo {title} {Classical
  and quantum chaos in a three-mode bosonic system},}\ }\href {\doibase
  10.1103/PhysRevA.101.053604} {\bibfield  {journal} {\bibinfo  {journal}
  {Phys. Rev. A}\ }\textbf {\bibinfo {volume} {101}},\ \bibinfo {pages}
  {053604} (\bibinfo {year} {2020})}\BibitemShut {NoStop}%
\bibitem [{\citenamefont {Hiller}\ \emph {et~al.}(2009)\citenamefont {Hiller},
  \citenamefont {Kottos},\ and\ \citenamefont
  {Geisel}}]{Hiller_Kottos_Geisel_BHtrimer_PRA2009}%
  \BibitemOpen
  \bibfield  {author} {\bibinfo {author} {\bibfnamefont {Moritz}\ \bibnamefont
  {Hiller}}, \bibinfo {author} {\bibfnamefont {Tsampikos}\ \bibnamefont
  {Kottos}}, \ and\ \bibinfo {author} {\bibfnamefont {Theo}\ \bibnamefont
  {Geisel}},\ }\bibfield  {title} {\enquote {\bibinfo {title} {Wave-packet
  dynamics in energy space of a chaotic trimeric bose-hubbard system},}\ }\href
  {\doibase 10.1103/PhysRevA.79.023621} {\bibfield  {journal} {\bibinfo
  {journal} {Phys. Rev. A}\ }\textbf {\bibinfo {volume} {79}},\ \bibinfo
  {pages} {023621} (\bibinfo {year} {2009})}\BibitemShut {NoStop}%
\bibitem [{\citenamefont {Arwas}\ \emph {et~al.}(2014)\citenamefont {Arwas},
  \citenamefont {Vardi},\ and\ \citenamefont
  {Cohen}}]{Vardi_DoronCohen_PRA2014_BHtrimer}%
  \BibitemOpen
  \bibfield  {author} {\bibinfo {author} {\bibfnamefont {Geva}\ \bibnamefont
  {Arwas}}, \bibinfo {author} {\bibfnamefont {Amichay}\ \bibnamefont {Vardi}},
  \ and\ \bibinfo {author} {\bibfnamefont {Doron}\ \bibnamefont {Cohen}},\
  }\bibfield  {title} {\enquote {\bibinfo {title} {Triangular bose-hubbard
  trimer as a minimal model for a superfluid circuit},}\ }\href {\doibase
  10.1103/PhysRevA.89.013601} {\bibfield  {journal} {\bibinfo  {journal} {Phys.
  Rev. A}\ }\textbf {\bibinfo {volume} {89}},\ \bibinfo {pages} {013601}
  (\bibinfo {year} {2014})}\BibitemShut {NoStop}%
\bibitem [{\citenamefont {Han}\ and\ \citenamefont
  {Wu}(2016)}]{Han_Wu_PRA2016_BHtrimer_ehrenfest}%
  \BibitemOpen
  \bibfield  {author} {\bibinfo {author} {\bibfnamefont {Xizhi}\ \bibnamefont
  {Han}}\ and\ \bibinfo {author} {\bibfnamefont {Biao}\ \bibnamefont {Wu}},\
  }\bibfield  {title} {\enquote {\bibinfo {title} {Ehrenfest breakdown of the
  mean-field dynamics of bose gases},}\ }\href {\doibase
  10.1103/PhysRevA.93.023621} {\bibfield  {journal} {\bibinfo  {journal} {Phys.
  Rev. A}\ }\textbf {\bibinfo {volume} {93}},\ \bibinfo {pages} {023621}
  (\bibinfo {year} {2016})}\BibitemShut {NoStop}%
\bibitem [{\citenamefont {B\"urkle}\ and\ \citenamefont
  {Anglin}(2019)}]{Buerkle_Anglin_4siteBH_PRA2019}%
  \BibitemOpen
  \bibfield  {author} {\bibinfo {author} {\bibfnamefont {R.}~\bibnamefont
  {B\"urkle}}\ and\ \bibinfo {author} {\bibfnamefont {J.~R.}\ \bibnamefont
  {Anglin}},\ }\bibfield  {title} {\enquote {\bibinfo {title} {Threshold
  coupling strength for equilibration between small systems},}\ }\href
  {\doibase 10.1103/PhysRevA.99.063617} {\bibfield  {journal} {\bibinfo
  {journal} {Phys. Rev. A}\ }\textbf {\bibinfo {volume} {99}},\ \bibinfo
  {pages} {063617} (\bibinfo {year} {2019})}\BibitemShut {NoStop}%
\bibitem [{\citenamefont {Bychek}\ \emph {et~al.}(2020)\citenamefont {Bychek},
  \citenamefont {Muraev}, \citenamefont {Maksimov}, \citenamefont {Bulgakov},\
  and\ \citenamefont {Kolovsky}}]{Kolovsky_russians_AIPConf2020}%
  \BibitemOpen
  \bibfield  {author} {\bibinfo {author} {\bibfnamefont {Anna~A.}\ \bibnamefont
  {Bychek}}, \bibinfo {author} {\bibfnamefont {Pavel~S.}\ \bibnamefont
  {Muraev}}, \bibinfo {author} {\bibfnamefont {Dmitrii~N.}\ \bibnamefont
  {Maksimov}}, \bibinfo {author} {\bibfnamefont {Evgeny~N.}\ \bibnamefont
  {Bulgakov}}, \ and\ \bibinfo {author} {\bibfnamefont {Andrey~R.}\
  \bibnamefont {Kolovsky}},\ }\bibfield  {title} {\enquote {\bibinfo {title}
  {Chaotic and regular dynamics in the three-site bose-hubbard model},}\ }\href
  {\doibase 10.1063/5.0011540} {\bibfield  {journal} {\bibinfo  {journal} {AIP
  Conference Proceedings}\ }\textbf {\bibinfo {volume} {2241}},\ \bibinfo
  {pages} {020007} (\bibinfo {year} {2020})}\BibitemShut {NoStop}%
\bibitem [{\citenamefont {Kolovsky}\ and\ \citenamefont
  {Buchleitner}(2004)}]{Kolovsky_Buchleitner_EPL2004_BHchaos}%
  \BibitemOpen
  \bibfield  {author} {\bibinfo {author} {\bibfnamefont {A.~R}\ \bibnamefont
  {Kolovsky}}\ and\ \bibinfo {author} {\bibfnamefont {A}~\bibnamefont
  {Buchleitner}},\ }\bibfield  {title} {\enquote {\bibinfo {title} {Quantum
  chaos in the bose-hubbard model},}\ }\href {\doibase
  10.1209/epl/i2004-10265-7} {\bibfield  {journal} {\bibinfo  {journal}
  {Europhysics Letters ({EPL})}\ }\textbf {\bibinfo {volume} {68}},\ \bibinfo
  {pages} {632} (\bibinfo {year} {2004})}\BibitemShut {NoStop}%
\bibitem [{\citenamefont {Kollath}\ \emph {et~al.}(2010)\citenamefont
  {Kollath}, \citenamefont {Roux}, \citenamefont {Biroli},\ and\ \citenamefont
  {Läuchli}}]{Kollath_Roux_Biroli_Laeuchli_JSM2010}%
  \BibitemOpen
  \bibfield  {author} {\bibinfo {author} {\bibfnamefont {Corinna}\ \bibnamefont
  {Kollath}}, \bibinfo {author} {\bibfnamefont {Guillaume}\ \bibnamefont
  {Roux}}, \bibinfo {author} {\bibfnamefont {Giulio}\ \bibnamefont {Biroli}}, \
  and\ \bibinfo {author} {\bibfnamefont {Andreas~M}\ \bibnamefont {Läuchli}},\
  }\bibfield  {title} {\enquote {\bibinfo {title} {Statistical properties of
  the spectrum of the extended bose{\textendash}hubbard model},}\ }\href
  {\doibase 10.1088/1742-5468/2010/08/p08011} {\bibfield  {journal} {\bibinfo
  {journal} {Journal of Statistical Mechanics: Theory and Experiment}\ }\textbf
  {\bibinfo {volume} {2010}},\ \bibinfo {pages} {P08011} (\bibinfo {year}
  {2010})}\BibitemShut {NoStop}%
\bibitem [{\citenamefont {de~la Cruz}\ \emph {et~al.}(2020)\citenamefont {de~la
  Cruz}, \citenamefont {Lerma-Hern\'andez},\ and\ \citenamefont
  {Hirsch}}]{Hirsch_PRE2020_quantumchaos_BH}%
  \BibitemOpen
  \bibfield  {author} {\bibinfo {author} {\bibfnamefont {Javier}\ \bibnamefont
  {de~la Cruz}}, \bibinfo {author} {\bibfnamefont {Sergio}\ \bibnamefont
  {Lerma-Hern\'andez}}, \ and\ \bibinfo {author} {\bibfnamefont {Jorge~G.}\
  \bibnamefont {Hirsch}},\ }\bibfield  {title} {\enquote {\bibinfo {title}
  {Quantum chaos in a system with high degree of symmetries},}\ }\href
  {\doibase 10.1103/PhysRevE.102.032208} {\bibfield  {journal} {\bibinfo
  {journal} {Phys. Rev. E}\ }\textbf {\bibinfo {volume} {102}},\ \bibinfo
  {pages} {032208} (\bibinfo {year} {2020})}\BibitemShut {NoStop}%
\bibitem [{\citenamefont {Pausch}\ \emph {et~al.}(2020)\citenamefont {Pausch},
  \citenamefont {Carnio}, \citenamefont {Rodr{\'\i}guez},\ and\ \citenamefont
  {Buchleitner}}]{Buchleitner_arxiv2020_BHchaos}%
  \BibitemOpen
  \bibfield  {author} {\bibinfo {author} {\bibfnamefont {Lukas}\ \bibnamefont
  {Pausch}}, \bibinfo {author} {\bibfnamefont {Edoardo~G}\ \bibnamefont
  {Carnio}}, \bibinfo {author} {\bibfnamefont {Alberto}\ \bibnamefont
  {Rodr{\'\i}guez}}, \ and\ \bibinfo {author} {\bibfnamefont {Andreas}\
  \bibnamefont {Buchleitner}},\ }\bibfield  {title} {\enquote {\bibinfo {title}
  {Chaos and ergodicity across the energy spectrum of interacting bosons},}\
  }\href {https://arxiv.org/abs/2009.05295} {\bibfield  {journal} {\bibinfo
  {journal} {arXiv preprint arXiv:2009.05295}\ } (\bibinfo {year}
  {2020})}\BibitemShut {NoStop}%
\bibitem [{\citenamefont {Beugeling}\ \emph {et~al.}(2018)\citenamefont
  {Beugeling}, \citenamefont {B\"acker}, \citenamefont {Moessner},\ and\
  \citenamefont {Haque}}]{Beugeling_coefficients_PRE2018}%
  \BibitemOpen
  \bibfield  {author} {\bibinfo {author} {\bibfnamefont {Wouter}\ \bibnamefont
  {Beugeling}}, \bibinfo {author} {\bibfnamefont {Arnd}\ \bibnamefont
  {B\"acker}}, \bibinfo {author} {\bibfnamefont {Roderich}\ \bibnamefont
  {Moessner}}, \ and\ \bibinfo {author} {\bibfnamefont {Masudul}\ \bibnamefont
  {Haque}},\ }\bibfield  {title} {\enquote {\bibinfo {title} {Statistical
  properties of eigenstate amplitudes in complex quantum systems},}\ }\href
  {\doibase 10.1103/PhysRevE.98.022204} {\bibfield  {journal} {\bibinfo
  {journal} {Phys. Rev. E}\ }\textbf {\bibinfo {volume} {98}},\ \bibinfo
  {pages} {022204} (\bibinfo {year} {2018})}\BibitemShut {NoStop}%
\bibitem [{\citenamefont {B{\"{a}}cker}\ \emph {et~al.}(2019)\citenamefont
  {B{\"{a}}cker}, \citenamefont {Haque},\ and\ \citenamefont
  {Khaymovich}}]{Baecker_Haque_Khaymovich_PRE2019}%
  \BibitemOpen
  \bibfield  {author} {\bibinfo {author} {\bibfnamefont {Arnd}\ \bibnamefont
  {B{\"{a}}cker}}, \bibinfo {author} {\bibfnamefont {Masudul}\ \bibnamefont
  {Haque}}, \ and\ \bibinfo {author} {\bibfnamefont {Ivan~M}\ \bibnamefont
  {Khaymovich}},\ }\bibfield  {title} {\enquote {\bibinfo {title}
  {{Multifractal dimensions for random matrices, chaotic quantum maps, and
  many-body systems}},}\ }\href {\doibase 10.1103/PhysRevE.100.032117}
  {\bibfield  {journal} {\bibinfo  {journal} {Phys. Rev. E}\ }\textbf {\bibinfo
  {volume} {100}},\ \bibinfo {pages} {032117} (\bibinfo {year}
  {2019})}\BibitemShut {NoStop}%
\bibitem [{\citenamefont {Luitz}\ \emph {et~al.}(2020)\citenamefont {Luitz},
  \citenamefont {Khaymovich},\ and\ \citenamefont
  {Lev}}]{Luitz_Khaymovich_BarLev_multifrac_SciPost2020}%
  \BibitemOpen
  \bibfield  {author} {\bibinfo {author} {\bibfnamefont {David~J.}\
  \bibnamefont {Luitz}}, \bibinfo {author} {\bibfnamefont {Ivan~M.}\
  \bibnamefont {Khaymovich}}, \ and\ \bibinfo {author} {\bibfnamefont
  {Yevgeny~Bar}\ \bibnamefont {Lev}},\ }\bibfield  {title} {\enquote {\bibinfo
  {title} {{Multifractality and its role in anomalous transport in the
  disordered XXZ spin-chain}},}\ }\href {\doibase
  10.21468/SciPostPhysCore.2.2.006} {\bibfield  {journal} {\bibinfo  {journal}
  {SciPost Phys. Core}\ }\textbf {\bibinfo {volume} {2}},\ \bibinfo {pages} {6}
  (\bibinfo {year} {2020})}\BibitemShut {NoStop}%
\bibitem [{\citenamefont {Beugeling}\ \emph
  {et~al.}(2015{\natexlab{b}})\citenamefont {Beugeling}, \citenamefont
  {Andreanov},\ and\ \citenamefont {Haque}}]{Beugeling_entanglement_JSM2015}%
  \BibitemOpen
  \bibfield  {author} {\bibinfo {author} {\bibfnamefont {W}~\bibnamefont
  {Beugeling}}, \bibinfo {author} {\bibfnamefont {A}~\bibnamefont {Andreanov}},
  \ and\ \bibinfo {author} {\bibfnamefont {Masudul}\ \bibnamefont {Haque}},\
  }\bibfield  {title} {\enquote {\bibinfo {title} {Global characteristics of
  all eigenstates of local many-body hamiltonians: participation ratio and
  entanglement entropy},}\ }\href
  {http://stacks.iop.org/1742-5468/2015/i=2/a=P02002} {\bibfield  {journal}
  {\bibinfo  {journal} {Journal of Statistical Mechanics: Theory and
  Experiment}\ }\textbf {\bibinfo {volume} {2015}},\ \bibinfo {pages} {P02002}
  (\bibinfo {year} {2015}{\natexlab{b}})}\BibitemShut {NoStop}%
\bibitem [{\citenamefont {Liu}\ \emph {et~al.}(2018)\citenamefont {Liu},
  \citenamefont {Chen},\ and\ \citenamefont
  {Balents}}]{Balents_SYK_entanglement_PRB2018}%
  \BibitemOpen
  \bibfield  {author} {\bibinfo {author} {\bibfnamefont {Chunxiao}\
  \bibnamefont {Liu}}, \bibinfo {author} {\bibfnamefont {Xiao}\ \bibnamefont
  {Chen}}, \ and\ \bibinfo {author} {\bibfnamefont {Leon}\ \bibnamefont
  {Balents}},\ }\bibfield  {title} {\enquote {\bibinfo {title} {Quantum
  entanglement of the sachdev-ye-kitaev models},}\ }\href {\doibase
  10.1103/PhysRevB.97.245126} {\bibfield  {journal} {\bibinfo  {journal} {Phys.
  Rev. B}\ }\textbf {\bibinfo {volume} {97}},\ \bibinfo {pages} {245126}
  (\bibinfo {year} {2018})}\BibitemShut {NoStop}%
\bibitem [{\citenamefont {Haque}\ \emph {et~al.}(2020)\citenamefont {Haque},
  \citenamefont {McClarty},\ and\ \citenamefont
  {Khaymovich}}]{Haque_McClarty_Khaymovich_entanglement_deviation}%
  \BibitemOpen
  \bibfield  {author} {\bibinfo {author} {\bibfnamefont {Masudul}\ \bibnamefont
  {Haque}}, \bibinfo {author} {\bibfnamefont {Paul~A}\ \bibnamefont
  {McClarty}}, \ and\ \bibinfo {author} {\bibfnamefont {Ivan~M}\ \bibnamefont
  {Khaymovich}},\ }\bibfield  {title} {\enquote {\bibinfo {title} {Entanglement
  of mid-spectrum eigenstates of chaotic many-body systems--deviation from
  random ensembles},}\ }\href {https://arxiv.org/abs/2008.12782} {\bibfield
  {journal} {\bibinfo  {journal} {arXiv preprint arXiv:2008.12782}\ } (\bibinfo
  {year} {2020})}\BibitemShut {NoStop}%
\bibitem [{\citenamefont {Gemmer}\ \emph {et~al.}(2009)\citenamefont {Gemmer},
  \citenamefont {Michel},\ and\ \citenamefont
  {Mahler}}]{Gemmer_book2009_QuantumThermo}%
  \BibitemOpen
  \bibfield  {author} {\bibinfo {author} {\bibfnamefont {J.}~\bibnamefont
  {Gemmer}}, \bibinfo {author} {\bibfnamefont {M.}~\bibnamefont {Michel}}, \
  and\ \bibinfo {author} {\bibfnamefont {G.}~\bibnamefont {Mahler}},\ }\href
  {https://www.springer.com/gp/book/9783540705093} {\emph {\bibinfo {title}
  {Quantum Thermodynamics: Emergence of Thermodynamic Behavior Within Composite
  Quantum Systems}}},\ Lecture Notes in Physics\ (\bibinfo  {publisher}
  {Springer},\ \bibinfo {year} {2009})\BibitemShut {NoStop}%
\bibitem [{\citenamefont {Reimann}(2007)}]{Reimann_PRL2007_typicality}%
  \BibitemOpen
  \bibfield  {author} {\bibinfo {author} {\bibfnamefont {Peter}\ \bibnamefont
  {Reimann}},\ }\bibfield  {title} {\enquote {\bibinfo {title} {Typicality for
  generalized microcanonical ensembles},}\ }\href {\doibase
  10.1103/PhysRevLett.99.160404} {\bibfield  {journal} {\bibinfo  {journal}
  {Phys. Rev. Lett.}\ }\textbf {\bibinfo {volume} {99}},\ \bibinfo {pages}
  {160404} (\bibinfo {year} {2007})}\BibitemShut {NoStop}%
\bibitem [{\citenamefont {Reimann}(2008)}]{Reimann_JStatPhys2008_typicality}%
  \BibitemOpen
  \bibfield  {author} {\bibinfo {author} {\bibfnamefont {Peter}\ \bibnamefont
  {Reimann}},\ }\bibfield  {title} {\enquote {\bibinfo {title} {Typicality of
  pure states randomly sampled according to the gaussian adjusted projected
  measure},}\ }\href {\doibase 10.1007/s10955-008-9576-1} {\bibfield  {journal}
  {\bibinfo  {journal} {Journal of Statistical Physics}\ }\textbf {\bibinfo
  {volume} {132}},\ \bibinfo {pages} {921} (\bibinfo {year}
  {2008})}\BibitemShut {NoStop}%
\bibitem [{\citenamefont {Lloyd}(2013)}]{Lloyd_arxiv2013_thesischapter}%
  \BibitemOpen
  \bibfield  {author} {\bibinfo {author} {\bibfnamefont {Seth}\ \bibnamefont
  {Lloyd}},\ }\bibfield  {title} {\enquote {\bibinfo {title} {Pure state
  quantum statistical mechanics and black holes},}\ }\href
  {https://arxiv.org/abs/1307.0378} {\bibfield  {journal} {\bibinfo  {journal}
  {arXiv preprint arXiv:1307.0378}\ } (\bibinfo {year} {2013})}\BibitemShut
  {NoStop}%
\bibitem [{\citenamefont {Feingold}\ and\ \citenamefont
  {Peres}(1983)}]{Feingold_Peres_PhysicaD1983_coupled_rotators}%
  \BibitemOpen
  \bibfield  {author} {\bibinfo {author} {\bibfnamefont {Mario}\ \bibnamefont
  {Feingold}}\ and\ \bibinfo {author} {\bibfnamefont {Asher}\ \bibnamefont
  {Peres}},\ }\bibfield  {title} {\enquote {\bibinfo {title} {Regular and
  chaotic motion of coupled rotators},}\ }\href {\doibase
  10.1016/0167-2789(83)90282-8} {\bibfield  {journal} {\bibinfo  {journal}
  {Physica D: Nonlinear Phenomena}\ }\textbf {\bibinfo {volume} {9}},\ \bibinfo
  {pages} {433} (\bibinfo {year} {1983})}\BibitemShut {NoStop}%
\bibitem [{\citenamefont {Feingold}\ \emph {et~al.}(1984)\citenamefont
  {Feingold}, \citenamefont {Moiseyev},\ and\ \citenamefont
  {Peres}}]{Feingold_Moiseyev_Peres_PRA1984}%
  \BibitemOpen
  \bibfield  {author} {\bibinfo {author} {\bibfnamefont {Mario}\ \bibnamefont
  {Feingold}}, \bibinfo {author} {\bibfnamefont {Nimrod}\ \bibnamefont
  {Moiseyev}}, \ and\ \bibinfo {author} {\bibfnamefont {Asher}\ \bibnamefont
  {Peres}},\ }\bibfield  {title} {\enquote {\bibinfo {title} {Ergodicity and
  mixing in quantum theory. ii},}\ }\href {\doibase 10.1103/PhysRevA.30.509}
  {\bibfield  {journal} {\bibinfo  {journal} {Phys. Rev. A}\ }\textbf {\bibinfo
  {volume} {30}},\ \bibinfo {pages} {509} (\bibinfo {year} {1984})}\BibitemShut
  {NoStop}%
\bibitem [{\citenamefont {Robb}\ and\ \citenamefont
  {Reichl}(1998)}]{Reichl_PRE1998_twospin}%
  \BibitemOpen
  \bibfield  {author} {\bibinfo {author} {\bibfnamefont {Daniel~T.}\
  \bibnamefont {Robb}}\ and\ \bibinfo {author} {\bibfnamefont {L.~E.}\
  \bibnamefont {Reichl}},\ }\bibfield  {title} {\enquote {\bibinfo {title}
  {Chaos in a two-spin system with applied magnetic field},}\ }\href {\doibase
  10.1103/PhysRevE.57.2458} {\bibfield  {journal} {\bibinfo  {journal} {Phys.
  Rev. E}\ }\textbf {\bibinfo {volume} {57}},\ \bibinfo {pages} {2458--2459}
  (\bibinfo {year} {1998})}\BibitemShut {NoStop}%
\bibitem [{\citenamefont {Emerson}\ and\ \citenamefont
  {Ballentine}(2001)}]{Emerson_Ballentine_PRA2001_twospin}%
  \BibitemOpen
  \bibfield  {author} {\bibinfo {author} {\bibfnamefont {J.}~\bibnamefont
  {Emerson}}\ and\ \bibinfo {author} {\bibfnamefont {L.E.}\ \bibnamefont
  {Ballentine}},\ }\bibfield  {title} {\enquote {\bibinfo {title}
  {Characteristics of quantum-classical correspondence for two interacting
  spins},}\ }\href {\doibase 10.1103/PhysRevA.63.052103} {\bibfield  {journal}
  {\bibinfo  {journal} {Phys. Rev. A}\ }\textbf {\bibinfo {volume} {63}},\
  \bibinfo {pages} {052103} (\bibinfo {year} {2001})}\BibitemShut {NoStop}%
\bibitem [{\citenamefont {Ballentine}(2004)}]{Ballentine_PRA2002_twospins}%
  \BibitemOpen
  \bibfield  {author} {\bibinfo {author} {\bibfnamefont {L.~E.}\ \bibnamefont
  {Ballentine}},\ }\bibfield  {title} {\enquote {\bibinfo {title}
  {Quantum-to-classical limit in a hamiltonian system},}\ }\href {\doibase
  10.1103/PhysRevA.70.032111} {\bibfield  {journal} {\bibinfo  {journal} {Phys.
  Rev. A}\ }\textbf {\bibinfo {volume} {70}},\ \bibinfo {pages} {032111}
  (\bibinfo {year} {2004})}\BibitemShut {NoStop}%
\bibitem [{\citenamefont {Fan}\ \emph {et~al.}(2017)\citenamefont {Fan},
  \citenamefont {Gnutzmann},\ and\ \citenamefont
  {Liang}}]{Fan_Gnutzmann_Liang_PRE2017_FeingoldPeres}%
  \BibitemOpen
  \bibfield  {author} {\bibinfo {author} {\bibfnamefont {Yiyun}\ \bibnamefont
  {Fan}}, \bibinfo {author} {\bibfnamefont {Sven}\ \bibnamefont {Gnutzmann}}, \
  and\ \bibinfo {author} {\bibfnamefont {Yuqi}\ \bibnamefont {Liang}},\
  }\bibfield  {title} {\enquote {\bibinfo {title} {Quantum chaos for
  nonstandard symmetry classes in the feingold-peres model of coupled tops},}\
  }\href {\doibase 10.1103/PhysRevE.96.062207} {\bibfield  {journal} {\bibinfo
  {journal} {Phys. Rev. E}\ }\textbf {\bibinfo {volume} {96}},\ \bibinfo
  {pages} {062207} (\bibinfo {year} {2017})}\BibitemShut {NoStop}%
\bibitem [{\citenamefont {Pappalardi}\ \emph {et~al.}(2018)\citenamefont
  {Pappalardi}, \citenamefont {Russomanno}, \citenamefont {\ifmmode
  \check{Z}\else \v{Z}\fi{}unkovi\ifmmode~\check{c}\else \v{c}\fi{}},
  \citenamefont {Iemini}, \citenamefont {Silva},\ and\ \citenamefont
  {Fazio}}]{Pappalardi_Silva_Fazio_PRB2018_longrange}%
  \BibitemOpen
  \bibfield  {author} {\bibinfo {author} {\bibfnamefont {Silvia}\ \bibnamefont
  {Pappalardi}}, \bibinfo {author} {\bibfnamefont {Angelo}\ \bibnamefont
  {Russomanno}}, \bibinfo {author} {\bibfnamefont {Bojan}\ \bibnamefont
  {\ifmmode \check{Z}\else \v{Z}\fi{}unkovi\ifmmode~\check{c}\else
  \v{c}\fi{}}}, \bibinfo {author} {\bibfnamefont {Fernando}\ \bibnamefont
  {Iemini}}, \bibinfo {author} {\bibfnamefont {Alessandro}\ \bibnamefont
  {Silva}}, \ and\ \bibinfo {author} {\bibfnamefont {Rosario}\ \bibnamefont
  {Fazio}},\ }\bibfield  {title} {\enquote {\bibinfo {title} {Scrambling and
  entanglement spreading in long-range spin chains},}\ }\href {\doibase
  10.1103/PhysRevB.98.134303} {\bibfield  {journal} {\bibinfo  {journal} {Phys.
  Rev. B}\ }\textbf {\bibinfo {volume} {98}},\ \bibinfo {pages} {134303}
  (\bibinfo {year} {2018})}\BibitemShut {NoStop}%
\bibitem [{\citenamefont {Piga}\ \emph {et~al.}(2019)\citenamefont {Piga},
  \citenamefont {Lewenstein},\ and\ \citenamefont
  {Quach}}]{Lewenstein_Quach_PRE2019_entanglement_kickedrotor}%
  \BibitemOpen
  \bibfield  {author} {\bibinfo {author} {\bibfnamefont {Angelo}\ \bibnamefont
  {Piga}}, \bibinfo {author} {\bibfnamefont {Maciej}\ \bibnamefont
  {Lewenstein}}, \ and\ \bibinfo {author} {\bibfnamefont {James~Q.}\
  \bibnamefont {Quach}},\ }\bibfield  {title} {\enquote {\bibinfo {title}
  {Quantum chaos and entanglement in ergodic and nonergodic systems},}\ }\href
  {\doibase 10.1103/PhysRevE.99.032213} {\bibfield  {journal} {\bibinfo
  {journal} {Phys. Rev. E}\ }\textbf {\bibinfo {volume} {99}},\ \bibinfo
  {pages} {032213} (\bibinfo {year} {2019})}\BibitemShut {NoStop}%
\bibitem [{\citenamefont {Lerose}\ and\ \citenamefont
  {Pappalardi}(2020)}]{Pappalardi_PRA2020_semiclassicalsystems}%
  \BibitemOpen
  \bibfield  {author} {\bibinfo {author} {\bibfnamefont {Alessio}\ \bibnamefont
  {Lerose}}\ and\ \bibinfo {author} {\bibfnamefont {Silvia}\ \bibnamefont
  {Pappalardi}},\ }\bibfield  {title} {\enquote {\bibinfo {title} {Bridging
  entanglement dynamics and chaos in semiclassical systems},}\ }\href {\doibase
  10.1103/PhysRevA.102.032404} {\bibfield  {journal} {\bibinfo  {journal}
  {Phys. Rev. A}\ }\textbf {\bibinfo {volume} {102}},\ \bibinfo {pages}
  {032404} (\bibinfo {year} {2020})}\BibitemShut {NoStop}%
\bibitem [{\citenamefont {Xu}\ \emph {et~al.}(2020)\citenamefont {Xu},
  \citenamefont {Scaffidi},\ and\ \citenamefont
  {Cao}}]{Scaffidi_Cao_PRL2020_scrambling}%
  \BibitemOpen
  \bibfield  {author} {\bibinfo {author} {\bibfnamefont {Tianrui}\ \bibnamefont
  {Xu}}, \bibinfo {author} {\bibfnamefont {Thomas}\ \bibnamefont {Scaffidi}}, \
  and\ \bibinfo {author} {\bibfnamefont {Xiangyu}\ \bibnamefont {Cao}},\
  }\bibfield  {title} {\enquote {\bibinfo {title} {Does scrambling equal
  chaos?}}\ }\href {\doibase 10.1103/PhysRevLett.124.140602} {\bibfield
  {journal} {\bibinfo  {journal} {Phys. Rev. Lett.}\ }\textbf {\bibinfo
  {volume} {124}},\ \bibinfo {pages} {140602} (\bibinfo {year}
  {2020})}\BibitemShut {NoStop}%
\bibitem [{\citenamefont {Pappalardi}\ \emph {et~al.}(2020)\citenamefont
  {Pappalardi}, \citenamefont {Polkovnikov},\ and\ \citenamefont
  {Silva}}]{Pappalardi_Polkovnikov_Silva_SciPost2020_SherringtonKirkpatrick}%
  \BibitemOpen
  \bibfield  {author} {\bibinfo {author} {\bibfnamefont {Silvia}\ \bibnamefont
  {Pappalardi}}, \bibinfo {author} {\bibfnamefont {Anatoli}\ \bibnamefont
  {Polkovnikov}}, \ and\ \bibinfo {author} {\bibfnamefont {Alessandro}\
  \bibnamefont {Silva}},\ }\bibfield  {title} {\enquote {\bibinfo {title}
  {{Quantum echo dynamics in the Sherrington-Kirkpatrick model}},}\ }\href
  {\doibase 10.21468/SciPostPhys.9.2.021} {\bibfield  {journal} {\bibinfo
  {journal} {SciPost Phys.}\ }\textbf {\bibinfo {volume} {9}},\ \bibinfo
  {pages} {21} (\bibinfo {year} {2020})}\BibitemShut {NoStop}%
\bibitem [{\citenamefont {Mondal}\ \emph {et~al.}(2020)\citenamefont {Mondal},
  \citenamefont {Sinha},\ and\ \citenamefont
  {Sinha}}]{Mondal_Sinha_PRE2020_twospins}%
  \BibitemOpen
  \bibfield  {author} {\bibinfo {author} {\bibfnamefont {Debabrata}\
  \bibnamefont {Mondal}}, \bibinfo {author} {\bibfnamefont {Sudip}\
  \bibnamefont {Sinha}}, \ and\ \bibinfo {author} {\bibfnamefont
  {S.}~\bibnamefont {Sinha}},\ }\bibfield  {title} {\enquote {\bibinfo {title}
  {Chaos and quantum scars in a coupled top model},}\ }\href {\doibase
  10.1103/PhysRevE.102.020101} {\bibfield  {journal} {\bibinfo  {journal}
  {Phys. Rev. E}\ }\textbf {\bibinfo {volume} {102}},\ \bibinfo {pages}
  {020101} (\bibinfo {year} {2020})}\BibitemShut {NoStop}%
\bibitem [{\citenamefont {Bastarrachea-Magnani}\ \emph
  {et~al.}(2014)\citenamefont {Bastarrachea-Magnani}, \citenamefont
  {Lerma-Hern\'andez},\ and\ \citenamefont
  {Hirsch}}]{Hirsch_PRA2014_Dicke_comparative2}%
  \BibitemOpen
  \bibfield  {author} {\bibinfo {author} {\bibfnamefont {M.~A.}\ \bibnamefont
  {Bastarrachea-Magnani}}, \bibinfo {author} {\bibfnamefont {S.}~\bibnamefont
  {Lerma-Hern\'andez}}, \ and\ \bibinfo {author} {\bibfnamefont {J.~G.}\
  \bibnamefont {Hirsch}},\ }\bibfield  {title} {\enquote {\bibinfo {title}
  {Comparative quantum and semiclassical analysis of atom-field systems. ii.
  chaos and regularity},}\ }\href {\doibase 10.1103/PhysRevA.89.032102}
  {\bibfield  {journal} {\bibinfo  {journal} {Phys. Rev. A}\ }\textbf {\bibinfo
  {volume} {89}},\ \bibinfo {pages} {032102} (\bibinfo {year}
  {2014})}\BibitemShut {NoStop}%
\bibitem [{\citenamefont {Bastarrachea-Magnani}\ \emph
  {et~al.}(2015)\citenamefont {Bastarrachea-Magnani}, \citenamefont {del
  Carpio}, \citenamefont {Lerma-Hern{\'{a}}ndez},\ and\ \citenamefont
  {Hirsch}}]{Hirsch_PhysicaScrpta2015_Dicke}%
  \BibitemOpen
  \bibfield  {author} {\bibinfo {author} {\bibfnamefont {Miguel~Angel}\
  \bibnamefont {Bastarrachea-Magnani}}, \bibinfo {author} {\bibfnamefont
  {Baldemar~L{\'{o}}pez}\ \bibnamefont {del Carpio}}, \bibinfo {author}
  {\bibfnamefont {Sergio}\ \bibnamefont {Lerma-Hern{\'{a}}ndez}}, \ and\
  \bibinfo {author} {\bibfnamefont {Jorge~G}\ \bibnamefont {Hirsch}},\
  }\bibfield  {title} {\enquote {\bibinfo {title} {Chaos in the dicke model:
  quantum and semiclassical analysis},}\ }\href {\doibase
  10.1088/0031-8949/90/6/068015} {\bibfield  {journal} {\bibinfo  {journal}
  {Physica Scripta}\ }\textbf {\bibinfo {volume} {90}},\ \bibinfo {pages}
  {068015} (\bibinfo {year} {2015})}\BibitemShut {NoStop}%
\bibitem [{\citenamefont {Ray}\ \emph {et~al.}(2016{\natexlab{b}})\citenamefont
  {Ray}, \citenamefont {Ghosh},\ and\ \citenamefont
  {Sinha}}]{Ghosh_Sinha_PRE2016_kicked_Dicke}%
  \BibitemOpen
  \bibfield  {author} {\bibinfo {author} {\bibfnamefont {S.}~\bibnamefont
  {Ray}}, \bibinfo {author} {\bibfnamefont {A.}~\bibnamefont {Ghosh}}, \ and\
  \bibinfo {author} {\bibfnamefont {S.}~\bibnamefont {Sinha}},\ }\bibfield
  {title} {\enquote {\bibinfo {title} {Quantum signature of chaos and
  thermalization in the kicked dicke model},}\ }\href {\doibase
  10.1103/PhysRevE.94.032103} {\bibfield  {journal} {\bibinfo  {journal} {Phys.
  Rev. E}\ }\textbf {\bibinfo {volume} {94}},\ \bibinfo {pages} {032103}
  (\bibinfo {year} {2016}{\natexlab{b}})}\BibitemShut {NoStop}%
\bibitem [{\citenamefont {Bastarrachea-Magnani}\ \emph
  {et~al.}(2016)\citenamefont {Bastarrachea-Magnani}, \citenamefont
  {L\'opez-del Carpio}, \citenamefont {Ch\'avez-Carlos}, \citenamefont
  {Lerma-Hern\'andez},\ and\ \citenamefont {Hirsch}}]{Hirsch_PRE2016_Dicke}%
  \BibitemOpen
  \bibfield  {author} {\bibinfo {author} {\bibfnamefont {M.~A.}\ \bibnamefont
  {Bastarrachea-Magnani}}, \bibinfo {author} {\bibfnamefont {B.}~\bibnamefont
  {L\'opez-del Carpio}}, \bibinfo {author} {\bibfnamefont {J.}~\bibnamefont
  {Ch\'avez-Carlos}}, \bibinfo {author} {\bibfnamefont {S.}~\bibnamefont
  {Lerma-Hern\'andez}}, \ and\ \bibinfo {author} {\bibfnamefont {J.~G.}\
  \bibnamefont {Hirsch}},\ }\bibfield  {title} {\enquote {\bibinfo {title}
  {Delocalization and quantum chaos in atom-field systems},}\ }\href {\doibase
  10.1103/PhysRevE.93.022215} {\bibfield  {journal} {\bibinfo  {journal} {Phys.
  Rev. E}\ }\textbf {\bibinfo {volume} {93}},\ \bibinfo {pages} {022215}
  (\bibinfo {year} {2016})}\BibitemShut {NoStop}%
\bibitem [{\citenamefont {Lewis-Swan}\ \emph {et~al.}(2019)\citenamefont
  {Lewis-Swan}, \citenamefont {Safavi-Naini}, \citenamefont {Bollinger},\ and\
  \citenamefont {Rey}}]{Bollinger_AMRey_NatureComm2019_Dicke}%
  \BibitemOpen
  \bibfield  {author} {\bibinfo {author} {\bibfnamefont {RJ}~\bibnamefont
  {Lewis-Swan}}, \bibinfo {author} {\bibfnamefont {Arghavan}\ \bibnamefont
  {Safavi-Naini}}, \bibinfo {author} {\bibfnamefont {John~J}\ \bibnamefont
  {Bollinger}}, \ and\ \bibinfo {author} {\bibfnamefont {Ana~M}\ \bibnamefont
  {Rey}},\ }\bibfield  {title} {\enquote {\bibinfo {title} {Unifying
  scrambling, thermalization and entanglement through measurement of fidelity
  out-of-time-order correlators in the dicke model},}\ }\href {\doibase
  10.1038/s41467-019-09436-y} {\bibfield  {journal} {\bibinfo  {journal}
  {Nature communications}\ }\textbf {\bibinfo {volume} {10}},\ \bibinfo {pages}
  {1--9} (\bibinfo {year} {2019})}\BibitemShut {NoStop}%
\bibitem [{\citenamefont {Ch\'avez-Carlos}\ \emph {et~al.}(2019)\citenamefont
  {Ch\'avez-Carlos}, \citenamefont {L\'opez-del Carpio}, \citenamefont
  {Bastarrachea-Magnani}, \citenamefont {Str\'ansk\'y}, \citenamefont
  {Lerma-Hern\'andez}, \citenamefont {Santos},\ and\ \citenamefont
  {Hirsch}}]{LSantos_Hirsch_PRL2019_Dicke_Lyapunovs}%
  \BibitemOpen
  \bibfield  {author} {\bibinfo {author} {\bibfnamefont {Jorge}\ \bibnamefont
  {Ch\'avez-Carlos}}, \bibinfo {author} {\bibfnamefont {B.}~\bibnamefont
  {L\'opez-del Carpio}}, \bibinfo {author} {\bibfnamefont {Miguel~A.}\
  \bibnamefont {Bastarrachea-Magnani}}, \bibinfo {author} {\bibfnamefont
  {Pavel}\ \bibnamefont {Str\'ansk\'y}}, \bibinfo {author} {\bibfnamefont
  {Sergio}\ \bibnamefont {Lerma-Hern\'andez}}, \bibinfo {author} {\bibfnamefont
  {Lea~F.}\ \bibnamefont {Santos}}, \ and\ \bibinfo {author} {\bibfnamefont
  {Jorge~G.}\ \bibnamefont {Hirsch}},\ }\bibfield  {title} {\enquote {\bibinfo
  {title} {Quantum and classical lyapunov exponents in atom-field interaction
  systems},}\ }\href {\doibase 10.1103/PhysRevLett.122.024101} {\bibfield
  {journal} {\bibinfo  {journal} {Phys. Rev. Lett.}\ }\textbf {\bibinfo
  {volume} {122}},\ \bibinfo {pages} {024101} (\bibinfo {year}
  {2019})}\BibitemShut {NoStop}%
\bibitem [{\citenamefont {Wang}\ and\ \citenamefont
  {Robnik}(2020)}]{Robnik_PRE2020_Dicke}%
  \BibitemOpen
  \bibfield  {author} {\bibinfo {author} {\bibfnamefont {Qian}\ \bibnamefont
  {Wang}}\ and\ \bibinfo {author} {\bibfnamefont {Marko}\ \bibnamefont
  {Robnik}},\ }\bibfield  {title} {\enquote {\bibinfo {title} {Statistical
  properties of the localization measure of chaotic eigenstates in the dicke
  model},}\ }\href {\doibase 10.1103/PhysRevE.102.032212} {\bibfield  {journal}
  {\bibinfo  {journal} {Phys. Rev. E}\ }\textbf {\bibinfo {volume} {102}},\
  \bibinfo {pages} {032212} (\bibinfo {year} {2020})}\BibitemShut {NoStop}%
\bibitem [{\citenamefont {Villase{\~{n}}or}\ \emph {et~al.}(2020)\citenamefont
  {Villase{\~{n}}or}, \citenamefont {Pilatowsky-Cameo}, \citenamefont
  {Bastarrachea-Magnani}, \citenamefont {Lerma-Hern{\'{a}}ndez}, \citenamefont
  {Santos},\ and\ \citenamefont {Hirsch}}]{LSantos_Hirsch_NJP2020_Dicke}%
  \BibitemOpen
  \bibfield  {author} {\bibinfo {author} {\bibfnamefont {D}~\bibnamefont
  {Villase{\~{n}}or}}, \bibinfo {author} {\bibfnamefont {S}~\bibnamefont
  {Pilatowsky-Cameo}}, \bibinfo {author} {\bibfnamefont {M~A}\ \bibnamefont
  {Bastarrachea-Magnani}}, \bibinfo {author} {\bibfnamefont {S}~\bibnamefont
  {Lerma-Hern{\'{a}}ndez}}, \bibinfo {author} {\bibfnamefont {L~F}\
  \bibnamefont {Santos}}, \ and\ \bibinfo {author} {\bibfnamefont {J~G}\
  \bibnamefont {Hirsch}},\ }\bibfield  {title} {\enquote {\bibinfo {title}
  {Quantum vs classical dynamics in a spin-boson system: manifestations of
  spectral correlations and scarring},}\ }\href {\doibase
  10.1088/1367-2630/ab8ef8} {\bibfield  {journal} {\bibinfo  {journal} {New
  Journal of Physics}\ }\textbf {\bibinfo {volume} {22}},\ \bibinfo {pages}
  {063036} (\bibinfo {year} {2020})}\BibitemShut {NoStop}%
\bibitem [{\citenamefont {Pilatowsky-Cameo}\ \emph {et~al.}(2020)\citenamefont
  {Pilatowsky-Cameo}, \citenamefont {Ch\'avez-Carlos}, \citenamefont
  {Bastarrachea-Magnani}, \citenamefont {Str\'ansk\'y}, \citenamefont
  {Lerma-Hern\'andez}, \citenamefont {Santos},\ and\ \citenamefont
  {Hirsch}}]{LSantos_Hirsch_PRE2020}%
  \BibitemOpen
  \bibfield  {author} {\bibinfo {author} {\bibfnamefont {Sa\'ul}\ \bibnamefont
  {Pilatowsky-Cameo}}, \bibinfo {author} {\bibfnamefont {Jorge}\ \bibnamefont
  {Ch\'avez-Carlos}}, \bibinfo {author} {\bibfnamefont {Miguel~A.}\
  \bibnamefont {Bastarrachea-Magnani}}, \bibinfo {author} {\bibfnamefont
  {Pavel}\ \bibnamefont {Str\'ansk\'y}}, \bibinfo {author} {\bibfnamefont
  {Sergio}\ \bibnamefont {Lerma-Hern\'andez}}, \bibinfo {author} {\bibfnamefont
  {Lea~F.}\ \bibnamefont {Santos}}, \ and\ \bibinfo {author} {\bibfnamefont
  {Jorge~G.}\ \bibnamefont {Hirsch}},\ }\bibfield  {title} {\enquote {\bibinfo
  {title} {Positive quantum lyapunov exponents in experimental systems with a
  regular classical limit},}\ }\href {\doibase 10.1103/PhysRevE.101.010202}
  {\bibfield  {journal} {\bibinfo  {journal} {Phys. Rev. E}\ }\textbf {\bibinfo
  {volume} {101}},\ \bibinfo {pages} {010202} (\bibinfo {year}
  {2020})}\BibitemShut {NoStop}%
\bibitem [{\citenamefont {Sinha}\ and\ \citenamefont
  {Sinha}(2020)}]{Sinha_PRL2020_BJJ_bosonicmode}%
  \BibitemOpen
  \bibfield  {author} {\bibinfo {author} {\bibfnamefont {Sudip}\ \bibnamefont
  {Sinha}}\ and\ \bibinfo {author} {\bibfnamefont {S.}~\bibnamefont {Sinha}},\
  }\bibfield  {title} {\enquote {\bibinfo {title} {Chaos and quantum scars in
  bose-josephson junction coupled to a bosonic mode},}\ }\href {\doibase
  10.1103/PhysRevLett.125.134101} {\bibfield  {journal} {\bibinfo  {journal}
  {Phys. Rev. Lett.}\ }\textbf {\bibinfo {volume} {125}},\ \bibinfo {pages}
  {134101} (\bibinfo {year} {2020})}\BibitemShut {NoStop}%
\bibitem [{\citenamefont {Oganesyan}\ and\ \citenamefont
  {Huse}(2007)}]{Oganesyan_Huse_PRB2007}%
  \BibitemOpen
  \bibfield  {author} {\bibinfo {author} {\bibfnamefont {Vadim}\ \bibnamefont
  {Oganesyan}}\ and\ \bibinfo {author} {\bibfnamefont {David~A.}\ \bibnamefont
  {Huse}},\ }\bibfield  {title} {\enquote {\bibinfo {title} {Localization of
  interacting fermions at high temperature},}\ }\href {\doibase
  10.1103/PhysRevB.75.155111} {\bibfield  {journal} {\bibinfo  {journal} {Phys.
  Rev. B}\ }\textbf {\bibinfo {volume} {75}},\ \bibinfo {pages} {155111}
  (\bibinfo {year} {2007})}\BibitemShut {NoStop}%
\bibitem [{\citenamefont {Atas}\ \emph {et~al.}(2013)\citenamefont {Atas},
  \citenamefont {Bogomolny}, \citenamefont {Giraud},\ and\ \citenamefont
  {Roux}}]{Atas_Bogomolny_Roux_PRL2013}%
  \BibitemOpen
  \bibfield  {author} {\bibinfo {author} {\bibfnamefont {Y.~Y.}\ \bibnamefont
  {Atas}}, \bibinfo {author} {\bibfnamefont {E.}~\bibnamefont {Bogomolny}},
  \bibinfo {author} {\bibfnamefont {O.}~\bibnamefont {Giraud}}, \ and\ \bibinfo
  {author} {\bibfnamefont {G.}~\bibnamefont {Roux}},\ }\bibfield  {title}
  {\enquote {\bibinfo {title} {Distribution of the ratio of consecutive level
  spacings in random matrix ensembles},}\ }\href {\doibase
  10.1103/PhysRevLett.110.084101} {\bibfield  {journal} {\bibinfo  {journal}
  {Phys. Rev. Lett.}\ }\textbf {\bibinfo {volume} {110}},\ \bibinfo {pages}
  {084101} (\bibinfo {year} {2013})}\BibitemShut {NoStop}%
\end{thebibliography}%

\end{document}